\documentclass[preprint]{elsarticle}

\makeatletter
\def\ps@pprintTitle{
 \let\@oddhead\@empty
 \let\@evenhead\@empty
 \def\@oddfoot{\centerline{\thepage}}
 \let\@evenfoot\@oddfoot}
\makeatother

\usepackage{hyperref}
\usepackage{amssymb,amsmath,amsthm,pifont,subcaption}

\theoremstyle{definition}
\newtheorem{theorem}{Theorem}[section]
\newtheorem{lemma}[theorem]{Lemma}
\newtheorem{corollary}[theorem]{Corollary}
\newtheorem{definition}[theorem]{Definition}
\newtheorem{example}[theorem]{Example}

\usepackage{float,proof,scalerel,tabto,tikz-cd}

\usepackage{tabularx,environ}
\makeatletter
\newcommand{\probinstance}[1]{\gdef\@probinstance{#1}}%
\newcommand{\probquestion}[1]{\gdef\@probquestion{#1}}
\NewEnviron{prob}{
  \BODY
  \par\addvspace{.25\baselineskip}
  \noindent
  \begin{tabularx}{\textwidth}{@{\hspace{\parindent}} l X c}
    \textbf{Instance:} & \@probinstance \\
    \textbf{Question:} & \@probquestion
  \end{tabularx}
  \par\addvspace{.25\baselineskip}
}
\makeatother

\begin{document}

\begin{frontmatter}

\title{Confluence up to Garbage in Graph Transformation}

\author[1]{Graham Campbell\fnref{fn1}}
\ead{g.j.campbell2@newcastle.ac.uk}

\author[2]{Detlef Plump}
\ead{detlef.plump@york.ac.uk}

\address[1]{School of Mathematics, Statistics and Physics, Newcastle University, Newcastle upon Tyne, United Kingdom}
\address[2]{Department of Computer Science, University of York, York, United Kingdom}

\fntext[fn1]{Supported by a Vacation Internship and a Doctoral Training Grant No. (2281162) from the Engineering and Physical Sciences Research Council (EPSRC) in the UK, while at University of York and Newcastle University, respectively.}

\begin{abstract}
The transformation of graphs and graph-like structures is ubiquitous in computer science. When a system is described by graph-transformation rules, it is often desirable that the rules are both terminating and confluent so that rule applications in an arbitrary order produce unique resulting graphs. However, there are application scenarios where the rules are not globally confluent but confluent on a subclass of graphs that are of interest. In other words, non-resolvable conflicts can only occur on graphs that are considered as ``garbage''. In this paper, we introduce the notion of confluence up to garbage and generalise Plump's critical pair lemma for double-pushout graph transformation, providing a sufficient condition for confluence up to garbage by non-garbage critical pair analysis. We apply our results in two case studies about efficient language recognition: we present backtracking-free graph reduction systems which recognise a class of flow diagrams and a class of labelled series-parallel graphs, respectively. Both systems are non-confluent but confluent up to garbage. We also give a critical pair condition for subcommutativity up to garbage which, together with closedness, implies confluence up to garbage even in non-terminating systems.
\end{abstract}


\end{frontmatter}

\section{Introduction} \label{sec:introduction}

Rule-based graph transformation and graph grammars date back to the late 1960s. The best developed theoretical framework is the so-called double-pushout (DPO) approach to graph transformation \cite{Ehrig-Pfender-Schneider73a,Ehrig-Ehrig-Prange-Taentzer06a}. When specifying systems in computer science by DPO graph transformation rules, it is often desirable that the rules are both terminating and confluent so that rule applications in an arbitrary order produce unique resulting graphs. For example, \cite{Bakewell-Plump-Runciman03a} contains 23 case studies of confluent and terminating graph reductions systems which specify pointer structures such as cyclic lists, balanced binary trees and red-black trees. Confluence is also important in the context of evaluating functional expressions by graph reduction, see for example \cite{Plump99b}.

However, there are application scenarios where the rules are not confluent but confluent on a subclass of graphs that are of interest. In other words, non-resolvable conflicts can only occur on graphs that are considered as ``garbage''. An example is the class of so-called extended flow diagrams discussed in Subsection \ref{subsec:flow_diagrams}. The reduction rules for these graphs give rise to ten critical pairs, nine of which are strongly joinable. But a single pair is not joinable and hence the rules are not confluent. The non-joinable pair represents a conflict in graphs containing a type of cycle that cannot occur in extended flow diagrams. Hence these graphs can be considered as garbage which in this case consists of all graphs that are not extended flow diagrams.  

In this paper, we introduce the notions of confluence up to garbage and termination up to garbage in graph transformation. We generalise Plump's Critical Pair Lemma \cite{Plump93b,Plump05a} and Newmann's Lemma \cite{Newman42a} and thereby allow to check confluence up to garbage via non-garbage critical pair analysis. We apply our results to language recognition by backtracking-free graph reduction, showing how to establish that a graph language can be decided by a system which is confluent up to garbage. We present two case studies with backtracking-free graph reduction systems which recognise a class of labelled series-parallel graphs and a class flow diagrams, respectively. Both systems are non-confluent but confluent up to garbage.

This paper an extended version of the ICGT 2020 paper \cite{Campbell-Plump20a}, which was in turn partly developed from Campbell's BSc Thesis \cite{Campbell19a}. In this paper, we are able to afford a proper treatment of isomorphism of critical pairs, and provide more detail and examples than previously, throughout. Section \ref{sec:confluence} is now presented at the level of abstract reduction systems, with a new subsection the relationship between confluence up to garbage and confluence modulo garbage. In Section \ref{sec:pair_lemma} we additionally discuss generation of non-garbage critical pairs, giving sufficient conditions for this process to be completely automatic. We also explicitly discuss joinability checking for pairs of direct derivations. Section \ref{sec:languages} has been revised with less confusing terminology and more details of the critical pair analyses included. Section \ref{sec:subcommuativity} is entirely new, looking at subcommutativity up to garbage. We give a second version of our generalised critical pair lemma in this setting, showing how critical pair analysis can be used to check for subcommutativity up to garbage. This property implies confluence up to garbage even in non-terminating systems, provided that non-garbage is closed under reduction. This is relevant for applications because confluence up to garbage in such systems implies that non-garbage graphs can be reduced to at most one irreducible graph.

\section{Preliminaries} \label{sec:preliminaries}

We review some terminology for binary relations, the DPO approach to graph transformation, graph languages, and confluence checking.

\subsection{Abstract Reduction Systems}

An \textit{abstract reduction system} (ARS) is a pair \((\mathcal{A}, \rightarrow)\) where \(\mathcal{A}\) is a \textit{class} and \(\rightarrow\) a \textit{binary relation} on \(\mathcal{A}\). Write \(\xrightarrow{=}\) for the reflexive closure of \(\rightarrow\), \(\xrightarrow{+}\) for the transitive closure, and \(\xrightarrow{*}\) for the reflexive transitive closure. Given \(x, x_i, y, y_1, y_2 \in \mathcal{A}\) (\(i \geq 0\)), we say that:

\begin{enumerate}
\item \(y\) is a \textit{successor} to \(x\) if \(x \xrightarrow{+} y\), and a \textit{direct successor} if \(x \rightarrow y\);
\item \(x\) and \(y\) are \textit{joinable} if there is a \(z\) such that \(x \xrightarrow{*} z \xleftarrow{*} y\);
\item \(x\) and \(y\) are \textit{subcommutative} if there is a \(z\) such that \(x \xrightarrow{=} z \xleftarrow{=} y\);
\item \(\rightarrow\) is \textit{confluent} if \(y_1 \xleftarrow{*} x \xrightarrow{*} y_2\) implies \(y_1, y_2\) are joinable;
\item \(\rightarrow\) is \textit{locally confluent} if \(y_1 \leftarrow x \rightarrow y_2\) implies \(y_1, y_2\) are joinable;
\item \(\rightarrow\) is \textit{subcommutative} if \(y_1 \leftarrow x \rightarrow y_2\) implies \(y_1, y_2\) are subcommutative;
\item \(\rightarrow\) is \textit{terminating} if there is no infinite sequence \(x_0 \rightarrow x_1 \rightarrow \dots\).
\end{enumerate}

The principle of \textit{Noetherian Induction} is:
\noindent
\begin{align*}
\infer{\forall x \in \mathcal{A}, P(x)}{\forall x \in \mathcal{A}, (\forall y \in \mathcal{A}, x \xrightarrow{+} y \Rightarrow P(y)) \Rightarrow P(x)}
\end{align*}

\begin{theorem}[Noetherian Induction \cite{Baader-Nipkow98a}] \label{thm:ninduct}
Given an ARS \((\mathcal{A}, \rightarrow)\), the principle of \textit{Noetherian induction} holds if and only if \(\rightarrow\) is \textit{terminating}.
\end{theorem}

\begin{lemma}
Subcommutativity implies confluence, and confluence implies local confluence.
\end{lemma}

\begin{theorem}[Newman's Lemma \cite{Newman42a}] \label{thm:newmanlem}
Let \(\rightarrow\) be a terminating relation. Then \(\rightarrow\) is confluent if and only if it is locally confluent.
\end{theorem}

\subsection{Labelled Graphs and Morphisms}

We will be working with \textit{directed labelled graphs} \cite{Habel-Muller-Plump01a}. A \textit{signature} is a pair \(\Sigma = (\Sigma_V, \Sigma_E)\) of finite sets of node and edge labels from which a graph can be labelled. A \textit{graph} over \(\Sigma\) is a tuple \(G = (V, E, s, t, l, m)\) where \(V\) is a finite set of nodes, \(E\) is a finite set of edges, \(s: E \to V\) is the source function, \(t: E \to V\) is the target function, \(l: V \to \Sigma_V\) is the node labelling function, and \(m: E \to \Sigma_E\) is the edge labelling function. We may write the components of \(G\) as $V_G$, $E_G$, $s_G$, $t_G$, $l_G$, and $m_G$.

A \textit{graph morphism} \(g: G \to H\) is a pair \(g = (g_V, g_E)\) of functions \(g_V: V_G \to V_H\) and \(g_E: E_G \to E_H\) such that \(g_V \circ s_G = s_H \circ g_E\), \(g_V \circ t_G = t_H \circ g_E\), \(l_G = l_H \circ g_V\) and \(m_G =  m_H \circ g_E\). We say \(g\) is \textit{injective} (\textit{surjective}, \textit{bijective}) if both functions \(g_V\) and \(g_E\) are. A graph \(H\) is a \textit{subgraph} of \(G\), denoted by \(H \subseteq G\), if there exists an \textit{inclusion} \(i\colon H \to G\) with $i(x) = x$ for all items $x$.

It is well known that graphs and morphisms over a fixed signature \(\Sigma\) form a category. Graph morphisms are injective (surjective, bijective) if and only if they are monomorphisms (epimorphisms, isomorphisms) in the categorical sense. We denote by \(\mathcal{G}(\Sigma)\) the class of all graphs over \(\Sigma\).

\subsection{Double-Pushout Graph Transformation} \label{sec:dpograph}

A \emph{rule} is a pair of inclusions \(r = \langle L \leftarrow K \rightarrow R \rangle\), where \(L\) is the left-hand side (LHS), \(K\) the interface, and \(R\) the right-hand side (RHS).  A \emph{match} of $r$ in a graph \(G\) is an injective morphism \(L \to G\). An application of rule r to \(G\) with match $g\colon L \to G$ requires to construct two pushouts as in Figure \ref{fig:directderivation}. We write \(G \Rightarrow_{r,g} H\) for this application and call the diagram in Figure \ref{fig:directderivation} a \emph{direct derivation}. 

\begin{figure}[!ht]
\centering
\begin{tikzpicture}
  \node[align=center] (a) at (0.0,1.5) {$L$};
  \node[align=center] (b) at (1.5,1.5) {$K$};
  \node[align=center] (c) at (3.0,1.5) {$R$};
  \node[align=center] (d) at (0.0,0) {$G$};
  \node[align=center] (e) at (1.5,0) {$D$};
  \node[align=center] (f) at (3.0,0) {$H$};

    \draw (b) edge[->] (a)
          (b) edge[->] (c)
          (e) edge[->] (d)
          (e) edge[->] (f)
          (a) edge[->] node[right] {$g$} (d)
          (b) edge[->] node[right] {$d$} (e)
          (c) edge[->] node[right] {$h$} (f);
\end{tikzpicture}
\caption{A direct derivation}
\label{fig:directderivation}
\end{figure}

Given $r$ and the match $g\colon L \to G$, the direct derivation of Figure \ref{fig:directderivation} exists if and only if the \emph{dangling condition} is satisfied: nodes in $g(L-K)$ must not be incident to edges in $G - g(L)$. In this case the graphs $D$ and $H$ are determined uniquely up to isomorphism \cite{Ehrig-Ehrig-Prange-Taentzer06a}. We call the injective morphism \(h\) the \emph{comatch} of the rule application.

Given a set of rules $\mathcal{R}$, we write \(G \Rightarrow_{\mathcal{R}} H\) if \(H\) is obtained from $G$ by applying any of the rules from \(\mathcal{R}\). Note that $\Rightarrow_{\mathcal{R}}$ is isomorphism-compatible. We write \(G \Rightarrow_{\mathcal{R}}^{+} H\) if \(H\) is obtained from \(G\) by one or more rule applications, and \(G \Rightarrow_{\mathcal{R}}^{*} H\) if \(G \cong H\) or \(G \Rightarrow_{\mathcal{R}}^{+} H\). We can view a graph transformation system as ARS \((\mathcal{G}(\Sigma),\Rightarrow_{\mathcal{R}})\), giving us the definition of local confluence, confluence, subcommutativity, and termination for graph transformation systems.

\subsection{Graph Languages}

A \emph{graph language} is an isomorphism-closed class of graphs, and the size of a graph language is defined to be the number of non-isomorphic graphs in the language. Just like we can define string languages using string grammars, we can define graph languages using graph grammars, where we rewrite some start graph using a set of graph transformation rules. Derived graphs are then defined to be in the language exactly when they are terminally labelled.

Given a \textit{graph transformation system} \(T = (\Sigma, \mathcal{R})\), a subsignature of \textit{non-terminals} \(N\), and a \textit{start graph} \(S\) over \(\Sigma\), then a \textit{graph grammar} is a tuple \(\mathcal{G} = (\Sigma, N, \mathcal{R}, S)\). We say that a graph \(G\) is \textit{terminally labelled} if \(l(V) \cap N_V = \emptyset\) and \(m(E) \cap N_E = \emptyset\). Thus, we can define the \textit{graph language} generated by \(\mathcal{G}\):
\begin{align*}
\mathrm{L}(\mathcal{G}) = \{G \mid S \Rightarrow_{\mathcal{R}}^{*} G, G \text{ terminally labelled}\}\mathrm{.}
\end{align*}

Given \(\mathcal{G} = (\Sigma, N, \mathcal{R}, S)\), we have \(G \Rightarrow_{r} H\) if and only if \(H \Rightarrow_{r^{-1}} G\), for some \(r \in \mathcal{R}\), by using the \textit{comatch}. Moreover, \(G \in \mathrm{L}(\mathcal{G})\) if and only if \(G \Rightarrow_{\mathcal{R}^{-1}}^* S\) and \(G\) is terminally labelled. So we have a non-deterministic membership checking algorithm, by running the rules in reverse.

\subsection{Confluence Checking} \label{subsec:confluence_checking}

In 1970, Knuth and Bendix showed that confluence checking of terminating term rewriting systems is decidable \cite{Knuth-Bendix70a}. Moreover, it suffices to compute all \textit{critical pairs} and check their joinability \cite{Huet80a,Baader-Nipkow98a}. Unfortunately, for terminating graph transformation systems, confluence is not decidable in general, and joinability of critical pairs does not imply local confluence. In 1993, Plump showed that \textit{strong joinability} of all critical pairs is sufficient but not necessary to show local confluence \cite{Plump93b,Plump05a}.

In order to define critical pairs and critical pair isomorphism, we first must define what we mean by an instance of a derivation based on a morphism and what it means for two derivations to be parallelly independent.

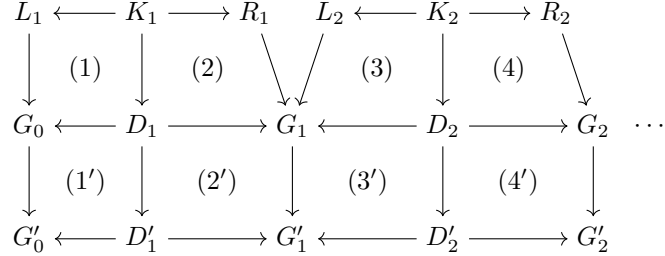
\begin{figure}[!ht]
\centering
\begin{tikzpicture}
    \node[align=center] (a0) at (0.0,3) {$L_1$};
    \node[align=center] (b0) at (1.5,3) {$K_1$};
    \node[align=center] (c0) at (3.0,3) {$R_1$};
    \node[align=center] (d0) at (4.0,3) {$L_2$};
    \node[align=center] (e0) at (5.5,3) {$K_2$};
    \node[align=center] (f0) at (7.0,3) {$R_2$};

    \node[align=center] (x) at (0.75,2.25) {$(1)$};
    \node[align=center] (x) at (2.375,2.25) {$(2)$};
    \node[align=center] (x) at (4.625,2.25) {$(3)$};
    \node[align=center] (x) at (6.375,2.25) {$(4)$};

    \node[align=center] (a1) at (0,1.5) {$G_0$};
    \node[align=center] (b1) at (1.5,1.5) {$D_1$};
    \node[align=center] (c1) at (3.5,1.5) {$G_1$};
    \node[align=center] (d1) at (5.5,1.5) {$D_2$};
    \node[align=center] (e1) at (7.5,1.5) {$G_2$};
    \node[align=center] (f1) at (8.275,1.5) {$\cdots$};

    \node[align=center] (x) at (0.75,0.75) {$(1')$};
    \node[align=center] (x) at (2.5,0.75) {$(2')$};
    \node[align=center] (x) at (4.5,0.75) {$(3')$};
    \node[align=center] (x) at (6.5,0.75) {$(4')$};

    \node[align=center] (a2) at (0.0,0) {$G'_0$};
    \node[align=center] (b2) at (1.5,0) {$D'_1$};
    \node[align=center] (c2) at (3.5,0) {$G'_1$};
    \node[align=center] (d2) at (5.5,0) {$D'_2$};
    \node[align=center] (e2) at (7.5,0) {$G'_2$};

    \draw (b0) edge[->] (a0)
          (b0) edge[->] (c0)
          (e0) edge[->] (d0)
          (e0) edge[->] (f0)
 
          (b1) edge[->] (a1)
          (b1) edge[->] (c1)
          (d1) edge[->] (c1)
          (d1) edge[->] (e1)

          (b2) edge[->] (a2)
          (b2) edge[->] (c2)
          (d2) edge[->] (c2)
          (d2) edge[->] (e2)

          (a0) edge[->] (a1)
          (b0) edge[->] (b1)
          (c0) edge[->] (c1)
          (d0) edge[->] (c1)
          (e0) edge[->] (d1)
          (f0) edge[->] (e1)

          (a1) edge[->] (a2)
          (b1) edge[->] (b2)
          (c1) edge[->] (c2)
          (d1) edge[->] (d2)
          (e1) edge[->] (e2);
\end{tikzpicture}
\caption{Derivation instances}
\label{fig:derivation_instances}
\end{figure}

Let the derivation \(\Delta: G_0 \Rightarrow^* G_n\) be given by pushouts \((1), (1'),\dots, (n), (n')\) and suppose there are pushouts  \((\underline{1}), (\underline{1'}),\dots, (\underline{n}), (\underline{n'})\) whose vertical morphisms are injective (Figure \ref{fig:derivation_instances}). Then, the derivation \(\Delta': G'_0 \Rightarrow^* G'_n\) consisting of the composed pushouts \((1 + \underline{1}),\dots, (n' + \underline{n'})\) is an instance of \(\Delta\) based on the morphism \(G_0 \to G'_0\). Moreover, we define the subgraph \(\mathrm{Use}_\Delta\) to be all items \(x\) such that there is some \(i \geq 0\) with \(G_0 \Rightarrow^* G_i(x) \in \mathrm{Match}(G_i \Rightarrow G_{i+1})\) where \(\mathrm{Match}(G_i \Rightarrow G_{i+1})\) is the image of the associated rule's left hand side graph under the match \(L \to G_i\).

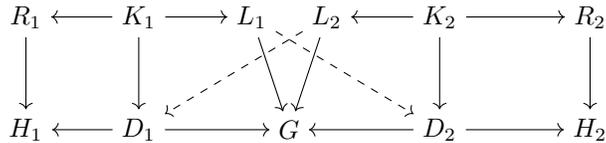
\begin{figure}[!ht]
\centering
\begin{tikzpicture}
    \node[align=center] (a0) at (0.0,3) {$R_1$};
    \node[align=center] (b0) at (1.5,3) {$K_1$};
    \node[align=center] (c0) at (3.0,3) {$L_1$};
    \node[align=center] (d0) at (4.0,3) {$L_2$};
    \node[align=center] (e0) at (5.5,3) {$K_2$};
    \node[align=center] (f0) at (7.5,3) {$R_2$};

    \node[align=center] (a1) at (0,1.5) {$H_1$};
    \node[align=center] (b1) at (1.5,1.5) {$D_1$};
    \node[align=center] (c1) at (3.5,1.5) {$G$};
    \node[align=center] (d1) at (5.5,1.5) {$D_2$};
    \node[align=center] (e1) at (7.5,1.5) {$H_2$};

    \draw (b0) edge[->] (a0)
          (b0) edge[->] (c0)
          (e0) edge[->] (d0)
          (e0) edge[->] (f0)
 
          (b1) edge[->] (a1)
          (b1) edge[->] (c1)
          (d1) edge[->] (c1)
          (d1) edge[->] (e1)

          (a0) edge[->] (a1)
          (b0) edge[->] (b1)
          (c0) edge[->] (c1)
          (d0) edge[->] (c1)
          (e0) edge[->] (d1)
          (f0) edge[->] (e1)

          (d0) edge[->,dashed] (b1)
          (c0) edge[->,dashed] (d1);
\end{tikzpicture}
\caption{Parallelly independent direct derivations}
\label{fig:parind}
\end{figure}

We say two direct derivations \(H_1 \Leftarrow_{r_1,g_1} G \Rightarrow_{r_2,g_2} H_2\) are \textit{parallelly independent} if \((g_1(L_1) \cap g_2(L_2)) \subseteq (g_1(K_1) \cap g_2(K_2))\), or equivalently, if there are morphisms \(L_1 \to D_2\) and \(L_2 \to D_1\) such that \(L_1 \to D_2 \to G = L_1 \to G\) and \(L_2 \to D_1 \to G = L_2 \to G\) (Figure \ref{fig:parind}).

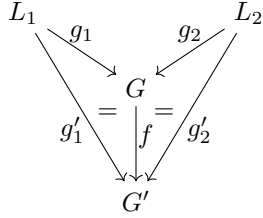
\begin{figure}[!ht]
\centering
\begin{tikzpicture}
    \node[align=center] (a) at (0.0,2) {$L_1$};
    \node[align=center] (b) at (3.0,2) {$L_2$};

    \node[align=center] (x) at (1.125,0.666667) {$=$};
    \node[align=center] (y) at (1.875,0.666667) {$=$};

    \node[align=center] (c) at (1.5,1) {$G$};

    \node[align=center] (d) at (1.5,-0.5) {$G'$};

    \draw (a) edge[->] node[above] {$g_1$} (c)
          (b) edge[->] node[above] {$g_2$} (c)
          (a) edge[->] node[below,xshift=-1mm] {$g'_1$} (d)
          (b) edge[->] node[below,xshift=1mm] {$g'_2$} (d)
          (c) edge[->] node[right,xshift=-1mm,yshift=1mm] {$f$}(d);
\end{tikzpicture}
\caption{Isomorphism of critical pairs}
\label{fig:isomatches}
\end{figure}

We say two \textit{parallelly independent} direct derivations \(H_1 \Leftarrow_{r_1,g_1} G \Rightarrow_{r_2,g_2} H_2\) are a \textit{critical pair} if additionally \(G = g_1(L_1) \cup g_2(L_2)\), and if \(r_1 = r_2\) then \(g_1 \neq g_2\). It is easy to see that every graph transformation system has only finitely many critical pairs. We call two critical pairs \(H_1 \Leftarrow_{r_1,g_1} G \Rightarrow_{r_2,g_2} H_2\) and \(H'_1 \Leftarrow_{r_1,g_1'} G' \Rightarrow_{r_2,g_2'} H'_2\) \textit{isomorphic} if there is a isomorphism \(f: G \to G'\) such that \(G' \Rightarrow H'_1\) is an instance of \(G \Rightarrow H_1\) based on \(f\) and \(G' \Rightarrow H'_2\) is an instance of \(G \Rightarrow H_2\) based on \(f\). Equivalently, the critical pairs \(H_1 \Leftarrow_{r_1,g_1} G \Rightarrow_{r_2,g_2} H_2\) and \(H'_1 \Leftarrow_{r_1,g'_1} G' \Rightarrow_{r_2,g'_2} H'_2\) are isomorphic if there is an isomorphism \(f: G \to G'\) such that \(g_1' = f \circ g_1\) and \(g_2' = f \circ g_2\) (Figure \ref{fig:isomatches}).

The \textit{track morphism} of a direct derivation \(G \Rightarrow H\) is defined to be the partial morphism \(\mathit{tr}_{G \Rightarrow H} = \mathit{in}' \circ \mathit{in}^{-1}\), where \(\mathit{in}\) and \(\mathit{in}'\) are the bottom left and right morphisms in Figure \ref{fig:directderivation}, respectively. We define \(\mathit{tr}_{G \Rightarrow^* H}\) inductively as the composition of track morphisms. The set of \textit{persistent nodes} of a critical pair \(\Phi : H_1 \Leftarrow G \Rightarrow H_2\) is \(\mathit{Persist}_{\Phi} = \{v \in G_V \mid \mathit{tr}_{G \Rightarrow H_1}(\{v\}), \mathit{tr}_{G \Rightarrow H_2}(\{v\}) \neq \emptyset\}\). That is, those nodes that are not deleted by the application of either rule.

A critical pair \(\Phi : H_1 \Leftarrow G \Rightarrow H_2\) is \textit{strongly joinable} (\textit{strongly subcommutative}) if it is \textit{joinable} (\textit{subcommutative}) without deleting any of the persistent nodes, and the persistent nodes are identified when joining. That is, there exists a graph \(M\) and derivations \(H_1 \Rightarrow_{\mathcal{R}}^* M \Leftarrow_{\mathcal{R}}^* H_2\) (\(H_1 \Rightarrow_{\mathcal{R}}^= M \Leftarrow_{\mathcal{R}}^= H_2\)) such that \(\forall v \in \mathit{Persist}_{\Phi}, \mathit{tr}_{G \Rightarrow H_1 \Rightarrow^* M}(\{v\}) = \mathit{tr}_{G \Rightarrow H_2 \Rightarrow^* M}(\{v\}) \neq \emptyset\).

\begin{theorem}[Critical Pair Lemma \cite{Plump93b,Plump05a}] \label{thm:critpairlem}
A graph transformation system \(T\) is \textit{locally confluent} if all its \textit{critical pairs} are \textit{strongly joinable}.
\end{theorem}

It's easy to see that the result also holds if one only considers non-isomorphic critical pairs, which can result in a large speedup in practice, since for large rules, there can often be many isomorphic critical pairs.

The original proof of the Critical Pair Lemma needs the Commutativity, Clipping and Embedding Theorems, which we shall now provide, and use in the proof of our Generalised Critical Pair Lemma (Theorem \ref{thm:ngcritpairlem}). 

\begin{theorem}[Commutativity \cite{Ehrig-Kreowski76a}] \label{thm:commute}
If \(H_1 \Leftarrow_{r_1,g_1} G \Rightarrow_{r_2,g_2} H_2\) are \textit{parallelly independent}, then there is a graph \(G'\) and derivations \(H_1 \Rightarrow_{r_2} G' \Leftarrow_{r_1} H_2\).
\end{theorem}

\begin{theorem}[Clipping \cite{Plump99a}] \label{thm:clipping}
Given a derivation \(\Delta': G' \Rightarrow^* H'\) and an injective morphism \(h: G \to G'\) such that \(\mathrm{Use}{\Delta'} \subseteq h(G)\), there exists a derivation \(\Delta: G \Rightarrow^* H\) such that \(\Delta'\) is an instance of \(\Delta\) based on \(h\).
\end{theorem}
 
Given a derivation \(\Delta: G \Rightarrow^* H\), the subgraph of \(G\), \(\mathrm{Persist}_{\Delta}\), consists of all items \(x\) such that \(\mathit{tr}_{G \Rightarrow^* H}(x)\) is defined.
  
\begin{theorem}[Embedding \cite{Plump99a}] \label{thm:embedding}
Let \(\Delta: G \Rightarrow^* H\) be a derivation, \(h: G \to G'\) an injective graph morphism, \(B_{\Delta}\) be the discrete subgraph of \(G\) consisting of all nodes \(x\) such that \(h(x)\) is incident to an edge in \(G' \setminus h(G)\). If \(B_{\Delta} \subseteq \mathrm{Persist}_{\Delta}\), then there exists a derivation \(\Delta': G' \Rightarrow^* H'\) such that \(\Delta'\) is an instance of \(\Delta\) based on \(h\). Moreover, there exists a pushout of \(t: B_{\Delta} \to H\) along \(h': B_{\Delta} \to C_{\Delta}\) where \(C_{\Delta} = (G' \setminus h(G)) \cup h(B_{\Delta})\) and \(t\) is the restriction of \(\mathit{tr}_{G \Rightarrow^* H}\) to \(B_{\Delta}\).
 \end{theorem}

\section{Closedness and Confluence up to Garbage} \label{sec:confluence}

The purpose of this section is to introduce the notion of ``up to garbage'' and lay some foundations that we can use in the remainder of the paper. We divide the section into three subsections, finishing by relating ``up to garbage'' with the existing notion of ``modulo garbage''.

\subsection{Closedness and Garbage}

We start with the definition of closedness, and what it means for an item to be considered garbage.

\begin{definition}
Let \(T = (\mathcal{A}, \rightarrow)\) be an ARS and \(\mathcal{D} \subseteq \mathcal{A}\). Then an object \(x \in \mathcal{A}\) is called \textit{garbage} if \(x \not\in \mathcal{D}\) and \(\mathcal{D}\) is \textit{closed} under \(T\) if for all \(x, y \in \mathcal{A}\) such that \(x \rightarrow y\), if \(x \in \mathcal{D}\) then \(y \in \mathcal{D}\).
\end{definition}

The idea is that \(\mathcal{D}\) represents the \textit{good input}, and the \textit{garbage} is the objects that are not in this class. In the context of graph transformation, \(\mathcal{D}\) will be a graph language, but need not be explicitly generated by a graph grammar. For example, it could be defined by some (monadic second-order \cite{Courcelle89a}) logical formula, a finite listing of graphs, or a type graph language (Subsection \ref{subsec:critpairgen}). Finite languages and type graph languages will be of particular interest to us due to the fact they they have decidable subgraph membership problem (Subsection \ref{subsec:critpairgen}).

\begin{figure}[!ht]
\centering
\begin{tikzpicture}[every node/.style={align=center}]
    \node (a) at (0.0,-0.05) {$r_1$:};
    \node (b) at (1.0,0.0)   [draw, circle, thick, fill=black, scale=0.3] {\,};
    \node (c) at (2.0,0.0)   [draw, circle, thick, fill=black, scale=0.3] {\,};
    \node (d) at (3.0,0.0)   {$\leftarrow$};
    \node (e) at (4.0,0.0)   [draw, circle, thick, fill=black, scale=0.3] {\,};
    \node (f) at (5.0,0.0)   [draw, circle, thick, fill=black, scale=0.3] {\,};
    \node (g) at (6.0,0.0)   {$\rightarrow$};
    \node (h) at (7.0,0.0)   [draw, circle, thick, fill=black, scale=0.3] {\,};
    \node (i) at (8.0,0.0)   [draw, circle, thick, fill=black, scale=0.3] {\,};

    \draw (b) edge[->,thick] (c);

    \node (B) at (1.0,-.18)  {\tiny{1}};
    \node (C) at (2.0,-.18)  {\tiny{2}};
    \node (E) at (4.0,-.18)  {\tiny{1}};
    \node (F) at (5.0,-.18)  {\tiny{2}};
    \node (H) at (7.0,-.18)  {\tiny{1}};
    \node (I) at (8.0,-.18)  {\tiny{2}};

    \node (j) at (0.0,-1.05)  {$r_2$:};
    \node (k) at (1.0,-1.0)  [draw, circle, thick, fill=black, scale=0.3] {\,};
    \node (l) at (2.0,-1.0)  [draw, circle, thick, fill=black, scale=0.3] {\,};
    \node (m) at (3.0,-1.0)  {$\leftarrow$};
    \node (n) at (4.0,-1.0)  [draw, circle, thick, fill=black, scale=0.3] {\,};
    \node (o) at (6.0,-1.0)  {$\rightarrow$};
    \node (p) at (7.0,-1.0)  [draw, circle, thick, fill=black, scale=0.3] {\,};

    \node (K) at (1.0,-1.18) {\tiny{1}};
    \node (N) at (4.0,-1.18) {\tiny{1}};
    \node (P) at (7.0,-1.18) {\tiny{1}};

    \draw (k) edge[->,thick] (l);
\end{tikzpicture}
\caption{Reduction rules for Example \ref{eg:a}}
\label{fig:eg-a-rules}
\end{figure}
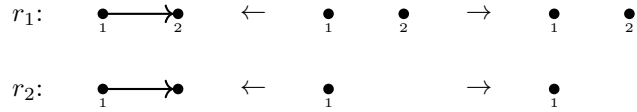

\begin{example} \label{eg:a}
Consider the reduction rules in Figure \ref{fig:eg-a-rules}. The language of acyclic graphs is \textit{closed} under the GT system \(((\{\square\}, \{\square\}), \{r_1\})\), and the language of trees (forests) and its complement are both \textit{closed} under \(((\{\square\}, \{\square\}), \{r_2\})\).
\end{example}

The closedness problem is defined in the obvious way:

\begin{definition}[Closedness Problem]~\\
\vspace{-1em}
\begin{prob}
\probinstance{A GT system \(T = (\Sigma, \mathcal{R})\) and a graph grammar \(\mathcal{G}\) over \(\Sigma\).}
\probquestion{Is \(\mathrm{L}(\mathcal{G})\) \textit{closed} under \(T\)?}
\end{prob}
\end{definition}

It turns out that this is undecidable in general, even if we restrict to recursive languages and terminating GT systems. In 1998, Fradet and Le M{\'e}tayer showed the following result:

\begin{theorem}[Undecidable Closedness \cite{Fradet-Metayer98a}]
The \textit{closedness problem} is \textit{undecidable} in general, even for terminating GT systems \(T\) with only one rule, and \(\mathcal{G}\) an edge replacement grammar.
\end{theorem}

\subsection{Confluence and Subcommutativity up to Garbage} \label{subsec:confluence_garbage}

In this subsection, we generalise the familiar definitions of local confluence, confluence, subcommutativity, and termination to permit ignoring \emph{garbage}.

\begin{definition}
Given an ARS \((\mathcal{A}, \rightarrow)\), \(\mathcal{D} \subseteq \mathcal{A}\), \(x, x_0 \in \mathcal{D}\), and \(x_i, y_1, y_2 \in \mathcal{A}\) (\(i \geq 1\)), we say that:

\begin{enumerate}
\item \(\rightarrow\) is \textit{confluent up to garbage} on \(\mathcal{D}\) if \(y_1 \xleftarrow{*} x \xrightarrow{*} y_2\) implies \(y_1, y_2\) are joinable;
\item \(\rightarrow\) is \textit{locally confluent up to garbage} on \(\mathcal{D}\) if \(y_1 \leftarrow x \rightarrow y_2\) implies \(y_1, y_2\) are joinable;
\item \(\rightarrow\) is \textit{subcommutative up to garbage} on \(\mathcal{D}\) if \(y_1 \leftarrow x \rightarrow y_2\) implies \(y_1, y_2\) are subcommutative;
\item \(\rightarrow\) is \textit{terminating up to garbage} on \(\mathcal{D}\) if there is no infinite sequence \(x_0 \rightarrow x_1 \rightarrow \dots\).
\end{enumerate}
\end{definition}

The following is an immediate consequence of inclusion:

\begin{lemma} \label{lem:transitiveconfluence}
Let \((\mathcal{A}, \rightarrow)\) be an ARS, \(\mathcal{D} \subseteq \mathcal{A}\), \(\mathcal{E} \subseteq \mathcal{D}\), and \(\mathcal{P}\) be the property of \textit{confluence up to garbage}, \textit{local confluence up to garbage}, \textit{subcommutativity up to garbage}, or \textit{termination up to garbage}. Then \(\mathcal{P}\) on \(\mathcal{D}\) implies \(\mathcal{P}\) on \(\mathcal{E}\).
\end{lemma}

Our next two examples show that confluence up to garbage need not correspond to confluence. That is, a system can be non-confluent, but confluent up to garbage.

\begin{example}
Consider again the rules in Figure \ref{fig:eg-a-rules}. It is easy to see that the GT system containing both rules is terminating, but not confluent. It is, however, both confluent up to garbage on the language of unlabelled discrete graphs and subcommutative up to garbage on the language of unlabelled discrete graphs.
\end{example}

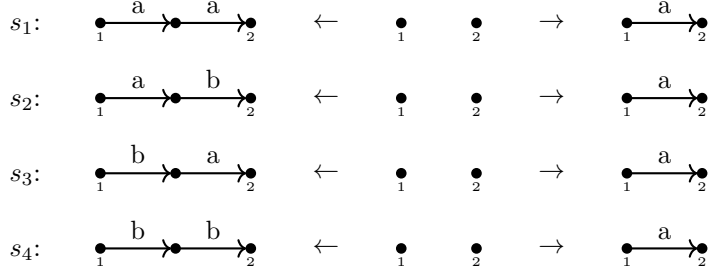
\begin{figure}[!ht]
\centering
\begin{tikzpicture}[every node/.style={align=center}]
    \node (a) at (0.0,-0.05) {$s_1$:};
    \node (b) at (1.0,0.0)   [draw, circle, thick, fill=black, scale=0.3] {\,};
    \node (c) at (2.0,0.0)   [draw, circle, thick, fill=black, scale=0.3] {\,};
    \node (d) at (3.0,0.0)   [draw, circle, thick, fill=black, scale=0.3] {\,};
    \node (e) at (4.0,0.0)   {$\leftarrow$};
    \node (f) at (5.0,0.0)   [draw, circle, thick, fill=black, scale=0.3] {\,};
    \node (g) at (6.0,0.0)   [draw, circle, thick, fill=black, scale=0.3] {\,};
    \node (h) at (7.0,0.0)   {$\rightarrow$};
    \node (i) at (8.0,0.0)   [draw, circle, thick, fill=black, scale=0.3] {\,};
    \node (j) at (9.0,0.0)   [draw, circle, thick, fill=black, scale=0.3] {\,};

    \node (B) at (1.0,-0.18)  {\tiny{1}};
    \node (D) at (3.0,-0.18)  {\tiny{2}};
    \node (F) at (5.0,-.18)  {\tiny{1}};
    \node (G) at (6.0,-00.18)  {\tiny{2}};
    \node (I) at (8.0,-0.18)  {\tiny{1}};
    \node (J) at (9.0,-0.18)  {\tiny{2}};

    \draw (b) edge[->,thick] node[above] {a} (c)
          (c) edge[->,thick] node[above] {a} (d)
          (i) edge[->,thick] node[above] {a} (j);

    \node (a) at (0.0,-1.05) {$s_2$:};
    \node (b) at (1.0,-1.0)  [draw, circle, thick, fill=black, scale=0.3] {\,};
    \node (c) at (2.0,-1.0)  [draw, circle, thick, fill=black, scale=0.3] {\,};
    \node (d) at (3.0,-1.0)  [draw, circle, thick, fill=black, scale=0.3] {\,};
    \node (e) at (4.0,-1.0)  {$\leftarrow$};
    \node (f) at (5.0,-1.0)  [draw, circle, thick, fill=black, scale=0.3] {\,};
    \node (g) at (6.0,-1.0)  [draw, circle, thick, fill=black, scale=0.3] {\,};
    \node (h) at (7.0,-1.0)  {$\rightarrow$};
    \node (i) at (8.0,-1.0)  [draw, circle, thick, fill=black, scale=0.3] {\,};
    \node (j) at (9.0,-1.0)  [draw, circle, thick, fill=black, scale=0.3] {\,};

    \node (B) at (1.0,-1.18)  {\tiny{1}};
    \node (D) at (3.0,-1.18)  {\tiny{2}};
    \node (F) at (5.0,-1.18)  {\tiny{1}};
    \node (G) at (6.0,-1.18)  {\tiny{2}};
    \node (I) at (8.0,-1.18)  {\tiny{1}};
    \node (J) at (9.0,-1.18)  {\tiny{2}};

    \draw (b) edge[->,thick] node[above] {a} (c)
          (c) edge[->,thick] node[above] {b} (d)
          (i) edge[->,thick] node[above] {a} (j);

    \node (a) at (0.0,-2.05) {$s_3$:};
    \node (b) at (1.0,-2.0)  [draw, circle, thick, fill=black, scale=0.3] {\,};
    \node (c) at (2.0,-2.0)  [draw, circle, thick, fill=black, scale=0.3] {\,};
    \node (d) at (3.0,-2.0)  [draw, circle, thick, fill=black, scale=0.3] {\,};
    \node (e) at (4.0,-2.0)  {$\leftarrow$};
    \node (f) at (5.0,-2.0)  [draw, circle, thick, fill=black, scale=0.3] {\,};
    \node (g) at (6.0,-2.0)  [draw, circle, thick, fill=black, scale=0.3] {\,};
    \node (h) at (7.0,-2.0)  {$\rightarrow$};
    \node (i) at (8.0,-2.0)  [draw, circle, thick, fill=black, scale=0.3] {\,};
    \node (j) at (9.0,-2.0)  [draw, circle, thick, fill=black, scale=0.3] {\,};

    \node (B) at (1.0,-2.18)  {\tiny{1}};
    \node (D) at (3.0,-2.18)  {\tiny{2}};
    \node (F) at (5.0,-2.18)  {\tiny{1}};
    \node (G) at (6.0,-2.18)  {\tiny{2}};
    \node (I) at (8.0,-2.18)  {\tiny{1}};
    \node (J) at (9.0,-2.18)  {\tiny{2}};

    \draw (b) edge[->,thick] node[above] {b} (c)
          (c) edge[->,thick] node[above] {a} (d)
          (i) edge[->,thick] node[above] {a} (j);

    \node (a) at (0.0,-3.05) {$s_4$:};
    \node (b) at (1.0,-3.0)  [draw, circle, thick, fill=black, scale=0.3] {\,};
    \node (c) at (2.0,-3.0)  [draw, circle, thick, fill=black, scale=0.3] {\,};
    \node (d) at (3.0,-3.0)  [draw, circle, thick, fill=black, scale=0.3] {\,};
    \node (e) at (4.0,-3.0)  {$\leftarrow$};
    \node (f) at (5.0,-3.0)  [draw, circle, thick, fill=black, scale=0.3] {\,};
    \node (g) at (6.0,-3.0)  [draw, circle, thick, fill=black, scale=0.3] {\,};
    \node (h) at (7.0,-3.0)  {$\rightarrow$};
    \node (i) at (8.0,-3.0)  [draw, circle, thick, fill=black, scale=0.3] {\,};
    \node (j) at (9.0,-3.0)  [draw, circle, thick, fill=black, scale=0.3] {\,};

    \node (B) at (1.0,-3.18)  {\tiny{1}};
    \node (D) at (3.0,-3.18)  {\tiny{2}};
    \node (F) at (5.0,-3.18)  {\tiny{1}};
    \node (G) at (6.0,-3.18)  {\tiny{2}};
    \node (I) at (8.0,-3.18)  {\tiny{1}};
    \node (J) at (9.0,-3.18)  {\tiny{2}};

    \draw (b) edge[->,thick] node[above] {b} (c)
          (c) edge[->,thick] node[above] {b} (d)
          (i) edge[->,thick] node[above] {a} (j);
\end{tikzpicture}
\caption{Reduction rules for Example \ref{eg:b}}
\label{fig:eg-b-rules}
\end{figure}

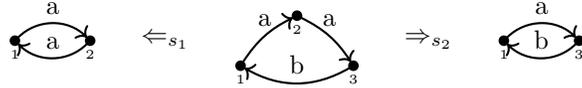
\begin{figure}[!ht]
\centering
\begin{tikzpicture}[every node/.style={align=center}]
    \node (a) at (0.0,0.0)    [draw, circle, thick, fill=black, scale=0.3] {\,};
    \node (b) at (1.0,0.0)    [draw, circle, thick, fill=black, scale=0.3] {\,};
    \node (c) at (2.0,0.0)    {$\Leftarrow_{s_1}$};
    \node (d) at (3.0,-0.333) [draw, circle, thick, fill=black, scale=0.3] {\,};
    \node (e) at (3.75,0.333) [draw, circle, thick, fill=black, scale=0.3] {\,};
    \node (f) at (4.5,-0.333) [draw, circle, thick, fill=black, scale=0.3] {\,};
    \node (g) at (5.5,0.0)    {$\Rightarrow_{s_2}$};
    \node (h) at (6.5,0.0)    [draw, circle, thick, fill=black, scale=0.3] {\,};
    \node (i) at (7.5,0.0)    [draw, circle, thick, fill=black, scale=0.3] {\,};

    \node (A) at (0.0,-0.18)  {\tiny{1}};
    \node (B) at (1.0,-0.18)  {\tiny{2}};
    \node (D) at (3.0,-0.513)  {\tiny{1}};
    \node (E) at (3.75,0.153) {\tiny{2}};
    \node (F) at (4.5,-0.513)  {\tiny{3}};
    \node (H) at (6.5,-0.18)  {\tiny{1}};
    \node (I) at (7.5,-0.18)  {\tiny{3}};

    \draw (a) edge[->,thick, bend left=45] node[above] {a} (b)
          (b) edge[->,thick, bend left=45] node[above] {a} (a)
          (d) edge[->,thick, bend left=15] node[above] {a} (e)
          (e) edge[->,thick, bend left=15] node[above] {a} (f)
          (f) edge[->,thick, bend left=30] node[above] {b} (d)
          (h) edge[->,thick, bend left=45] node[above] {a} (i)
          (i) edge[->,thick, bend left=45] node[above] {b} (h);
\end{tikzpicture}
\caption{Non-joinable derivations for Example \ref{eg:b}}
\label{fig:eg-b-nj-derivations}
\end{figure}

\begin{example} \label{eg:b}
Consider the rules in Figure \ref{fig:eg-b-rules}. They are terminating, since they are size reducing. Moreover, the language of all linked lists with edge labels \(a\) or \(b\) and its complement (over the same signature) are closed under the rules. These rules are confluent up to garbage on linked lists, since any non-trivial linked list is necessarily reduced to the length one linked list labelled by \(a\), and the length zero and one linked lists are already in normal form. By comparison, the rules are not locally confluent due to the counter example in Figure \ref{fig:eg-b-nj-derivations}. 
\end{example}

It is easy to see that confluence up to garbage always implies local confluence up to garbage, and subcommutativity up to garbage implies local confluence up to garbage, however subcommutativity up to garbage need not imply confluence up to garbage. Similarly, in the presence of termination, local confluence up to garbage need not imply confluence up to garbage. Our next example demonstrates this.

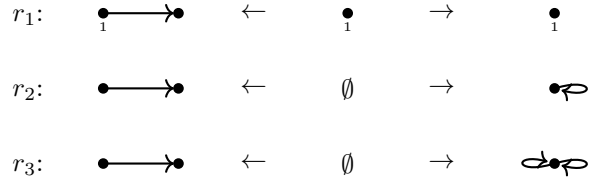
\begin{figure}[!ht]
\centering
\begin{tikzpicture}[every node/.style={align=center}]
    \node (a) at (0.0,-0.05)  {$r_1$:};
    \node (b) at (1.0,0.0)    [draw, circle, thick, fill=black, scale=0.3] {\,};
    \node (c) at (2.0,0.0)    [draw, circle, thick, fill=black, scale=0.3] {\,};
    \node (d) at (3.0,0.0)    {$\leftarrow$};
    \node (e) at (4.25,0.0)    [draw, circle, thick, fill=black, scale=0.3] {\,};
    \node (g) at (5.5,0.0)    {$\rightarrow$};
    \node (h) at (7.0,0.0)    [draw, circle, thick, fill=black, scale=0.3] {\,};

    \node (B) at (1.0,-0.18)  {\tiny{1}};
    \node (E) at (4.25,-0.18) {\tiny{1}};
    \node (H) at (7.0,-0.18)  {\tiny{1}};

    \draw (b) edge[->,thick] (c);

    \node (a) at (0.0,-1.05)  {$r_2$:};
    \node (b) at (1.0,-1.0)   [draw, circle, thick, fill=black, scale=0.3] {\,};
    \node (c) at (2.0,-1.0)   [draw, circle, thick, fill=black, scale=0.3] {\,};
    \node (d) at (3.0,-1.0)   {$\leftarrow$};
    \node (e) at (4.25,-1.0)  {$\emptyset$};;
    \node (f) at (5.5,-1.0)   {$\rightarrow$};
    \node (g) at (7.0,-1.0)   [draw, circle, thick, fill=black, scale=0.3] {\,};

    \draw (b) edge[->,thick] (c)
          (g) edge[->,thick,loop right] (g);

    \node (a) at (0.0,-2.05)  {$r_3$:};
    \node (b) at (1.0,-2.0)   [draw, circle, thick, fill=black, scale=0.3] {\,};
    \node (c) at (2.0,-2.0)   [draw, circle, thick, fill=black, scale=0.3] {\,};
    \node (d) at (3.0,-2.0)   {$\leftarrow$};
    \node (e) at (4.25,-2.0)  {$\emptyset$};
    \node (f) at (5.5,-2.0)   {$\rightarrow$};
    \node (g) at (7.0,-2.0)   [draw, circle, thick, fill=black, scale=0.3] {\,};

    \draw (b) edge[->,thick] (c)
          (g) edge[->,thick,loop left] (g)
          (g) edge[->,thick,loop right] (g);
\end{tikzpicture}
\caption{Rules for Example \ref{eg:0}}
\label{fig:eg-0-rules}
\end{figure}

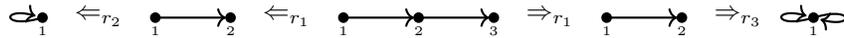
\begin{figure}[!ht]
\centering
\begin{tikzpicture}[every node/.style={align=center}]
    \node (a) at (0.0,0.0)     [draw, circle, thick, fill=black, scale=0.3] {\,};
    \node (b) at (0.75,0.0)    {$\Leftarrow_{r_2}$};
    \node (c) at (1.5,0.0)     [draw, circle, thick, fill=black, scale=0.3] {\,};
    \node (d) at (2.5,0.0)     [draw, circle, thick, fill=black, scale=0.3] {\,};
    \node (e) at (3.25,0.0)    {$\Leftarrow_{r_1}$};
    \node (f) at (4.0,0.0)     [draw, circle, thick, fill=black, scale=0.3] {\,};
    \node (g) at (5.0,0.0)     [draw, circle, thick, fill=black, scale=0.3] {\,};
    \node (h) at (6.0,0.0)     [draw, circle, thick, fill=black, scale=0.3] {\,};
    \node (i) at (6.75,0.0)    {$\Rightarrow_{r_1}$};
    \node (j) at (7.5,0.0)     [draw, circle, thick, fill=black, scale=0.3] {\,};
    \node (k) at (8.5,0.0)     [draw, circle, thick, fill=black, scale=0.3] {\,};
    \node (l) at (9.25,0.0)    {$\Rightarrow_{r_3}$};
    \node (m) at (10.25,0.0)   [draw, circle, thick, fill=black, scale=0.3] {\,};

    \node (A) at (0.0,-0.18)   {\tiny{1}};
    \node (C) at (1.5,-0.18)   {\tiny{1}};
    \node (D) at (2.5,-0.18)   {\tiny{2}};
    \node (F) at (4.0,-0.18)   {\tiny{1}};
    \node (G) at (5.0,-0.18)   {\tiny{2}};
    \node (H) at (6.0,-0.18)   {\tiny{3}};
    \node (J) at (7.5,-0.18)   {\tiny{1}};
    \node (K) at (8.5,-0.18)   {\tiny{2}};
    \node (M) at (10.25,-0.18) {\tiny{1}};

    \draw (a) edge[->,thick,loop left] (a)
          (c) edge[->,thick] (d)
          (f) edge[->,thick] (g)
          (g) edge[->,thick] (h)
          (j) edge[->,thick] (k)
          (m) edge[->,thick,loop right] (m)
          (m) edge[->,thick,loop left] (m);
\end{tikzpicture}
\caption{Non-joinable derivations for Example \ref{eg:0}}
\label{fig:eg-0-nj-derivations}
\end{figure}

\begin{example} \label{eg:0}
Let \(\mathcal{D}\) be the language of linked lists containing at least two edges and \(T\) be the GT system with rules from Figure \ref{fig:eg-0-rules}. Then \(r_2\) and \(r_3\) cannot be applied to any graph in \(\mathcal{D}\), and \(r_1\) will always be applicable in a unique way, with the effect of deleting the last node and its edge. It is thus immediate that \(T\) is subcommutative up to garbage on \(\mathcal{D}\), and thus also locally confluent up to garbage on \(\mathcal{D}\). \(T\) is not, however, confluent up to garbage on \(\mathcal{D}\) due to the following two-step counter example in Figure \ref{fig:eg-0-nj-derivations}.
\end{example}

Our next two results show that closedness is the missing ingredient to recovering the familiar relationships between local confluence, confluence, subcommutativity, and termination, in our generalised setting of ``up to garbage''.

\begin{lemma} \label{lem:confimpl}
Let \((\mathcal{A}, \rightarrow)\) be an ARS and \(\mathcal{D} \subseteq \mathcal{A}\).
\begin{enumerate}
    \item If \(\mathcal{D}\) is \textit{closed} under \(\rightarrow\) and \(\rightarrow\) is \textit{subcommutative up to garbage} on \(\mathcal{D}\), then \(\rightarrow\) is \textit{confluent up to garbage} on \(\mathcal{D}\);
    \item If \(\rightarrow\) is \textit{confluent up to garbage} on \(\mathcal{D}\), then \(\rightarrow\) is \textit{locally confluent up to garbage} on \(\mathcal{D}\).
\end{enumerate}
\end{lemma}

\begin{proof}
The first part can be seen by Noetherian Induction, due to the fact that closedness ensures applicability of the induction hypothesis, and the second part follows immediately from the definitions.
\end{proof}

\begin{theorem}[Generalised Newman's Lemma] \label{thm:newmangarbage}
Let \((\mathcal{A}, \rightarrow)\) be an ARS and \(\mathcal{D} \subseteq \mathcal{A}\). If \(\rightarrow\) is \textit{terminating up to garbage} on \(\mathcal{D}\) and \(\mathcal{D}\) is \textit{closed} under \(\rightarrow\), then \(\rightarrow\) is \textit{confluent up to garbage} on \(\mathcal{D}\) if and only \(\rightarrow\) it is \textit{locally confluent up to garbage} on \(\mathcal{D}\).
\end{theorem}

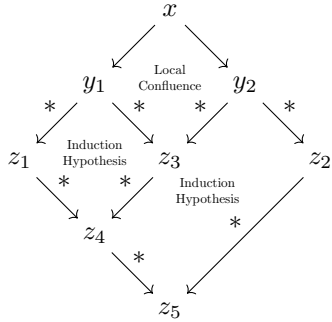
\begin{figure}[!ht]
\centering
\begin{tikzpicture}
    \node[align=center] (a) at (0,0) {$x$};
    \node[align=center] (b) at (-1,-1) {$y_1$};
    \node[align=center] (c) at (1,-1) {$y_2$};
    \node[align=center] (d) at (-2,-2) {$z_1$};
    \node[align=center] (e) at (0,-2) {$z_3$};
    \node[align=center] (f) at (2,-2) {$z_2$};
    \node[align=center] (g) at (-1,-3) {$z_4$};
    \node[align=center] (h) at (0,-4) {$z_5$};

    \node[align=center] (x) at (0,-1) {\scalebox{.5}{\parbox{2cm}{\centering{Local \\ Confluence \\ \,}}}};
    \node[align=center] (y) at (-1,-2) {\scalebox{.5}{\parbox{2cm}{\centering{Induction \\ Hypothesis \\ \,}}}};
    \node[align=center] (z) at (0.5,-2.5) {\scalebox{.5}{\parbox{2cm}{\centering{Induction \\ Hypothesis \\ \,}}}};

    \draw (a) edge[->] node[above,pos=0.666667] {\,} (b)
          (a) edge[->] node[above,pos=0.666667] {\,} (c)
          (b) edge[->] node[above,pos=0.666667] {*} (d)
          (b) edge[->] node[above,pos=0.666667] {*} (e)
          (c) edge[->] node[above,pos=0.666667] {*} (e)
          (c) edge[->] node[above,pos=0.666667] {*} (f)
          (d) edge[->] node[above,pos=0.666667] {*} (g)
          (e) edge[->] node[above,pos=0.666667] {*} (g)
          (f) edge[->] node[above,pos=0.583333] {*} (h)
          (g) edge[->] node[above,pos=0.666667] {*} (h);
\end{tikzpicture}
\caption{Diagram for the proof of Theorem \ref{thm:newmangarbage}}
\label{fig:inductionstep}
\end{figure}

\begin{proof}
One direction can be seen by Noetherian Induction (Figure \ref{fig:inductionstep}), due to the fact that closedness ensures applicability of the induction hypothesis, and the other follows from the second part of Lemma \ref{lem:confimpl}.
\end{proof}

\subsection{Confluence and Subcommutativity Modulo Garbage}

In this subsection, we show that our notion of confluence up to garbage can be related to the existing notion of confluence modulo.

First, we recall the definition of local confluence (confluence, subcommutativity) modulo an equivalence relation. If the relation is the identity relation, then we recover the standard definitions of local confluence (confluence, subcommutativity).

\begin{definition}
Given an ARS \((\mathcal{A}, \rightarrow)\), an equivalence \(\sim\) on \(\mathcal{A}\), and \(x, x_i, y,\) \(y_1, y_2 \in \mathcal{A}\) (\(i \geq 0\)), we say that:

\begin{enumerate}
\item \(x\) and \(y\) are \(\sim\)-\textit{joinable} if there is are \(z_1, z_2 \in \mathcal{A}\) such that \(x \xrightarrow{*} z_1 \sim z_2 \xleftarrow{*} y\);
\item \(x\) and \(y\) are \(\sim\)-\textit{subcommutative} if there are \(z_1, z_2 \in \mathcal{A}\) such that \(x \xrightarrow{=} z_1 \sim z_2 \xleftarrow{=} y\);
\item \(\rightarrow\) is \textit{confluent modulo} \(\sim\) if \(y_1 \xleftarrow{*} x \xrightarrow{*} y_2\) implies \(y_1, y_2\) are \(\sim\)-joinable;
\item \(\rightarrow\) is \textit{locally confluent modulo} \(\sim\)  if \(y_1 \leftarrow x \rightarrow y_2\) implies \(y_1, y_2\) are \(\sim\)-joinable;
\item \(\rightarrow\) is \textit{subcommutative modulo} \(\sim\)  if \(y_1 \leftarrow x \rightarrow y_2\) implies \(y_1, y_2\) are \(\sim\)-subcommutative.
\end{enumerate}
\end{definition}

We now show that confluence up to garbage as confluence up to garbage, and vice versa.

\begin{theorem}[Encoding Confluence up to Garbage]
Let \((\mathcal{A}, \rightarrow)\) be an ARS, \(\mathcal{D} \subseteq \mathcal{A}\), and define the equivalence \(\sim\) on \(\mathcal{A}\) by \(x \sim y\) exactly when \(x = y\) or \(x, y \in \mathcal{A} \setminus \mathcal{D}\). Then:

\begin{enumerate}
\item if \(\mathcal{A} \setminus \mathcal{D}\) is closed under \(\rightarrow\) and \(\rightarrow\) is \(\mathcal{P}\) up to garbage on \(\mathcal{D}\), then \(\rightarrow\) is \(\mathcal{P}\) modulo \(\sim\);
\item if \(\mathcal{D}\) is closed under \(\rightarrow\) and \(\rightarrow\) is \(\mathcal{P}\) modulo \(\sim\), then \(\rightarrow\) is \(\mathcal{P}\) up to garbage on \(\mathcal{D}\).
\end{enumerate}

where \(\mathcal{P}\) is the property \textit{confluence}, \textit{local confluence}, or \textit{subcommutativity}.
\end{theorem}

\begin{proof}
We deal only with confluence. Local confluence and subcommutativity are trivial modifications of the same argument.

Suppose \(\mathcal{A} \setminus \mathcal{D}\) is closed under \(\rightarrow\) and \(\rightarrow\) is confluent up to garbage on \(\mathcal{D}\). Then, for all derivations \(x \xrightarrow{*} y\), \(x \xrightarrow{*} z\) such that \(x \in \mathcal{D}\), there is a \(t \in \mathcal{A}\) such that \(y \xrightarrow{*} t\) and \(z \xrightarrow{*} t\). Clearly \(t \sim t\), so all such derivations are \(\sim\)-joinable. Suppose now that \(x \not\in \mathcal{D}\). Then by closedness, \(y, z \not\in \mathcal{D}\) too, so \(y \sim z\), so all such derivations are \(\sim\)-joinable. Thus \(\rightarrow\) is confluent modulo \(\sim\), as required.

Suppose \(\mathcal{D}\) is closed under \(\rightarrow\) and \(\rightarrow\) is confluent modulo \(\sim\). Then, for all all derivations \(x \xrightarrow{*} y\), \(x \xrightarrow{*} z\) such that \(x \in \mathcal{D}\), there are \(t_1, t_2 \in \mathcal{A}\) such that \(y \xrightarrow{*} t_1\), \(z \xrightarrow{*} t_2\), and \(t_1 \sim t_2\). Due to closedness, we have \(y, t_1, z, t_2 \mathcal{D}\). Putting this together with the fact that \(t_1 \sim t_2\) tells us that in fact \(t_1 = t_2\). Thus all such derivations are joinable. Finally, if \(x \not\in \mathcal{D}\), we don't need to consider joinability. Thus, \(\rightarrow\) is confluent up to garbage on \(\mathcal{D}\), as required.
\end{proof}

Thus, if both \(\mathcal{A} \setminus \mathcal{D}\) and \(\mathcal{D}\) are closed under \(\rightarrow\), then the notions of confluence (local confluence, subcommutativity) up to garbage and modulo garbage exactly correspond:

\begin{corollary}
Let \((\mathcal{A}, \rightarrow)\) be an ARS, \(\mathcal{D} \subseteq \mathcal{A}\), and define the equivalence \(\sim\) on \(\mathcal{A}\) by \(x \sim y\) exactly when \(x = y\) or \(x, y \in \mathcal{A} \setminus \mathcal{D}\). If \(\mathcal{A} \setminus \mathcal{D}\) and \(\mathcal{D}\) are closed under \(\rightarrow\), then \(\rightarrow\) is \(\mathcal{P}\) up to garbage on \(\mathcal{D}\) if and only if \(\rightarrow\) is \(\mathcal{P}\) modulo \(\sim\), where \(\mathcal{P}\) is the property \textit{confluence}, \textit{local confluence}, or \textit{subcommutativity}.
\end{corollary}

\section{Generalised Critical Pair Lemma} \label{sec:pair_lemma}

\begin{figure}[!ht]
\centering
\begin{tikzpicture}[every node/.style={align=center}]
    \node (a) at (0.0,0.0)   [draw, circle, thick, fill=black, scale=0.3] {\,};
    \node (b) at (1.0,0.0)   [draw, circle, thick, fill=black, scale=0.3] {\,};
    \node (c) at (2.0,0.0)   {$\Leftarrow_{r_1}$};
    \node (d) at (3.0,0.0)   [draw, circle, thick, fill=black, scale=0.3] {\,};
    \node (e) at (4.0,0.0)   [draw, circle, thick, fill=black, scale=0.3] {\,};
    \node (f) at (5.0,0.0)   {$\Rightarrow_{r_2}$};
    \node (g) at (6.0,0.0)   [draw, circle, thick, fill=black, scale=0.3] {\,};
    \node (h) at (7.0,0.0)   [draw, circle, thick, fill=black, scale=0.3] {\,};

    \node (A) at (0.0,-0.18)  {\tiny{1}};
    \node (B) at (1.0,-0.18)  {\tiny{2}};
    \node (D) at (3.0,-0.18)  {\tiny{1}};
    \node (E) at (4.0,-0.18)  {\tiny{2}};
    \node (G) at (6.0,-0.18)  {\tiny{1}};
    \node (H) at (7.0,-0.18)  {\tiny{2}};

    \draw (a) edge[->,thick,loop left] node[left] {b} (a)
          (d) edge[->,thick] node[above] {a} (e)
          (h) edge[->,thick,loop right] node[right] {b} (h);
\end{tikzpicture}
\caption{Critical pair for Example \ref{eg:1}}
\label{fig:eg-1-crit-pair}
\end{figure}
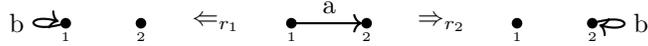

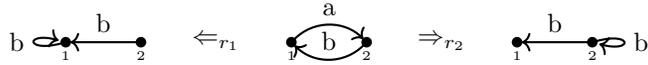
\begin{figure}[!ht]
\centering
\begin{tikzpicture}[every node/.style={align=center}]
    \node (a) at (0.0,0.0)   [draw, circle, thick, fill=black, scale=0.3] {\,};
    \node (b) at (1.0,0.0)   [draw, circle, thick, fill=black, scale=0.3] {\,};
    \node (c) at (2.0,0.0)   {$\Leftarrow_{r_1}$};
    \node (d) at (3.0,0.0)   [draw, circle, thick, fill=black, scale=0.3] {\,};
    \node (e) at (4.0,0.0)   [draw, circle, thick, fill=black, scale=0.3] {\,};
    \node (f) at (5.0,0.0)   {$\Rightarrow_{r_2}$};
    \node (g) at (6.0,0.0)   [draw, circle, thick, fill=black, scale=0.3] {\,};
    \node (h) at (7.0,0.0)   [draw, circle, thick, fill=black, scale=0.3] {\,};

    \node (A) at (0.0,-0.18)  {\tiny{1}};
    \node (B) at (1.0,-0.18)  {\tiny{2}};
    \node (D) at (3.0,-0.18)  {\tiny{1}};
    \node (E) at (4.0,-0.18)  {\tiny{2}};
    \node (G) at (6.0,-0.18)  {\tiny{1}};
    \node (H) at (7.0,-0.18)  {\tiny{2}};

    \draw (a) edge[->,thick,loop left] node[left] {b} (a)
          (b) edge[->,thick] node[above] {b} (a)
          (d) edge[->,thick, bend left=45] node[above] {a} (e)
          (e) edge[->,thick, bend left=45] node[above] {b} (d)
          (h) edge[->,thick] node[above] {b} (g)
          (h) edge[->,thick,loop right] node[right] {b} (h);
\end{tikzpicture}
\caption{Non-strongly joinable pair for Example \ref{eg:1}}
\label{fig:eg-1-nsj-pair}
\end{figure}

Recall that strong joinability of all critical pairs is a sufficient condition for local confluence (Theorem \ref{thm:critpairlem}). This was first shown by Plump in 1993 \cite{Plump93b}. Combining this with Newman's Lemma (Theorem \ref{thm:newmanlem}), we have a checkable condition for confluence of a GT system. Unlike for string and term rewriting, joinability is not sufficient to show local confluence, as demonstrated by the following example due to Plump \cite{Plump05a}:

\begin{example} \label{eg:1}
Consider the terminating GT system with the rules from Figure \ref{fig:eg-1-rules}. The only critical pair (Figure \ref{fig:eg-1-crit-pair}) is joinable, but not strongly, and the system is not locally confluent due to the counter example in Figure \ref{fig:eg-1-nsj-pair}.
\end{example}

\begin{figure}[!ht]
\centering
\begin{tikzpicture}[every node/.style={align=center}]
    \node (a) at (0.0,-0.05) {$r_1$:};
    \node (b) at (1.0,0.0)   [draw, circle, thick, fill=black, scale=0.3] {\,};
    \node (c) at (2.0,0.0)   [draw, circle, thick, fill=black, scale=0.3] {\,};
    \node (d) at (3.0,0.0)   {$\leftarrow$};
    \node (e) at (4.0,0.0)   [draw, circle, thick, fill=black, scale=0.3] {\,};
    \node (f) at (5.0,0.0)   [draw, circle, thick, fill=black, scale=0.3] {\,};
    \node (g) at (6.0,0.0)   {$\rightarrow$};
    \node (h) at (7.5,0.0)   [draw, circle, thick, fill=black, scale=0.3] {\,};
    \node (i) at (8.5,0.0)   [draw, circle, thick, fill=black, scale=0.3] {\,};

    \node (B) at (1.0,-0.18)  {\tiny{1}};
    \node (D) at (2.0,-0.18)  {\tiny{2}};
    \node (F) at (4.0,-0.18)  {\tiny{1}};
    \node (G) at (5.0,-0.18)  {\tiny{2}};
    \node (I) at (7.5,-0.18)  {\tiny{1}};
    \node (J) at (8.5,-0.18)  {\tiny{2}};

    \draw (b) edge[->,thick] node[above] {a} (c)
          (h) edge[->,thick,loop left] node[left] {b} (h);

    \node (a) at (0.0,-1.05) {$r_2$:};
    \node (b) at (1.0,-1.0)   [draw, circle, thick, fill=black, scale=0.3] {\,};
    \node (c) at (2.0,-1.0)   [draw, circle, thick, fill=black, scale=0.3] {\,};
    \node (d) at (3.0,-1.0)   {$\leftarrow$};
    \node (e) at (4.0,-1.0)   [draw, circle, thick, fill=black, scale=0.3] {\,};
    \node (f) at (5.0,-1.0)   [draw, circle, thick, fill=black, scale=0.3] {\,};
    \node (g) at (6.0,-1.0)   {$\rightarrow$};
    \node (h) at (7.0,-1.0)   [draw, circle, thick, fill=black, scale=0.3] {\,};
    \node (i) at (8.0,-1.0)   [draw, circle, thick, fill=black, scale=0.3] {\,};

    \node (B) at (1.0,-1.18)  {\tiny{1}};
    \node (D) at (2.0,-1.18)  {\tiny{2}};
    \node (F) at (4.0,-1.18)  {\tiny{1}};
    \node (G) at (5.0,-1.18)  {\tiny{2}};
    \node (I) at (7.0,-1.18)  {\tiny{1}};
    \node (J) at (8.0,-1.18)  {\tiny{2}};

    \draw (b) edge[->,thick] node[above] {a} (c)
          (i) edge[->,thick,loop right] node[right] {b} (i);
\end{tikzpicture}
\caption{Rules for Example \ref{eg:1}}
\label{fig:eg-1-rules}
\end{figure}
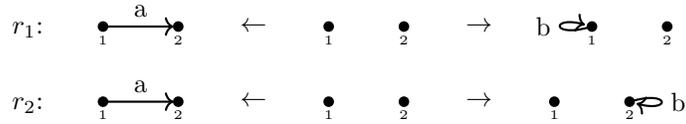

In this section, we generalise Plump's critical pair analysis for confluence up to garbage. We delay the treatment of subcommutativity up to garbage to Section \ref{sec:subcommuativity}, for ease of reading.

\subsection{Subgraph Closure and Subgraph Closed Languages}

In the original proof of the Critical Pair Lemma for (hyper)graphs \cite{Plump93b}, the argument is that if a pair of derivations is not parallelly independent, then it must be the case that a critical pair can be embedded within it. In our new setting, the possible start graphs will be restricted, since some of the graphs will be \textit{garbage}. We are only interested in those critical pairs with start graphs that can be embedded in non-garbage graphs. This is exactly the statement that the start graph of the critical pair is in the subgraph closure of the non-garbage graphs. We start this subsection by defining subgraph closure.

\begin{definition} \label{dfn:subgraphclosure}
Let \(\mathcal{D} \subseteq \mathcal{G}(\Sigma)\) be a language over some signature \(\Sigma\). Then \(\mathcal{D}\) is \textit{subgraph closed} if for all graphs \(G\), \(H\), such that \(H \subseteq G\), if \(G \in \mathcal{D}\), then \(H \in \mathcal{D}\). The \textit{subgraph closure} of \(\mathcal{D}\), denoted \(\widehat{\mathcal{D}}\), is the smallest language (with respect to inclusion) containing \(\mathcal{D}\) that is \textit{subgraph closed}.
\end{definition}

\begin{lemma}
Given a language \(\mathcal{D} \subseteq \mathcal{G}(\Sigma)\), \(\widehat{\mathcal{D}}\) always \textit{exists}, and is \textit{unique}. Moreover, \(\mathcal{D} = \widehat{\mathcal{D}}\) if and only if \(\mathcal{D}\) is \textit{subgraph closed}.
\end{lemma}

\begin{proof}
The key observations are that the subgraph relation is transitive, and each graph has only finitely many subgraphs. Clearly, the smallest possible set containing \(\mathcal{D}\) is just the union of all subgraphs of the elements of \(\mathcal{D}\), up to isomorphism. This is the unique subgraph closure of \(\mathcal{D}\).
\end{proof}

\(\widehat{\mathcal{D}}\) always exists, however it need not be decidable, even when \(\mathcal{D}\) is! It is not obvious what conditions on \(\mathcal{D}\) ensure that \(\widehat{\mathcal{D}}\) is decidable. If we move to the setting of string rewriting, there are some known cases where this can be solved. The classes of regular and context-free string languages are closed under substring closure, and the substring membership problem is decidable for context-free grammars \cite{Berstel79a} due to the fact that showing closure is constructive. We return to the issue of deciding subgraph membership in Subsection \ref{subsec:critpairgen}.

\begin{example}
The following graph languages are subgraph closed, over any signature \(\Sigma\):

\begin{enumerate}
    \item the empty language \(\emptyset\) and the language of all graphs \(\mathcal{G}(\Sigma)\);
    \item the language of discrete graphs;
    \item the language of acyclic graphs;
    \item the language of planar graphs;
    \item the language of \(k\)-colourable graphs for any fixed \(k \geq 2\);
    \item the language of bounded degree graphs for any fixed bound;
    \item the language of bounded treewidth graphs for any fixed bound.
\end{enumerate}
\end{example}

\begin{example}
The subgraph closure of the language of trees is the language of forests. The subgraph closure of the language of connected graphs is the language of all graphs.
\end{example}

\subsection{Generalising the Critical Pair Lemma}

We now define non-garbage critical pairs, which allow us to ignore certain pairs, which if all are strongly joinable, will allow us to conclude local confluence up to garbage, even in the presence of (local) non-confluence on all graphs.

\begin{definition}
Let \(T = (\Sigma, \mathcal{R})\) be a GT system and \(\mathcal{D} \subseteq \mathcal{G}(\Sigma)\) a language. A \textit{critical pair} \(H_1 \Leftarrow G \Rightarrow H_2\) is \(\mathcal{D} \)-\textit{non-garbage} if \(G \in \widehat{\mathcal{D}}\).
\end{definition}

\begin{lemma} \label{lem:fincritpairs}
Given a GT system \(T = (\Sigma, \mathcal{R})\) and a language \(\mathcal{D} \subseteq \mathcal{G}(\Sigma)\), then there are only finitely many \(\mathcal{D} \)-\textit{non-garbage critical pairs} (up to isomorphism).
\end{lemma}

\begin{proof}
Recall from Subsection \ref{subsec:confluence_checking} that any GT system has only finitely many critical pairs. Filtering out those that are garbage or isomorphic is certainly only going to leave us with a finite number of critical pairs.
\end{proof}

Of course, just because there are only finitely many non-garbage critical pairs, it doesn't mean that generation of them is effective, in general. In order to avoid interrupting the flow, we will discuss this further in Subsection \ref{subsec:critpairgen}. We now proceed to present the main result:

\begin{theorem}[Generalised Critical Pair Lemma] \label{thm:ngcritpairlem}
Let \(T = (\Sigma, \mathcal{R})\) be a GT system and \(\mathcal{D} \subseteq \mathcal{G}(\Sigma)\) a language. If all \(T\)'s \(\mathcal{D}\)-\textit{non-garbage critical pairs} are \textit{strongly joinable}, then \(T\) is \textit{locally confluent up to garbage} on \(\mathcal{D}\).
\end{theorem}

\begin{proof}
Our proof is a generalisation of Plump's original proof of the Critical Pair Lemma for (hyper)graph transformation systems (Theorem \ref{thm:critpairlem}). We need to show that every pair of derivations \(H_1 \Leftarrow_{r_1,g_1} G \Rightarrow_{r_2,g_2} H_2\) such that \(G\) is non-garbage can be joined. There are two cases to consider. Firstly, if the derivations are parallelly independent, then by Theorem \ref{thm:commute}, the result is immediate. Otherwise, we must consider the case that they are not parallelly independent.

By Theorem \ref{thm:clipping}, we can factor out a pair \(T_1 \Leftarrow S \Rightarrow T_2\). Since critical pairs are, by construction, the overlaps of rule left hand sides, it must be the case that this pair is actually a critical pair. Moreover, since \(G \in \mathcal{D}\), then \(S \in \widehat{\mathcal{D}}\) and so the critical pair must be non-garbage, and must be strongly joinable to \(U\). We can now apply Theorem \ref{thm:embedding} to \(T_1 \Rightarrow^* U\) and \(T_2 \Rightarrow^* U\), separately, giving result graphs \(M_1\) and \(M_2\) (applicability of the theorem is a consequence of strong joinability). To see that \(M_1\) and \(M_2\) are isomorphic follows from elementary properties of pushouts along monomorphisms \cite{Plump05a}.
\end{proof}

\begin{figure}[!ht]
\centering
\includegraphics[totalheight=4.0cm]{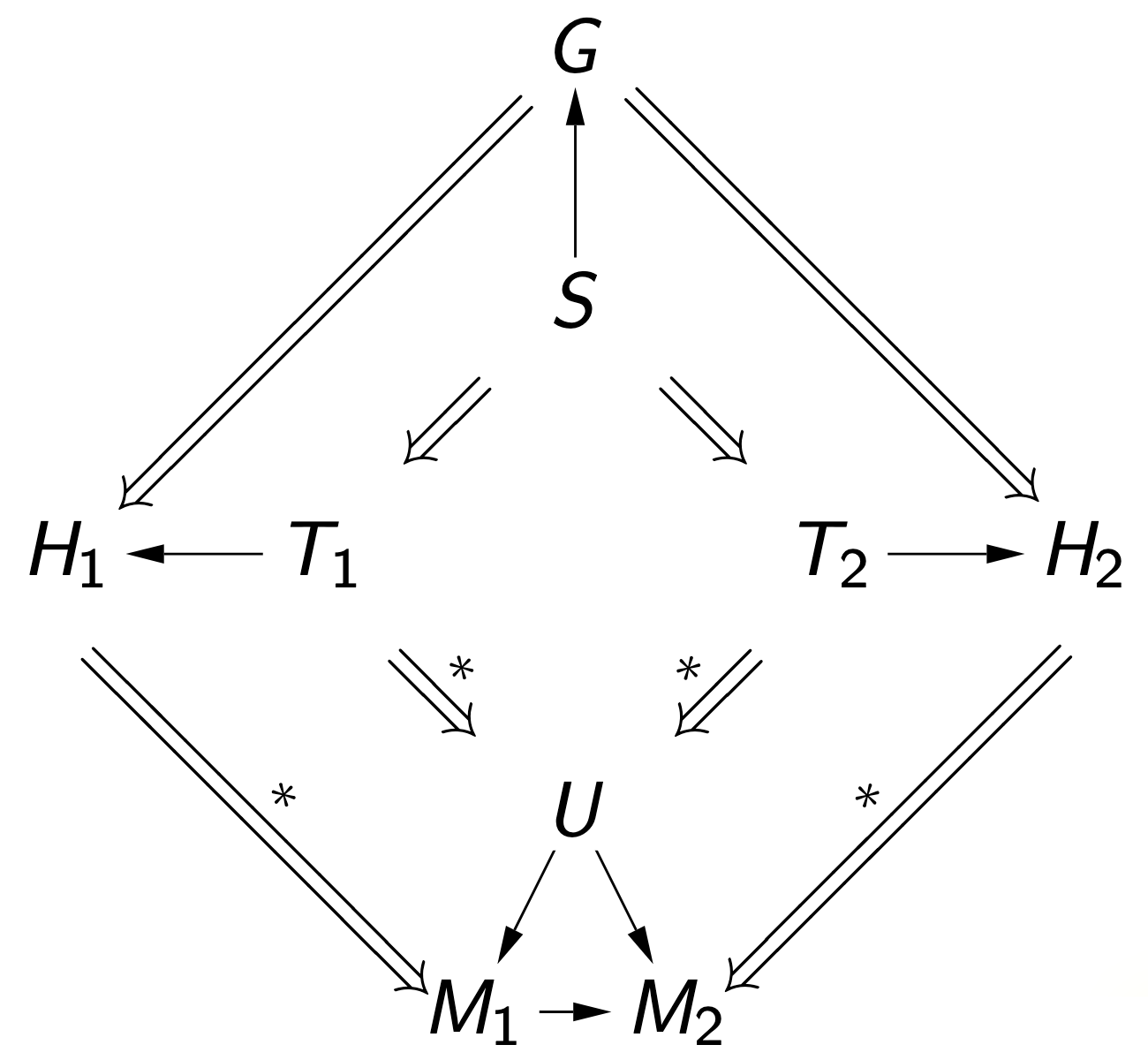}
\caption{Diagram for the proof of Theorem \ref{thm:ngcritpairlem}}
\label{fig:critpairdiag}
\end{figure}

Just as with the original Critical Pair Lemma in Subsection \ref{subsec:confluence_checking}, it is sufficient to only check the non-isomorphic critical pairs for strong joinability due to the fact that derivations based on a critical pair can be reset as derivations based on any other isomorphic critical pair simply by passing through the isomorphism. Thus, if all \(T\)'s non-isomorphic \(\mathcal{D}\)-non-garbage critical pairs are strongly joinable, then \(T\) is locally confluent up to garbage on \(\mathcal{D}\).

The most common use case of this generalised critical pair will be the following corollary, where the aim is not to show local confluence up to garbage, but confluence up to garbage, given termination up to garbage:

\begin{corollary} \label{cor:ngcritpairlem}
Let \(T = (\Sigma, \mathcal{R})\) be a GT system and \(\mathcal{D} \subseteq \mathcal{G}(\Sigma)\) a language. If \(\mathcal{D}\) is \textit{closed} under \(T\), \(T\) is \textit{terminating up to garbage} on \(\mathcal{D}\), and all \(T\)'s \textit{non-isomorphic} \(\mathcal{D}\)-\textit{non-garbage critical pairs} are \textit{strongly joinable}, then \(T\) is \textit{confluent up to garbage} on \(\mathcal{D}\).
\end{corollary}

\begin{proof}
By the above theorem, \(T\) is \textit{locally confluent up to garbage}, so by the Generalised Newman's Lemma (Theorem \ref{thm:newmangarbage}), \(T\) is \textit{confluent up to garbage}.
\end{proof}

\begin{example} \label{eg:4}
Recall from Subsection \ref{subsec:confluence_garbage}, the non-confluent GT system from Example \ref{eg:b} (Figure \ref{fig:eg-b-rules}). We can use Corollary \ref{cor:ngcritpairlem} to show that this system is confluent up to garbage on the language of acyclic graphs with edge labels \(a\) and \(b\), \(\mathcal{D}\). First, we observe that the rules (Figure \ref{eg:b}) are terminating, and that \(\mathcal{D}\) is closed under the rules. Next, we observe that the system has \(16\) non-isomorphic critical pairs. Figure \ref{fig:critpairs1} shows, for each of the pairs, if they are joinable, strongly joinable, or \(\mathcal{D}\)-non-garbage. From this, we can see that every non-garbage critical pair is strongly joinable, and so the system is confluent up to garbage on \(\mathcal{D}\).
\end{example}

\begin{figure}[!ht]
\centering
\scalebox{0.825}{
\begin{tabular}{|l|c|c|c|c|}
\hline
\multicolumn{1}{|>{\centering\arraybackslash}p{6.4cm}|}{Pair/Property} & \multicolumn{1}{>{\centering\arraybackslash}p{1.666667cm}|}{Joinable} & \multicolumn{1}{>{\centering\arraybackslash}p{1.666667cm}|}{Strongly Joinable} & \multicolumn{1}{>{\centering\arraybackslash}p{1.666667cm}|}{Non-Garbage} \\ \hline
\scalebox{0.666667}{\begin{tikzpicture}[every node/.style={align=center}]
    \node (x) at (-0.1,-0.25) {\,};
    \node (y) at (7.1,1.25)   {\,};

    \node (a) at (0.0,0.25)   [draw, circle, thick, fill=black, scale=0.3] {\,};
    \node (b) at (1.5,0.25)   [draw, circle, thick, fill=black, scale=0.3] {\,};
    \node (c) at (2.5,0.4)    {$\Leftarrow_{s_1}$};
    \node (d) at (3.5,0.0)    [draw, circle, thick, fill=black, scale=0.3] {\,};
    \node (e) at (4.5,1.0)    [draw, circle, thick, fill=black, scale=0.3] {\,};
    \node (f) at (5.5,0.0)    [draw, circle, thick, fill=black, scale=0.3] {\,};
    \node (g) at (6.5,0.4)    {$\Rightarrow_{s_1}$};
    \node (h) at (7.5,0.25)   [draw, circle, thick, fill=black, scale=0.3] {\,};
    \node (i) at (9.0,0.25)   [draw, circle, thick, fill=black, scale=0.3] {\,};

    \node (A) at (0.0,0.07)   {\tiny{1}};
    \node (B) at (1.5,0.07)   {\tiny{2}};
    \node (D) at (3.5,-0.18)  {\tiny{1}};
    \node (E) at (4.5,0.82)   {\tiny{2}};
    \node (F) at (5.5,-0.18)  {\tiny{3}};
    \node (H) at (7.5,0.07)   {\tiny{1}};
    \node (I) at (9.0,0.07)   {\tiny{3}};

    \draw (a) edge[->,thick,bend left=60] node[above] {a} (b)
          (b) edge[->,thick] node[above] {a} (a)
          (d) edge[->,thick] node[above,xshift=-1.5mm,yshift=-1mm] {a} (e)
          (e) edge[->,thick] node[above,xshift=1.5mm,yshift=-1mm] {a} (f)
          (f) edge[->,thick] node[above] {a} (d)
          (h) edge[->,thick,bend left=60] node[above] {a} (i)
          (i) edge[->,thick] node[above] {a} (h);
\end{tikzpicture}} & \ding{51} & \ding{51} & \ding{55} \\ \hline
\scalebox{0.666667}{\begin{tikzpicture}[every node/.style={align=center}]
    \node (x) at (-0.1,-0.25) {\,};
    \node (y) at (7.1,1.25)   {\,};

    \node (a) at (0.0,0.333333)      [draw, circle, thick, fill=black, scale=0.3] {\,};
    \node (b) at (0.75,0.333333)     [draw, circle, thick, fill=black, scale=0.3] {\,};
    \node (c) at (1.5,0.333333)      [draw, circle, thick, fill=black, scale=0.3] {\,};
    \node (d) at (2.5,0.4)           {$\Leftarrow_{s_1}$};
    \node (e) at (3.5,0.333333)      [draw, circle, thick, fill=black, scale=0.3] {\,};
    \node (f) at (4.166667,0.333333) [draw, circle, thick, fill=black, scale=0.3] {\,};
    \node (g) at (4.866667,0.333333) [draw, circle, thick, fill=black, scale=0.3] {\,};
    \node (h) at (5.5,0.333333)      [draw, circle, thick, fill=black, scale=0.3] {\,};
    \node (i) at (6.5,0.4)           {$\Rightarrow_{s_1}$};
    \node (j) at (7.5,0.333333)      [draw, circle, thick, fill=black, scale=0.3] {\,};
    \node (k) at (8.25,0.333333)     [draw, circle, thick, fill=black, scale=0.3] {\,};
    \node (l) at (9.0,0.333333)      [draw, circle, thick, fill=black, scale=0.3] {\,};

    \node (A) at (0.0,0.153333)      {\tiny{1}};
    \node (B) at (0.75,0.153333)     {\tiny{2}};
    \node (C) at (1.5,0.153333)      {\tiny{4}};
    \node (E) at (3.5,0.153333)      {\tiny{1}};
    \node (F) at (4.166667,0.153333) {\tiny{2}};
    \node (G) at (4.866667,0.153333) {\tiny{3}};
    \node (H) at (5.5,0.153333)      {\tiny{4}};
    \node (J) at (7.5,0.153333)      {\tiny{1}};
    \node (K) at (8.25,0.153333)     {\tiny{3}};
    \node (L) at (9.0,0.153333)      {\tiny{4}};

    \draw (a) edge[->,thick] node[above] {a} (b)
          (b) edge[->,thick] node[above] {a} (c)
          (e) edge[->,thick] node[above] {a} (f)
          (f) edge[->,thick] node[above] {a} (g)
          (g) edge[->,thick] node[above] {a} (h)
          (j) edge[->,thick] node[above] {a} (k)
          (k) edge[->,thick] node[above] {a} (l);
\end{tikzpicture}} & \ding{51} & \ding{51} & \ding{51} \\ \hline
\scalebox{0.666667}{\begin{tikzpicture}[every node/.style={align=center}]
    \node (x) at (-0.1,-0.25) {\,};
    \node (y) at (7.1,1.25)   {\,};

    \node (a) at (0.0,0.25)   [draw, circle, thick, fill=black, scale=0.3] {\,};
    \node (b) at (1.5,0.25)   [draw, circle, thick, fill=black, scale=0.3] {\,};
    \node (c) at (2.5,0.4)    {$\Leftarrow_{s_1}$};
    \node (d) at (3.5,0.0)    [draw, circle, thick, fill=black, scale=0.3] {\,};
    \node (e) at (4.5,1.0)    [draw, circle, thick, fill=black, scale=0.3] {\,};
    \node (f) at (5.5,0.0)    [draw, circle, thick, fill=black, scale=0.3] {\,};
    \node (g) at (6.5,0.4)    {$\Rightarrow_{s_2}$};
    \node (h) at (7.5,0.25)   [draw, circle, thick, fill=black, scale=0.3] {\,};
    \node (i) at (9.0,0.25)   [draw, circle, thick, fill=black, scale=0.3] {\,};

    \node (A) at (0.0,0.07)   {\tiny{1}};
    \node (B) at (1.5,0.07)   {\tiny{3}};
    \node (D) at (3.5,-0.18)  {\tiny{1}};
    \node (E) at (4.5,0.82)   {\tiny{2}};
    \node (F) at (5.5,-0.18)  {\tiny{3}};
    \node (H) at (7.5,0.07)   {\tiny{1}};
    \node (I) at (9.0,0.07)   {\tiny{2}};

    \draw (a) edge[->,thick,bend left=60] node[above] {a} (b)
          (b) edge[->,thick] node[above] {b} (a)
          (d) edge[->,thick] node[above,xshift=-1.5mm,yshift=-1mm] {a} (e)
          (e) edge[->,thick] node[above,xshift=1.5mm,yshift=-1mm] {a} (f)
          (f) edge[->,thick] node[above] {b} (d)
          (h) edge[->,thick,bend left=60] node[above] {a} (i)
          (i) edge[->,thick] node[above] {a} (h);
\end{tikzpicture}} & \ding{55} & \ding{55} & \ding{55} \\ \hline
\scalebox{0.666667}{\begin{tikzpicture}[every node/.style={align=center}]
    \node (x) at (-0.1,-0.25) {\,};
    \node (y) at (7.1,1.25)   {\,};

    \node (a) at (0.0,0.333333)      [draw, circle, thick, fill=black, scale=0.3] {\,};
    \node (b) at (0.75,0.333333)     [draw, circle, thick, fill=black, scale=0.3] {\,};
    \node (c) at (1.5,0.333333)      [draw, circle, thick, fill=black, scale=0.3] {\,};
    \node (d) at (2.5,0.4)           {$\Leftarrow_{s_1}$};
    \node (e) at (3.5,0.333333)      [draw, circle, thick, fill=black, scale=0.3] {\,};
    \node (f) at (4.166667,0.333333) [draw, circle, thick, fill=black, scale=0.3] {\,};
    \node (g) at (4.866667,0.333333) [draw, circle, thick, fill=black, scale=0.3] {\,};
    \node (h) at (5.5,0.333333)      [draw, circle, thick, fill=black, scale=0.3] {\,};
    \node (i) at (6.5,0.4)           {$\Rightarrow_{s_2}$};
    \node (j) at (7.5,0.333333)      [draw, circle, thick, fill=black, scale=0.3] {\,};
    \node (k) at (8.25,0.333333)     [draw, circle, thick, fill=black, scale=0.3] {\,};
    \node (l) at (9.0,0.333333)      [draw, circle, thick, fill=black, scale=0.3] {\,};

    \node (A) at (0.0,0.153333)      {\tiny{1}};
    \node (B) at (0.75,0.153333)     {\tiny{3}};
    \node (C) at (1.5,0.153333)      {\tiny{4}};
    \node (E) at (3.5,0.153333)      {\tiny{1}};
    \node (F) at (4.166667,0.153333) {\tiny{2}};
    \node (G) at (4.866667,0.153333) {\tiny{3}};
    \node (H) at (5.5,0.153333)      {\tiny{4}};
    \node (J) at (7.5,0.153333)      {\tiny{1}};
    \node (K) at (8.25,0.153333)     {\tiny{2}};
    \node (L) at (9.0,0.153333)      {\tiny{4}};

    \draw (a) edge[->,thick] node[above] {a} (b)
          (b) edge[->,thick] node[above] {b} (c)
          (e) edge[->,thick] node[above] {a} (f)
          (f) edge[->,thick] node[above] {a} (g)
          (g) edge[->,thick] node[above] {b} (h)
          (j) edge[->,thick] node[above] {a} (k)
          (k) edge[->,thick] node[above] {a} (l);
\end{tikzpicture}} & \ding{51} & \ding{51} & \ding{51} \\ \hline
\scalebox{0.666667}{\begin{tikzpicture}[every node/.style={align=center}]
    \node (x) at (-0.1,-0.25) {\,};
    \node (y) at (7.1,1.25)   {\,};

    \node (a) at (0.0,0.25)   [draw, circle, thick, fill=black, scale=0.3] {\,};
    \node (b) at (1.5,0.25)   [draw, circle, thick, fill=black, scale=0.3] {\,};
    \node (c) at (2.5,0.4)    {$\Leftarrow_{s_1}$};
    \node (d) at (3.5,0.0)    [draw, circle, thick, fill=black, scale=0.3] {\,};
    \node (e) at (4.5,1.0)    [draw, circle, thick, fill=black, scale=0.3] {\,};
    \node (f) at (5.5,0.0)    [draw, circle, thick, fill=black, scale=0.3] {\,};
    \node (g) at (6.5,0.4)    {$\Rightarrow_{s_3}$};
    \node (h) at (7.5,0.25)   [draw, circle, thick, fill=black, scale=0.3] {\,};
    \node (i) at (9.0,0.25)   [draw, circle, thick, fill=black, scale=0.3] {\,};

    \node (A) at (0.0,0.07)   {\tiny{1}};
    \node (B) at (1.5,0.07)   {\tiny{3}};
    \node (D) at (3.5,-0.18)  {\tiny{1}};
    \node (E) at (4.5,0.82)   {\tiny{2}};
    \node (F) at (5.5,-0.18)  {\tiny{3}};
    \node (H) at (7.5,0.07)   {\tiny{2}};
    \node (I) at (9.0,0.07)   {\tiny{3}};

    \draw (a) edge[->,thick,bend left=60] node[above] {a} (b)
          (b) edge[->,thick] node[above] {b} (a)
          (d) edge[->,thick] node[above,xshift=-1.5mm,yshift=-1mm] {a} (e)
          (e) edge[->,thick] node[above,xshift=1.5mm,yshift=-1mm] {a} (f)
          (f) edge[->,thick] node[above] {b} (d)
          (h) edge[->,thick,bend left=60] node[above] {a} (i)
          (i) edge[->,thick] node[above] {a} (h);
\end{tikzpicture}} & \ding{55} & \ding{55} & \ding{55} \\ \hline
\scalebox{0.666667}{\begin{tikzpicture}[every node/.style={align=center}]
    \node (x) at (-0.1,-0.25) {\,};
    \node (y) at (7.1,1.25)   {\,};

    \node (a) at (0.0,0.333333)      [draw, circle, thick, fill=black, scale=0.3] {\,};
    \node (b) at (0.75,0.333333)     [draw, circle, thick, fill=black, scale=0.3] {\,};
    \node (c) at (1.5,0.333333)      [draw, circle, thick, fill=black, scale=0.3] {\,};
    \node (d) at (2.5,0.4)           {$\Leftarrow_{s_1}$};
    \node (e) at (3.5,0.333333)      [draw, circle, thick, fill=black, scale=0.3] {\,};
    \node (f) at (4.166667,0.333333) [draw, circle, thick, fill=black, scale=0.3] {\,};
    \node (g) at (4.866667,0.333333) [draw, circle, thick, fill=black, scale=0.3] {\,};
    \node (h) at (5.5,0.333333)      [draw, circle, thick, fill=black, scale=0.3] {\,};
    \node (i) at (6.5,0.4)           {$\Rightarrow_{s_3}$};
    \node (j) at (7.5,0.333333)      [draw, circle, thick, fill=black, scale=0.3] {\,};
    \node (k) at (8.25,0.333333)     [draw, circle, thick, fill=black, scale=0.3] {\,};
    \node (l) at (9.0,0.333333)      [draw, circle, thick, fill=black, scale=0.3] {\,};

    \node (A) at (0.0,0.153333)      {\tiny{1}};
    \node (B) at (0.75,0.153333)     {\tiny{2}};
    \node (C) at (1.5,0.153333)      {\tiny{4}};
    \node (E) at (3.5,0.153333)      {\tiny{1}};
    \node (F) at (4.166667,0.153333) {\tiny{2}};
    \node (G) at (4.866667,0.153333) {\tiny{3}};
    \node (H) at (5.5,0.153333)      {\tiny{4}};
    \node (J) at (7.5,0.153333)      {\tiny{1}};
    \node (K) at (8.25,0.153333)     {\tiny{3}};
    \node (L) at (9.0,0.153333)      {\tiny{4}};

    \draw (a) edge[->,thick] node[above] {b} (b)
          (b) edge[->,thick] node[above] {a} (c)
          (e) edge[->,thick] node[above] {b} (f)
          (f) edge[->,thick] node[above] {a} (g)
          (g) edge[->,thick] node[above] {a} (h)
          (j) edge[->,thick] node[above] {a} (k)
          (k) edge[->,thick] node[above] {a} (l);
\end{tikzpicture}} & \ding{51} & \ding{51} & \ding{51} \\ \hline
\scalebox{0.666667}{\begin{tikzpicture}[every node/.style={align=center}]
    \node (x) at (-0.1,-0.25) {\,};
    \node (y) at (7.1,1.25)   {\,};

    \node (a) at (0.0,0.25)   [draw, circle, thick, fill=black, scale=0.3] {\,};
    \node (b) at (1.5,0.25)   [draw, circle, thick, fill=black, scale=0.3] {\,};
    \node (c) at (2.5,0.4)    {$\Leftarrow_{s_2}$};
    \node (d) at (3.5,0.0)    [draw, circle, thick, fill=black, scale=0.3] {\,};
    \node (e) at (4.5,1.0)    [draw, circle, thick, fill=black, scale=0.3] {\,};
    \node (f) at (5.5,0.0)    [draw, circle, thick, fill=black, scale=0.3] {\,};
    \node (g) at (6.5,0.4)    {$\Rightarrow_{s_3}$};
    \node (h) at (7.5,0.25)   [draw, circle, thick, fill=black, scale=0.3] {\,};
    \node (i) at (9.0,0.25)   [draw, circle, thick, fill=black, scale=0.3] {\,};

    \node (A) at (0.0,0.07)   {\tiny{1}};
    \node (B) at (1.5,0.07)   {\tiny{2}};
    \node (D) at (3.5,-0.18)  {\tiny{1}};
    \node (E) at (4.5,0.82)   {\tiny{2}};
    \node (F) at (5.5,-0.18)  {\tiny{3}};
    \node (H) at (7.5,0.07)   {\tiny{2}};
    \node (I) at (9.0,0.07)   {\tiny{3}};

    \draw (a) edge[->,thick,bend left=60] node[above] {a} (b)
          (b) edge[->,thick] node[above] {a} (a)
          (d) edge[->,thick] node[above,xshift=-1.5mm,yshift=-1mm] {a} (e)
          (e) edge[->,thick] node[above,xshift=1.5mm,yshift=-1mm] {a} (f)
          (f) edge[->,thick] node[above] {b} (d)
          (h) edge[->,thick,bend left=60] node[above] {a} (i)
          (i) edge[->,thick] node[above] {a} (h);
\end{tikzpicture}} & \ding{51} & \ding{51} & \ding{55} \\ \hline
\scalebox{0.666667}{\begin{tikzpicture}[every node/.style={align=center}]
    \node (x) at (-0.1,-0.25) {\,};
    \node (y) at (7.1,1.25)   {\,};

    \node (a) at (0.0,0.333333)      [draw, circle, thick, fill=black, scale=0.3] {\,};
    \node (b) at (0.75,0.333333)     [draw, circle, thick, fill=black, scale=0.3] {\,};
    \node (c) at (1.5,0.333333)      [draw, circle, thick, fill=black, scale=0.3] {\,};
    \node (d) at (2.5,0.4)           {$\Leftarrow_{s_2}$};
    \node (e) at (3.5,0.333333)      [draw, circle, thick, fill=black, scale=0.3] {\,};
    \node (f) at (4.166667,0.333333) [draw, circle, thick, fill=black, scale=0.3] {\,};
    \node (g) at (4.866667,0.333333) [draw, circle, thick, fill=black, scale=0.3] {\,};
    \node (h) at (5.5,0.333333)      [draw, circle, thick, fill=black, scale=0.3] {\,};
    \node (i) at (6.5,0.4)           {$\Rightarrow_{s_3}$};
    \node (j) at (7.5,0.333333)      [draw, circle, thick, fill=black, scale=0.3] {\,};
    \node (k) at (8.25,0.333333)     [draw, circle, thick, fill=black, scale=0.3] {\,};
    \node (l) at (9.0,0.333333)      [draw, circle, thick, fill=black, scale=0.3] {\,};

    \node (A) at (0.0,0.153333)      {\tiny{1}};
    \node (B) at (0.75,0.153333)     {\tiny{3}};
    \node (C) at (1.5,0.153333)      {\tiny{4}};
    \node (E) at (3.5,0.153333)      {\tiny{1}};
    \node (F) at (4.166667,0.153333) {\tiny{2}};
    \node (G) at (4.866667,0.153333) {\tiny{3}};
    \node (H) at (5.5,0.153333)      {\tiny{4}};
    \node (J) at (7.5,0.153333)      {\tiny{1}};
    \node (K) at (8.25,0.153333)     {\tiny{2}};
    \node (L) at (9.0,0.153333)      {\tiny{4}};

    \draw (a) edge[->,thick] node[above] {b} (b)
          (b) edge[->,thick] node[above] {a} (c)
          (e) edge[->,thick] node[above] {a} (f)
          (f) edge[->,thick] node[above] {b} (g)
          (g) edge[->,thick] node[above] {a} (h)
          (j) edge[->,thick] node[above] {a} (k)
          (k) edge[->,thick] node[above] {a} (l);
\end{tikzpicture}} & \ding{51} & \ding{51} & \ding{51} \\ \hline
\scalebox{0.666667}{\begin{tikzpicture}[every node/.style={align=center}]
    \node (x) at (-0.1,-0.25) {\,};
    \node (y) at (7.1,1.25)   {\,};

    \node (a) at (0.0,0.25)   [draw, circle, thick, fill=black, scale=0.3] {\,};
    \node (b) at (1.5,0.25)   [draw, circle, thick, fill=black, scale=0.3] {\,};
    \node (c) at (2.5,0.4)    {$\Leftarrow_{s_2}$};
    \node (d) at (3.5,0.0)    [draw, circle, thick, fill=black, scale=0.3] {\,};
    \node (e) at (4.5,1.0)    [draw, circle, thick, fill=black, scale=0.3] {\,};
    \node (f) at (5.5,0.0)    [draw, circle, thick, fill=black, scale=0.3] {\,};
    \node (g) at (6.5,0.4)    {$\Rightarrow_{s_3}$};
    \node (h) at (7.5,0.25)   [draw, circle, thick, fill=black, scale=0.3] {\,};
    \node (i) at (9.0,0.25)   [draw, circle, thick, fill=black, scale=0.3] {\,};

    \node (A) at (0.0,0.07)   {\tiny{1}};
    \node (B) at (1.5,0.07)   {\tiny{3}};
    \node (D) at (3.5,-0.18)  {\tiny{1}};
    \node (E) at (4.5,0.82)   {\tiny{2}};
    \node (F) at (5.5,-0.18)  {\tiny{3}};
    \node (H) at (7.5,0.07)   {\tiny{2}};
    \node (I) at (9.0,0.07)   {\tiny{3}};

    \draw (a) edge[->,thick,bend left=60] node[above] {a} (b)
          (b) edge[->,thick] node[above] {b} (a)
          (d) edge[->,thick] node[above,xshift=-1.5mm,yshift=-1mm] {a} (e)
          (e) edge[->,thick] node[above,xshift=1.5mm,yshift=-1mm] {b} (f)
          (f) edge[->,thick] node[above] {b} (d)
          (h) edge[->,thick,bend left=60] node[above] {b} (i)
          (i) edge[->,thick] node[above] {a} (h);
\end{tikzpicture}} & \ding{51} & \ding{55} & \ding{55} \\ \hline
\scalebox{0.666667}{\begin{tikzpicture}[every node/.style={align=center}]
    \node (x) at (-0.1,-0.25) {\,};
    \node (y) at (7.1,1.25)   {\,};

    \node (a) at (0.0,0.333333)      [draw, circle, thick, fill=black, scale=0.3] {\,};
    \node (b) at (0.75,0.333333)     [draw, circle, thick, fill=black, scale=0.3] {\,};
    \node (c) at (1.5,0.333333)      [draw, circle, thick, fill=black, scale=0.3] {\,};
    \node (d) at (2.5,0.4)           {$\Leftarrow_{s_2}$};
    \node (e) at (3.5,0.333333)      [draw, circle, thick, fill=black, scale=0.3] {\,};
    \node (f) at (4.166667,0.333333) [draw, circle, thick, fill=black, scale=0.3] {\,};
    \node (g) at (4.866667,0.333333) [draw, circle, thick, fill=black, scale=0.3] {\,};
    \node (h) at (5.5,0.333333)      [draw, circle, thick, fill=black, scale=0.3] {\,};
    \node (i) at (6.5,0.4)           {$\Rightarrow_{s_3}$};
    \node (j) at (7.5,0.333333)      [draw, circle, thick, fill=black, scale=0.3] {\,};
    \node (k) at (8.25,0.333333)     [draw, circle, thick, fill=black, scale=0.3] {\,};
    \node (l) at (9.0,0.333333)      [draw, circle, thick, fill=black, scale=0.3] {\,};

    \node (A) at (0.0,0.153333)      {\tiny{1}};
    \node (B) at (0.75,0.153333)     {\tiny{2}};
    \node (C) at (1.5,0.153333)      {\tiny{4}};
    \node (E) at (3.5,0.153333)      {\tiny{1}};
    \node (F) at (4.166667,0.153333) {\tiny{2}};
    \node (G) at (4.866667,0.153333) {\tiny{3}};
    \node (H) at (5.5,0.153333)      {\tiny{4}};
    \node (J) at (7.5,0.153333)      {\tiny{1}};
    \node (K) at (8.25,0.153333)     {\tiny{3}};
    \node (L) at (9.0,0.153333)      {\tiny{4}};

    \draw (a) edge[->,thick] node[above] {b} (b)
          (b) edge[->,thick] node[above] {a} (c)
          (e) edge[->,thick] node[above] {b} (f)
          (f) edge[->,thick] node[above] {a} (g)
          (g) edge[->,thick] node[above] {b} (h)
          (j) edge[->,thick] node[above] {a} (k)
          (k) edge[->,thick] node[above] {b} (l);
\end{tikzpicture}} & \ding{51} & \ding{51} & \ding{51} \\ \hline
\scalebox{0.666667}{\begin{tikzpicture}[every node/.style={align=center}]
    \node (x) at (-0.1,-0.25) {\,};
    \node (y) at (7.1,1.25)   {\,};

    \node (a) at (0.0,0.25)   [draw, circle, thick, fill=black, scale=0.3] {\,};
    \node (b) at (1.5,0.25)   [draw, circle, thick, fill=black, scale=0.3] {\,};
    \node (c) at (2.5,0.4)    {$\Leftarrow_{s_2}$};
    \node (d) at (3.5,0.0)    [draw, circle, thick, fill=black, scale=0.3] {\,};
    \node (e) at (4.5,1.0)    [draw, circle, thick, fill=black, scale=0.3] {\,};
    \node (f) at (5.5,0.0)    [draw, circle, thick, fill=black, scale=0.3] {\,};
    \node (g) at (6.5,0.4)    {$\Rightarrow_{s_4}$};
    \node (h) at (7.5,0.25)   [draw, circle, thick, fill=black, scale=0.3] {\,};
    \node (i) at (9.0,0.25)   [draw, circle, thick, fill=black, scale=0.3] {\,};

    \node (A) at (0.0,0.07)   {\tiny{1}};
    \node (B) at (1.5,0.07)   {\tiny{3}};
    \node (D) at (3.5,-0.18)  {\tiny{1}};
    \node (E) at (4.5,0.82)   {\tiny{2}};
    \node (F) at (5.5,-0.18)  {\tiny{3}};
    \node (H) at (7.5,0.07)   {\tiny{1}};
    \node (I) at (9.0,0.07)   {\tiny{2}};

    \draw (a) edge[->,thick,bend left=60] node[above] {a} (b)
          (b) edge[->,thick] node[above] {b} (a)
          (d) edge[->,thick] node[above,xshift=-1.5mm,yshift=-1mm] {a} (e)
          (e) edge[->,thick] node[above,xshift=1.5mm,yshift=-1mm] {b} (f)
          (f) edge[->,thick] node[above] {b} (d)
          (h) edge[->,thick,bend left=60] node[above] {a} (i)
          (i) edge[->,thick] node[above] {a} (h);
\end{tikzpicture}} & \ding{55} & \ding{55} & \ding{55} \\ \hline
\scalebox{0.666667}{\begin{tikzpicture}[every node/.style={align=center}]
    \node (x) at (-0.1,-0.25) {\,};
    \node (y) at (7.1,1.25)   {\,};

    \node (a) at (0.0,0.333333)      [draw, circle, thick, fill=black, scale=0.3] {\,};
    \node (b) at (0.75,0.333333)     [draw, circle, thick, fill=black, scale=0.3] {\,};
    \node (c) at (1.5,0.333333)      [draw, circle, thick, fill=black, scale=0.3] {\,};
    \node (d) at (2.5,0.4)           {$\Leftarrow_{s_2}$};
    \node (e) at (3.5,0.333333)      [draw, circle, thick, fill=black, scale=0.3] {\,};
    \node (f) at (4.166667,0.333333) [draw, circle, thick, fill=black, scale=0.3] {\,};
    \node (g) at (4.866667,0.333333) [draw, circle, thick, fill=black, scale=0.3] {\,};
    \node (h) at (5.5,0.333333)      [draw, circle, thick, fill=black, scale=0.3] {\,};
    \node (i) at (6.5,0.4)           {$\Rightarrow_{s_4}$};
    \node (j) at (7.5,0.333333)      [draw, circle, thick, fill=black, scale=0.3] {\,};
    \node (k) at (8.25,0.333333)     [draw, circle, thick, fill=black, scale=0.3] {\,};
    \node (l) at (9.0,0.333333)      [draw, circle, thick, fill=black, scale=0.3] {\,};

    \node (A) at (0.0,0.153333)      {\tiny{1}};
    \node (B) at (0.75,0.153333)     {\tiny{3}};
    \node (C) at (1.5,0.153333)      {\tiny{4}};
    \node (E) at (3.5,0.153333)      {\tiny{1}};
    \node (F) at (4.166667,0.153333) {\tiny{2}};
    \node (G) at (4.866667,0.153333) {\tiny{3}};
    \node (H) at (5.5,0.153333)      {\tiny{4}};
    \node (J) at (7.5,0.153333)      {\tiny{1}};
    \node (K) at (8.25,0.153333)     {\tiny{2}};
    \node (L) at (9.0,0.153333)      {\tiny{4}};

    \draw (a) edge[->,thick] node[above] {a} (b)
          (b) edge[->,thick] node[above] {b} (c)
          (e) edge[->,thick] node[above] {a} (f)
          (f) edge[->,thick] node[above] {b} (g)
          (g) edge[->,thick] node[above] {b} (h)
          (j) edge[->,thick] node[above] {a} (k)
          (k) edge[->,thick] node[above] {a} (l);
\end{tikzpicture}} & \ding{51} & \ding{51} & \ding{51} \\ \hline
\scalebox{0.666667}{\begin{tikzpicture}[every node/.style={align=center}]
    \node (x) at (-0.1,-0.25) {\,};
    \node (y) at (7.1,1.25)   {\,};

    \node (a) at (0.0,0.25)   [draw, circle, thick, fill=black, scale=0.3] {\,};
    \node (b) at (1.5,0.25)   [draw, circle, thick, fill=black, scale=0.3] {\,};
    \node (c) at (2.5,0.4)    {$\Leftarrow_{s_3}$};
    \node (d) at (3.5,0.0)    [draw, circle, thick, fill=black, scale=0.3] {\,};
    \node (e) at (4.5,1.0)    [draw, circle, thick, fill=black, scale=0.3] {\,};
    \node (f) at (5.5,0.0)    [draw, circle, thick, fill=black, scale=0.3] {\,};
    \node (g) at (6.5,0.4)    {$\Rightarrow_{s_4}$};
    \node (h) at (7.5,0.25)   [draw, circle, thick, fill=black, scale=0.3] {\,};
    \node (i) at (9.0,0.25)   [draw, circle, thick, fill=black, scale=0.3] {\,};

    \node (A) at (0.0,0.07)   {\tiny{2}};
    \node (B) at (1.5,0.07)   {\tiny{3}};
    \node (D) at (3.5,-0.18)  {\tiny{1}};
    \node (E) at (4.5,0.82)   {\tiny{2}};
    \node (F) at (5.5,-0.18)  {\tiny{3}};
    \node (H) at (7.5,0.07)   {\tiny{1}};
    \node (I) at (9.0,0.07)   {\tiny{2}};

    \draw (a) edge[->,thick,bend left=60] node[above] {b} (b)
          (b) edge[->,thick] node[above] {a} (a)
          (d) edge[->,thick] node[above,xshift=-1.5mm,yshift=-1mm] {a} (e)
          (e) edge[->,thick] node[above,xshift=1.5mm,yshift=-1mm] {b} (f)
          (f) edge[->,thick] node[above] {b} (d)
          (h) edge[->,thick,bend left=60] node[above] {a} (i)
          (i) edge[->,thick] node[above] {a} (h);
\end{tikzpicture}} & \ding{55} & \ding{55} & \ding{55} \\ \hline
\scalebox{0.666667}{\begin{tikzpicture}[every node/.style={align=center}]
    \node (x) at (-0.1,-0.25) {\,};
    \node (y) at (7.1,1.25)   {\,};

    \node (a) at (0.0,0.333333)      [draw, circle, thick, fill=black, scale=0.3] {\,};
    \node (b) at (0.75,0.333333)     [draw, circle, thick, fill=black, scale=0.3] {\,};
    \node (c) at (1.5,0.333333)      [draw, circle, thick, fill=black, scale=0.3] {\,};
    \node (d) at (2.5,0.4)           {$\Leftarrow_{s_3}$};
    \node (e) at (3.5,0.333333)      [draw, circle, thick, fill=black, scale=0.3] {\,};
    \node (f) at (4.166667,0.333333) [draw, circle, thick, fill=black, scale=0.3] {\,};
    \node (g) at (4.866667,0.333333) [draw, circle, thick, fill=black, scale=0.3] {\,};
    \node (h) at (5.5,0.333333)      [draw, circle, thick, fill=black, scale=0.3] {\,};
    \node (i) at (6.5,0.4)           {$\Rightarrow_{s_4}$};
    \node (j) at (7.5,0.333333)      [draw, circle, thick, fill=black, scale=0.3] {\,};
    \node (k) at (8.25,0.333333)     [draw, circle, thick, fill=black, scale=0.3] {\,};
    \node (l) at (9.0,0.333333)      [draw, circle, thick, fill=black, scale=0.3] {\,};

    \node (A) at (0.0,0.153333)      {\tiny{1}};
    \node (B) at (0.75,0.153333)     {\tiny{2}};
    \node (C) at (1.5,0.153333)      {\tiny{4}};
    \node (E) at (3.5,0.153333)      {\tiny{1}};
    \node (F) at (4.166667,0.153333) {\tiny{2}};
    \node (G) at (4.866667,0.153333) {\tiny{3}};
    \node (H) at (5.5,0.153333)      {\tiny{4}};
    \node (J) at (7.5,0.153333)      {\tiny{1}};
    \node (K) at (8.25,0.153333)     {\tiny{3}};
    \node (L) at (9.0,0.153333)      {\tiny{4}};

    \draw (a) edge[->,thick] node[above] {b} (b)
          (b) edge[->,thick] node[above] {a} (c)
          (e) edge[->,thick] node[above] {b} (f)
          (f) edge[->,thick] node[above] {b} (g)
          (g) edge[->,thick] node[above] {a} (h)
          (j) edge[->,thick] node[above] {a} (k)
          (k) edge[->,thick] node[above] {a} (l);
\end{tikzpicture}} & \ding{51} & \ding{51} & \ding{51} \\ \hline
\scalebox{0.666667}{\begin{tikzpicture}[every node/.style={align=center}]
    \node (x) at (-0.1,-0.25) {\,};
    \node (y) at (7.1,1.25)   {\,};

    \node (a) at (0.0,0.25)   [draw, circle, thick, fill=black, scale=0.3] {\,};
    \node (b) at (1.5,0.25)   [draw, circle, thick, fill=black, scale=0.3] {\,};
    \node (c) at (2.5,0.4)    {$\Leftarrow_{s_4}$};
    \node (d) at (3.5,0.0)    [draw, circle, thick, fill=black, scale=0.3] {\,};
    \node (e) at (4.5,1.0)    [draw, circle, thick, fill=black, scale=0.3] {\,};
    \node (f) at (5.5,0.0)    [draw, circle, thick, fill=black, scale=0.3] {\,};
    \node (g) at (6.5,0.4)    {$\Rightarrow_{s_4}$};
    \node (h) at (7.5,0.25)   [draw, circle, thick, fill=black, scale=0.3] {\,};
    \node (i) at (9.0,0.25)   [draw, circle, thick, fill=black, scale=0.3] {\,};

    \node (A) at (0.0,0.07)   {\tiny{1}};
    \node (B) at (1.5,0.07)   {\tiny{3}};
    \node (D) at (3.5,-0.18)  {\tiny{1}};
    \node (E) at (4.5,0.82)   {\tiny{2}};
    \node (F) at (5.5,-0.18)  {\tiny{3}};
    \node (H) at (7.5,0.07)   {\tiny{1}};
    \node (I) at (9.0,0.07)   {\tiny{2}};

    \draw (a) edge[->,thick,bend left=60] node[above] {a} (b)
          (b) edge[->,thick] node[above] {b} (a)
          (d) edge[->,thick] node[above,xshift=-1.5mm,yshift=-1mm] {b} (e)
          (e) edge[->,thick] node[above,xshift=1.5mm,yshift=-1mm] {b} (f)
          (f) edge[->,thick] node[above] {b} (d)
          (h) edge[->,thick,bend left=60] node[above] {b} (i)
          (i) edge[->,thick] node[above] {a} (h);
\end{tikzpicture}} & \ding{51} & \ding{55} & \ding{55} \\ \hline
\scalebox{0.666667}{\begin{tikzpicture}[every node/.style={align=center}]
    \node (x) at (-0.1,-0.25) {\,};
    \node (y) at (7.1,1.25)   {\,};

    \node (a) at (0.0,0.333333)      [draw, circle, thick, fill=black, scale=0.3] {\,};
    \node (b) at (0.75,0.333333)     [draw, circle, thick, fill=black, scale=0.3] {\,};
    \node (c) at (1.5,0.333333)      [draw, circle, thick, fill=black, scale=0.3] {\,};
    \node (d) at (2.5,0.4)           {$\Leftarrow_{s_4}$};
    \node (e) at (3.5,0.333333)      [draw, circle, thick, fill=black, scale=0.3] {\,};
    \node (f) at (4.166667,0.333333) [draw, circle, thick, fill=black, scale=0.3] {\,};
    \node (g) at (4.866667,0.333333) [draw, circle, thick, fill=black, scale=0.3] {\,};
    \node (h) at (5.5,0.333333)      [draw, circle, thick, fill=black, scale=0.3] {\,};
    \node (i) at (6.5,0.4)           {$\Rightarrow_{s_4}$};
    \node (j) at (7.5,0.333333)      [draw, circle, thick, fill=black, scale=0.3] {\,};
    \node (k) at (8.25,0.333333)     [draw, circle, thick, fill=black, scale=0.3] {\,};
    \node (l) at (9.0,0.333333)      [draw, circle, thick, fill=black, scale=0.3] {\,};

    \node (A) at (0.0,0.153333)      {\tiny{1}};
    \node (B) at (0.75,0.153333)     {\tiny{2}};
    \node (C) at (1.5,0.153333)      {\tiny{4}};
    \node (E) at (3.5,0.153333)      {\tiny{1}};
    \node (F) at (4.166667,0.153333) {\tiny{2}};
    \node (G) at (4.866667,0.153333) {\tiny{3}};
    \node (H) at (5.5,0.153333)      {\tiny{4}};
    \node (J) at (7.5,0.153333)      {\tiny{1}};
    \node (K) at (8.25,0.153333)     {\tiny{3}};
    \node (L) at (9.0,0.153333)      {\tiny{4}};

    \draw (a) edge[->,thick] node[above] {b} (b)
          (b) edge[->,thick] node[above] {a} (c)
          (e) edge[->,thick] node[above] {b} (f)
          (f) edge[->,thick] node[above] {b} (g)
          (g) edge[->,thick] node[above] {b} (h)
          (j) edge[->,thick] node[above] {a} (k)
          (k) edge[->,thick] node[above] {b} (l);
\end{tikzpicture}} & \ding{51} & \ding{51} & \ding{51} \\ \hline
\end{tabular}
}
\caption{Critical pair analysis for Example \ref{eg:4}}
\label{fig:critpairs1}
\end{figure}

Checking for local confluence up to garbage is undecidable in general, even when \(\widehat{\mathcal{D}}\) is decidable and the system is terminating and closed. Moreover, local confluence up to garbage is actually undecidable in general for a terminating non-length-increasing string rewriting systems and \(\mathcal{D}\) a regular string language \cite{Caron91a}. The following (corrected) example due to Plump \cite{Plump05a} demonstrates that a GT system can be confluent and terminating, with all critical pairs joinable, and at least one not strongly joinable:

\begin{example} \label{eg:2}
The GT system with rules in Figure \ref{fig:eg-2-rules} is terminating and confluent, and all its critical pairs are strongly joinable apart from the pair in Figure \ref{fig:eg-2-nsj-pair} which is only joinable.
\end{example}

\begin{figure}[!ht]
\centering
\begin{tikzpicture}[every node/.style={align=center}]
    \node (a) at (0.0,-0.05) {$r_1$:};
    \node (b) at (1.0,0.0)   [draw, circle, thick, fill=black, scale=0.3] {\,};
    \node (c) at (2.0,0.0)   [draw, circle, thick, fill=black, scale=0.3] {\,};
    \node (d) at (3.0,0.0)   {$\leftarrow$};
    \node (e) at (4.0,0.0)   [draw, circle, thick, fill=black, scale=0.3] {\,};
    \node (f) at (5.0,0.0)   [draw, circle, thick, fill=black, scale=0.3] {\,};
    \node (g) at (6.0,0.0)   {$\rightarrow$};
    \node (h) at (7.0,0.0)   [draw, circle, thick, fill=black, scale=0.3] {\,};
    \node (i) at (8.0,0.0)   [draw, circle, thick, fill=black, scale=0.3] {\,};

    \node (B) at (1.0,-0.18)  {\tiny{1}};
    \node (D) at (2.0,-0.18)  {\tiny{2}};
    \node (F) at (4.0,-0.18)  {\tiny{1}};
    \node (G) at (5.0,-0.18)  {\tiny{2}};
    \node (I) at (7.0,-0.18)  {\tiny{1}};
    \node (J) at (8.0,-0.18)  {\tiny{2}};

    \draw (b) edge[->,thick,loop above] (b)
          (c) edge[->,thick,loop above] (c)
          (h) edge[->,thick,loop above] (h);

    \node (a) at (0.0,-1.05)  {$r_2$:};
    \node (b) at (1.5,-1.0)   [draw, circle, thick, fill=black, scale=0.3] {\,};
    \node (c) at (3.0,-1.0)   {$\leftarrow$};
    \node (d) at (4.5,-1.0)   [draw, circle, thick, fill=black, scale=0.3] {\,};
    \node (e) at (6.0,-1.0)   {$\rightarrow$};
    \node (f) at (7.5,-1.0)   [draw, circle, thick, fill=black, scale=0.3] {\,};

    \node (B) at (1.5,-1.18)  {\tiny{1}};
    \node (D) at (4.5,-1.18)  {\tiny{1}};
    \node (F) at (7.5,-1.18)  {\tiny{1}};

    \draw (b) edge[->,thick,loop left] (b)
          (b) edge[->,thick,loop right] (b)
          (f) edge[->,thick,loop right] (f);

    \node (a) at (0.0,-2.05)  {$r_3$:};
    \node (b) at (1.0,-2.0)  [draw, circle, thick, fill=black, scale=0.3] {\,};
    \node (c) at (2.0,-2.0)  [draw, circle, thick, fill=black, scale=0.3] {\,};
    \node (d) at (3.0,-2.0)   {$\leftarrow$};
    \node (e) at (4.0,-2.0)   [draw, circle, thick, fill=black, scale=0.3] {\,};
    \node (f) at (5.0,-2.0)   [draw, circle, thick, fill=black, scale=0.3] {\,};
    \node (g) at (6.0,-2.0)   {$\rightarrow$};
    \node (h) at (7.0,-2.0)   [draw, circle, thick, fill=black, scale=0.3] {\,};
    \node (i) at (8.0,-2.0)   [draw, circle, thick, fill=black, scale=0.3] {\,};

    \node (B) at (1.0,-2.18)  {\tiny{1}};
    \node (D) at (2.0,-2.18)  {\tiny{2}};
    \node (F) at (4.0,-2.18)  {\tiny{1}};
    \node (G) at (5.0,-2.18)  {\tiny{2}};
    \node (I) at (7.0,-2.18)  {\tiny{1}};
    \node (J) at (8.0,-2.18)  {\tiny{2}};

    \draw (b) edge[->,thick] (c);
\end{tikzpicture}
\caption{Rules for Example \ref{eg:2}}
\label{fig:eg-2-rules}
\end{figure}

\begin{figure}[!ht]
\centering
\begin{tikzpicture}[every node/.style={align=center}]
    \node (a) at (0.0,0.0)   [draw, circle, thick, fill=black, scale=0.3] {\,};
    \node (b) at (1.0,0.0)   [draw, circle, thick, fill=black, scale=0.3] {\,};
    \node (c) at (2.0,0.0)   {$\Leftarrow_{r_1}$};
    \node (d) at (3.0,0.0)   [draw, circle, thick, fill=black, scale=0.3] {\,};
    \node (e) at (4.0,0.0)   [draw, circle, thick, fill=black, scale=0.3] {\,};
    \node (f) at (5.0,0.0)   {$\Rightarrow_{r_1}$};
    \node (g) at (6.0,0.0)   [draw, circle, thick, fill=black, scale=0.3] {\,};
    \node (h) at (7.0,0.0)   [draw, circle, thick, fill=black, scale=0.3] {\,};

    \node (A) at (0.0,-0.18)  {\tiny{1}};
    \node (B) at (1.0,-0.18)  {\tiny{2}};
    \node (D) at (3.0,-0.18)  {\tiny{1}};
    \node (E) at (4.0,-0.18)  {\tiny{2}};
    \node (G) at (6.0,-0.18)  {\tiny{1}};
    \node (H) at (7.0,-0.18)  {\tiny{2}};

    \draw (a) edge[->,thick,loop above] (a)
          (d) edge[->,thick,loop above] (d)
          (e) edge[->,thick,loop above] (e)
          (h) edge[->,thick,loop above] (h);
\end{tikzpicture}
\caption{Non-strongly joinable pair for Example \ref{eg:2}}
\label{fig:eg-2-nsj-pair}
\end{figure}

We now extend this even further, showing that there is a GT system \(T\) that is not confluent and an infinite language of graphs \(\mathcal{D}\) such that \(\mathcal{D}\) is closed under \(T\), \(T\) is terminating on \(\mathcal{D}\), \(T\) is confluent on \(\mathcal{D}\), all \(T\)'s \(\mathcal{D}\)-non-garbage critical pairs are joinable, and at least one of them is not strongly joinable:

\begin{example} \label{eg:3}
Let \(\mathcal{D}\) be the language of all graphs that are trees with exactly one looped edge added to one of the nodes, and \(T\) be the GT system with rules in Figure \ref{fig:eg-3-rules}. Figure \ref{fig:eg-3-crit-pairs} shows the four non-isomorphic critical pairs of the system. We can see there is a garbage pair which is non-joinable, which tells us that \(T\) is not locally confluent. By direct argument, one can see that \(\mathcal{D}\) is closed under \(T\), \(T\) is terminating on \(\mathcal{D}\), and \(T\) is confluent on \(\mathcal{D}\), however there is a non-strongly joinable non-garbage critical pair.
\end{example}

\begin{figure}[!ht]
\centering
\begin{tikzpicture}[every node/.style={align=center}]
    \node (a) at (0.0,-0.05) {$r_1$:};
    \node (b) at (1.25,0.0)  [draw, circle, thick, fill=black, scale=0.3] {\,};
    \node (c) at (2.25,0.0)  [draw, circle, thick, fill=black, scale=0.3] {\,};
    \node (d) at (3.5,0.0)   {$\leftarrow$};
    \node (e) at (4.5,0.0)   [draw, circle, thick, fill=black, scale=0.3] {\,};
    \node (f) at (5.5,0.0)   [draw, circle, thick, fill=black, scale=0.3] {\,};
    \node (g) at (6.5,0.0)   {$\rightarrow$};
    \node (h) at (7.5,0.0)   [draw, circle, thick, fill=black, scale=0.3] {\,};
    \node (i) at (8.5,0.0)   [draw, circle, thick, fill=black, scale=0.3] {\,};

    \node (B) at (1.25,-0.18) {\tiny{1}};
    \node (C) at (2.25,-0.18) {\tiny{2}};
    \node (E) at (4.5,-0.18)  {\tiny{1}};
    \node (F) at (5.5,-0.18)  {\tiny{2}};
    \node (H) at (7.5,-0.18)  {\tiny{1}};
    \node (I) at (8.5,-0.18)  {\tiny{2}};

    \draw (b) edge[->,thick,loop left] (b)
          (b) edge[->,thick] (c)
          (h) edge[->,thick] (i)
          (i) edge[->,thick,loop right] (i);

    \node (a) at (0.0,-1.05) {$r_2$:};
    \node (b) at (1.25,-1.0)  [draw, circle, thick, fill=black, scale=0.3] {\,};
    \node (c) at (2.25,-1.0)  [draw, circle, thick, fill=black, scale=0.3] {\,};
    \node (d) at (3.5,-1.0)   {$\leftarrow$};
    \node (e) at (4.5,-1.0)   [draw, circle, thick, fill=black, scale=0.3] {\,};
    \node (g) at (6.5,-1.0)   {$\rightarrow$};
    \node (h) at (7.5,-1.0)   [draw, circle, thick, fill=black, scale=0.3] {\,};

    \node (B) at (1.25,-1.18) {\tiny{1}};
    \node (E) at (4.5,-1.18)  {\tiny{1}};
    \node (H) at (7.5,-1.18)  {\tiny{1}};

    \draw (c) edge[->,thick,loop right] (c)
          (b) edge[->,thick] (c)
          (h) edge[->,thick,loop right] (h);
\end{tikzpicture}
\caption{Rules for Example \ref{eg:3}}
\label{fig:eg-3-rules}
\end{figure}

\begin{figure}[!ht]
\centering
\scalebox{0.825}{
\begin{tabular}{|l|c|c|c|c|}
\hline
\multicolumn{1}{|>{\centering\arraybackslash}p{6cm}|}{Pair/Property} & \multicolumn{1}{>{\centering\arraybackslash}p{1.666667cm}|}{Joinable} & \multicolumn{1}{>{\centering\arraybackslash}p{1.666667cm}|}{Strongly Joinable} & \multicolumn{1}{>{\centering\arraybackslash}p{1.666667cm}|}{Non-Garbage} \\ \hline
\scalebox{0.75}{\begin{tikzpicture}[every node/.style={align=center}]
    \node (x) at (-0.1,-1.25) {\,};
    \node (y) at (6.8,0.25)   {\,};

    \node (a) at (0.5,0.0)   [draw, circle, thick, fill=black, scale=0.3] {\,};
    \node (b) at (0.5,-1.0)  [draw, circle, thick, fill=black, scale=0.3] {\,};
    \node (d) at (2.0,-0.5)  {$\Leftarrow_{r_1}$};
    \node (e) at (3.5,0.0)   [draw, circle, thick, fill=black, scale=0.3] {\,};
    \node (f) at (3.5,-1.0)  [draw, circle, thick, fill=black, scale=0.3] {\,};
    \node (h) at (5.0,-0.5)  {$\Rightarrow_{r_2}$};
    \node (i) at (6.5,0.0)   [draw, circle, thick, fill=black, scale=0.3] {\,};
    \node (j) at (6.5,-1.0)  [draw, circle, thick, fill=black, scale=0.3] {\,};

    \node (A) at (0.5,0.18)   {\tiny{1}};
    \node (B) at (0.5,-1.18)  {\tiny{2}};
    \node (E) at (3.5,0.18)   {\tiny{1}};
    \node (F) at (3.5,-1.18)  {\tiny{2}};
    \node (I) at (6.5,0.18)   {\tiny{1}};
    \node (J) at (6.5,-1.18)  {\tiny{2}};

    \draw (a) edge[->,thick,bend left=30] (b)
          (a) edge[->,thick,bend right=30] (b)
          (b) edge[->,thick,loop right] (b)
          (e) edge[->,thick,loop right] (e)
          (e) edge[->,thick,bend left=30] (f)
          (e) edge[->,thick,bend right=30] (f)
          (i) edge[->,thick,bend left=30] (j)
          (i) edge[->,thick,bend right=30] (j)
          (j) edge[->,thick,loop right] (j);
\end{tikzpicture}} & \ding{51} & \ding{51} & \ding{55} \\ \hline
\scalebox{0.75}{\begin{tikzpicture}[every node/.style={align=center}]
    \node (x) at (-0.1,-1.25) {\,};
    \node (y) at (6.8,0.25)   {\,};

    \node (a) at (0.5,0.0)   [draw, circle, thick, fill=black, scale=0.3] {\,};
    \node (b) at (0.5,-1.0)  [draw, circle, thick, fill=black, scale=0.3] {\,};
    \node (d) at (2.0,-0.5)  {$\Leftarrow_{r_1}$};
    \node (e) at (3.5,0.0)   [draw, circle, thick, fill=black, scale=0.3] {\,};
    \node (f) at (3.5,-1.0)  [draw, circle, thick, fill=black, scale=0.3] {\,};
    \node (h) at (5.0,-0.5)  {$\Rightarrow_{r_2}$};
    \node (i) at (6.5,0.0)   [draw, circle, thick, fill=black, scale=0.3] {\,};
    \node (j) at (6.5,-1.0)  [draw, circle, thick, fill=black, scale=0.3] {\,};

    \node (A) at (0.5,0.18)   {\tiny{1}};
    \node (B) at (0.5,-1.18)  {\tiny{2}};
    \node (E) at (3.5,0.18)   {\tiny{1}};
    \node (F) at (3.5,-1.18)  {\tiny{2}};
    \node (I) at (6.5,0.18)   {\tiny{1}};
    \node (J) at (6.5,-1.18)  {\tiny{2}};

    \draw (a) edge[->,thick,loop right] (a)
          (a) edge[->,thick] (b)
          (b) edge[->,thick,loop left] (b)
          (e) edge[->,thick,loop left] (e)
          (e) edge[->,thick,loop right] (e)
          (e) edge[->,thick] (f)
          (i) edge[->,thick,loop left] (i)
          (i) edge[->,thick] (j)
          (j) edge[->,thick,loop right] (j);
\end{tikzpicture}} & \ding{51} & \ding{51} & \ding{55} \\ \hline
\scalebox{0.75}{\begin{tikzpicture}[every node/.style={align=center}]
    \node (x) at (-0.1,-1.25) {\,};
    \node (y) at (6.8,0.25)   {\,};

    \node (a) at (0.5,0.0)   [draw, circle, thick, fill=black, scale=0.3] {\,};
    \node (b) at (0.0,-1.0)  [draw, circle, thick, fill=black, scale=0.3] {\,};
    \node (c) at (1.0,-1.0)  [draw, circle, thick, fill=black, scale=0.3] {\,};
    \node (d) at (2.0,-0.5)  {$\Leftarrow_{r_1}$};
    \node (e) at (3.5,0.0)   [draw, circle, thick, fill=black, scale=0.3] {\,};
    \node (f) at (3.0,-1.0)  [draw, circle, thick, fill=black, scale=0.3] {\,};
    \node (g) at (4.0,-1.0)  [draw, circle, thick, fill=black, scale=0.3] {\,};
    \node (h) at (5.0,-0.5)  {$\Rightarrow_{r_1}$};
    \node (i) at (6.5,0.0)   [draw, circle, thick, fill=black, scale=0.3] {\,};
    \node (j) at (6.0,-1.0)  [draw, circle, thick, fill=black, scale=0.3] {\,};
    \node (k) at (7.0,-1.0)  [draw, circle, thick, fill=black, scale=0.3] {\,};

    \node (A) at (0.5,0.18)   {\tiny{1}};
    \node (B) at (0.0,-1.18)  {\tiny{2}};
    \node (C) at (1.0,-1.18)  {\tiny{3}};
    \node (E) at (3.5,0.18)   {\tiny{1}};
    \node (F) at (3.0,-1.18)  {\tiny{2}};
    \node (G) at (4.0,-1.18)  {\tiny{3}};
    \node (I) at (6.5,0.18)   {\tiny{1}};
    \node (J) at (6.0,-1.18)  {\tiny{2}};
    \node (K) at (7.0,-1.18)  {\tiny{3}};

    \draw (a) edge[->,thick] (b)
          (a) edge[->,thick] (c)
          (b) edge[->,thick,loop right] (b)
          (e) edge[->,thick] (f)
          (e) edge[->,thick] (g)
          (e) edge[->,thick,loop right] (e)
          (i) edge[->,thick] (j)
          (i) edge[->,thick] (k)
          (k) edge[->,thick,loop right] (k);
\end{tikzpicture}} & \ding{51} & \ding{55} & \ding{51} \\ \hline
\scalebox{0.75}{\begin{tikzpicture}[every node/.style={align=center}]
    \node (x) at (-0.1,-1.25) {\,};
    \node (y) at (6.8,0.25)   {\,};

    \node (a) at (0.5,0.0)   [draw, circle, thick, fill=black, scale=0.3] {\,};
    \node (b) at (0.5,-1.0)  [draw, circle, thick, fill=black, scale=0.3] {\,};
    \node (d) at (2.0,-0.5)  {$\Leftarrow_{r_1}$};
    \node (e) at (3.5,0.0)   [draw, circle, thick, fill=black, scale=0.3] {\,};
    \node (f) at (3.5,-1.0)  [draw, circle, thick, fill=black, scale=0.3] {\,};
    \node (h) at (5.0,-0.5)  {$\Rightarrow_{r_2}$};
    \node (i) at (6.5,0.0)   [draw, circle, thick, fill=black, scale=0.3] {\,};

    \node (A) at (0.5,0.18)   {\tiny{1}};
    \node (B) at (0.5,-1.18)  {\tiny{2}};
    \node (E) at (3.5,0.18)   {\tiny{1}};
    \node (F) at (3.5,-1.18)  {\tiny{2}};
    \node (I) at (6.5,0.18)   {\tiny{1}};

    \draw (a) edge[->,thick] (b)
          (b) edge[->,thick,loop left] (b)
          (b) edge[->,thick,loop right] (b)
          (e) edge[->,thick,loop right] (e)
          (e) edge[->,thick] (f)
          (f) edge[->,thick,loop right] (f)
          (i) edge[->,thick,loop left] (i)
          (i) edge[->,thick,loop right] (i);
\end{tikzpicture}} & \ding{55} & \ding{55} & \ding{55} \\ \hline
\end{tabular}
}
\caption{Critical pair analysis for Example \ref{eg:3}}
\label{fig:eg-3-crit-pairs}
\end{figure}

In Subsection \ref{subsec:critpairgen}, we will discuss generation of the set of non-isomorphic non-garbage critical pairs of a given GT system, discussing sufficient conditions on \(\mathcal{D}\) for this process to be effective. In Subsection \ref{subsec:critpairjoinability} we discuss testing for strong joinability of a given non-garbage critical pair.

\subsection{Generation of Non-Garbage Critical Pairs} \label{subsec:critpairgen} \label{sec:confnorm}

In general, there is no algorithm that, when given a DPO grammar and a graph, can decide if the graph is contained in the language generated by the grammar. That is, the universal membership problem is undecidable. It is easy to see that similar problem of whether a graph is in the subgraph closure of the language generated by a DPO grammar is undecidable in general too.

This means that, unlike for critical pairs, generation of all the non-garbage critical pairs is not possible in general, due the impossibility of deciding subgraph membership. Though, if we are provided with an algorithm for testing if a graph is a subgraph of a graph of \(D\), then we can generate the set of \textit{non-isomorphic} \(\mathcal{D}\)-\textit{non-garbage critical pairs}.

\begin{definition}[Universal Subgraph Membership Problem]~\\
\vspace{-1em}
\begin{prob}
\probinstance{A graph grammar \(\mathcal{G}\) over \(\Sigma\) and a graph \(G\) over \(\Sigma\).}
\probquestion{Is \(G \in \widehat{\mathrm{L}(\mathcal{G})}\)?}
\end{prob}
\end{definition}

\begin{lemma}
The \textit{universal subgraph membership problem} is undecidable.
\end{lemma}

\begin{proof}
By reduction of undecidability of the emptiness problem, since \(\emptyset \not\in \widehat{\mathrm{L}(\mathcal{G})}\) if and only if \(\mathrm{L}(\mathcal{G}) = \emptyset\).
\end{proof}

In practice, it is often the case that one can determine if graph is contained in the subgraph closure of a language, and so undecidability is not too much of a concern. For example, if a graph language is known to only contain acyclic graphs, critical pairs with start graphs containing a cycle can be discarded as garbage. Moreover, it may not even be necessary to decide if a critical pair is garbage, if one can show that it is strongly joinable instead.

Here are some types of graph languages for which membership in the subgraph closure is decidable:

\begin{enumerate}
    \item If \(D\) is finite, then membership in the subgraph closure can be decided simply by checking if the given graph is a subgraph of any graph in the language.
    \item If $D$ is subgraph-closed, then membership in the subgraph closure is the same as membership in $D$. Hence, membership is decidable if $D$ is the class of discrete graphs, bounded degree graphs for some fixed bound, acyclic graphs, $k$-colourable graphs or planar graphs (see \cite{Skiena08a} for how to decide membership in the latter three classes). If $D$ is the class of bounded treewidth graphs, for a fixed bound, membership can be decided by the algorithm in \cite{Bodlaender96a}.
    \item If \(D\) is specified by a so-called type graph (see below), then $D$ is also subgraph-closed and membership is decidable. Type graph languages are studied by Corradini, K{\"o}nig, and Nolte in \cite{Corradini-Konig-Nolte19a}.
\end{enumerate}

\begin{definition}[Type Graph Language]
Given a signature \(\Sigma\) and a graph \(G \in \mathcal{G}(\Sigma)\), define the type graph language \(\mathrm{L}_{\Sigma}(G) = \{H \in \mathcal{G}(\Sigma) \mid H \to G\}\).
\end{definition}

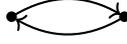
\begin{figure}[!ht]
\centering
\begin{tikzpicture}[every node/.style={align=center}]
    \node (a) at (0.0,0.0) [draw, circle, thick, fill=black, scale=0.3] {\,};
    \node (b) at (1.5,0.0) [draw, circle, thick, fill=black, scale=0.3] {\,};

    \draw (a) edge[->,thick, bend left=30] (b)
          (b) edge[->,thick, bend left=30] (a);
\end{tikzpicture}
\caption{Type graph for Example \ref{eg:6}}
\label{fig:eg-6-type-graph}
\end{figure}

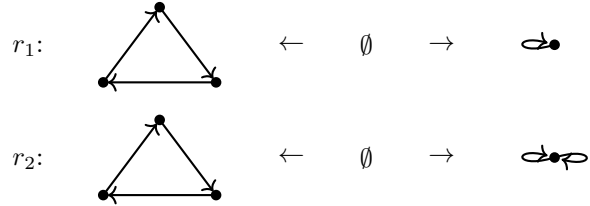
\begin{figure}[!ht]
\centering
\begin{tikzpicture}[every node/.style={align=center}]
    \node (a) at (0.0,0.45) {$r_1$:};
    \node (b) at (1.0,0.0)   [draw, circle, thick, fill=black, scale=0.3] {\,};
    \node (c) at (1.75,1.0)  [draw, circle, thick, fill=black, scale=0.3] {\,};
    \node (d) at (2.5,0.0)   [draw, circle, thick, fill=black, scale=0.3] {\,};
    \node (e) at (3.5,0.5)   {$\leftarrow$};
    \node (f) at (4.5,0.5)   {$\emptyset$};
    \node (g) at (5.5,0.5)   {$\rightarrow$};
    \node (h) at (7.0,0.5)   [draw, circle, thick, fill=black, scale=0.3] {\,};

    \draw (b) edge[->,thick] (c)
          (c) edge[->,thick] (d)
          (d) edge[->,thick] (b)
          (h) edge[->,thick,loop left] (h);

    \node (a) at (0.0,-1.05)  {$r_2$:};
    \node (b) at (1.0,-1.5)   [draw, circle, thick, fill=black, scale=0.3] {\,};
    \node (c) at (1.75,-0.5)  [draw, circle, thick, fill=black, scale=0.3] {\,};
    \node (d) at (2.5,-1.5)   [draw, circle, thick, fill=black, scale=0.3] {\,};
    \node (e) at (3.5,-1.0)   {$\leftarrow$};
    \node (f) at (4.5,-1.0)   {$\emptyset$};
    \node (g) at (5.5,-1.0)   {$\rightarrow$};
    \node (h) at (7.0,-1.0)   [draw, circle, thick, fill=black, scale=0.3] {\,};

    \draw (b) edge[->,thick] (c)
          (c) edge[->,thick] (d)
          (d) edge[->,thick] (b)
          (h) edge[->,thick,loop left] (h)
          (h) edge[->,thick,loop right] (h);
\end{tikzpicture}
\caption{Rules for Example \ref{eg:6}}
\label{fig:eg-6-rules}
\end{figure}

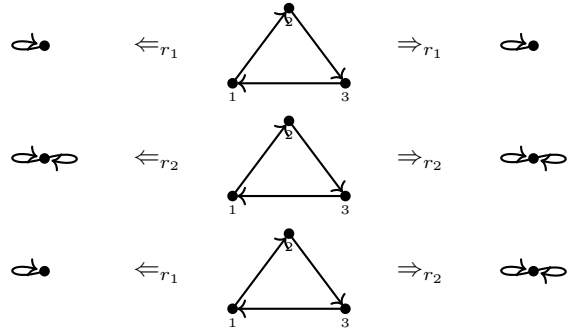
\begin{figure}[!ht]
\centering
\begin{tikzpicture}[every node/.style={align=center}]
    \node (a) at (0.0,-0.5)   [draw, circle, thick, fill=black, scale=0.3] {\,};
    \node (b) at (1.5,-0.55)   {$\Leftarrow_{r_1}$};
    \node (c) at (2.5,-1.0)   [draw, circle, thick, fill=black, scale=0.3] {\,};
    \node (d) at (3.25,0.0)   [draw, circle, thick, fill=black, scale=0.3] {\,};
    \node (e) at (4.0,-1.0)   [draw, circle, thick, fill=black, scale=0.3] {\,};
    \node (f) at (5.0,-0.55)   {$\Rightarrow_{r_1}$};
    \node (g) at (6.5,-0.5)   [draw, circle, thick, fill=black, scale=0.3] {\,};

    \node (C) at (2.5,-1.18)  {\tiny{1}};
    \node (D) at (3.25,-0.18) {\tiny{2}};
    \node (E) at (4.0,-1.18)  {\tiny{3}};

    \draw (a) edge[->,thick,loop left] (a)
          (c) edge[->,thick] (d)
          (d) edge[->,thick] (e)
          (e) edge[->,thick] (c)
          (g) edge[->,thick,loop left] (g);

    \node (a) at (0.0,-2.0)   [draw, circle, thick, fill=black, scale=0.3] {\,};
    \node (b) at (1.5,-2.05)   {$\Leftarrow_{r_2}$};
    \node (c) at (2.5,-2.5)   [draw, circle, thick, fill=black, scale=0.3] {\,};
    \node (d) at (3.25,-1.5)  [draw, circle, thick, fill=black, scale=0.3] {\,};
    \node (e) at (4.0,-2.5)   [draw, circle, thick, fill=black, scale=0.3] {\,};
    \node (f) at (5.0,-2.05)   {$\Rightarrow_{r_2}$};
    \node (g) at (6.5,-2.0)   [draw, circle, thick, fill=black, scale=0.3] {\,};

    \node (C) at (2.5,-2.68)  {\tiny{1}};
    \node (D) at (3.25,-1.68) {\tiny{2}};
    \node (E) at (4.0,-2.68)  {\tiny{3}};

    \draw (a) edge[->,thick,loop left] (a)
          (a) edge[->,thick,loop right] (a)
          (c) edge[->,thick] (d)
          (d) edge[->,thick] (e)
          (e) edge[->,thick] (c)
          (g) edge[->,thick,loop left] (g)
          (g) edge[->,thick,loop right] (g);

    \node (a) at (0.0,-3.5)   [draw, circle, thick, fill=black, scale=0.3] {\,};
    \node (b) at (1.5,-3.55)   {$\Leftarrow_{r_1}$};
    \node (c) at (2.5,-4.0)   [draw, circle, thick, fill=black, scale=0.3] {\,};
    \node (d) at (3.25,-3.0)  [draw, circle, thick, fill=black, scale=0.3] {\,};
    \node (e) at (4.0,-4.0)   [draw, circle, thick, fill=black, scale=0.3] {\,};
    \node (f) at (5.0,-3.55)   {$\Rightarrow_{r_2}$};
    \node (g) at (6.5,-3.5)   [draw, circle, thick, fill=black, scale=0.3] {\,};

    \node (C) at (2.5,-4.18)  {\tiny{1}};
    \node (D) at (3.25,-3.18) {\tiny{2}};
    \node (E) at (4.0,-4.18)  {\tiny{3}};

    \draw (a) edge[->,thick,loop left] (a)
          (c) edge[->,thick] (d)
          (d) edge[->,thick] (e)
          (e) edge[->,thick] (c)
          (g) edge[->,thick,loop left] (g)
          (g) edge[->,thick,loop right] (g);
\end{tikzpicture}
\caption{Critical pairs for Example \ref{eg:6}}
\label{fig:eg-6-crit-pairs}
\end{figure}

\begin{example} \label{eg:6}
It is easy to see that the language of \(2\)-colourable unlabelled graphs, \(\mathcal{D}\), can be specified by the type graph in Figure \ref{fig:eg-6-type-graph}. Consider the GT system with the two rules in Figure \ref{fig:eg-6-rules}. It is easy to see that \(\mathcal{D}\) is closed under these rules (due to the fact that they are never applicable), that they are terminating (due to the fact that they are size reducing), and that there are five non-isomorphic critical pairs (Figure \ref{fig:eg-6-crit-pairs}, where the third pair is repeated twice more, formally with different matches), all of which are garbage (which we can machine check because \(\mathcal{D}\) is specified by a type graph). Thus, by Corollary \ref{cor:ngcritpairlem}, the rules are confluent up to garbage on \(\mathcal{D}\).
\end{example}

We are not aware of any other general families of grammars, or otherwise, for which we can solve the subgraph membership problem. We do not believe this problem is even decidable for hyperedge replacement grammars. This conjecture is not incompatible with this problem being easy for type graph languages, since the the class of graph languages generated by hyperedge replacement grammars is incomparable with the class of graph languages specified by type graphs.

\subsection{Checking for Strong Joinability} \label{subsec:critpairjoinability}

We briefly, explicitly discuss the process for checking if a pair of direct derivations is strongly joinable in a given GT system.

If a GT system is terminating, then it is easy to check if a  pair of direct derivations is joinable or strongly joinable due to the fact that GT systems are finitely branching up to isomorphism, so there can only be finitely many successor graphs, up to isomorphism. It is then simply a matter of checking if there is an isomorphism between any of the successor graphs which behaves correctly with respect to the preserved nodes, as in the definition of strong joinability.

Alternatively, if a GT system is only terminating up to garbage on some language \(\mathcal{D}\) and \(\mathcal{D}\) is closed under \(T\), then similarly, one can test joinability and strong joinability due to the fact that closedness ensures only finitely many successor graphs, as above.

\subsection{Summary} \label{subsec:critpairsumary}

We have presented our Generalised Critical Pair Lemma and Generalised Newman's Lemma, which together, allow one to check for confluence up to garbage on some language \(\mathcal{D}\) in the presence of termination and closedness. If there is an algorithm for solving the subgraph membership problem of \(\mathcal{D}\), then we can effectively generate the set of non-isomorphic \(\mathcal{D}\)-non-garbage critical pairs and effectively test each of them for strong joinability. This process will always terminate, however may not provide a conclusive answer.

If the analysis completes with all the non-isomorphic \(\mathcal{D}\)-non-garbage critical pairs being strongly joinable, then we can conclude the system is confluent up to garbage on \(\mathcal{D}\). If the analysis completes with a non-joinable \(\mathcal{D}\)-non-garbage critical pair that has its start graph in \(\mathcal{D}\), then we can conclude the system is not confluent up to garbage on \(\mathcal{D}\). In any other scenario, we cannot directly make a conclusion.

Finally, sometimes one might want to show confluence of a GT system \(T\) on a language \(\mathcal{D}\) which is not necessarily closed under \(T\). That is, either \(\mathcal{D}\) is not closed under \(T\), or indeed closure is simply unknown. In this scenario, one should attempt to show confluence on some larger language \(\mathcal{E}\) containing \(\mathcal{D}\). For example, if \(\mathcal{D}\) contains only acyclic graphs, a good choice for \(\mathcal{E}\) could be the language of acyclic graphs over the same signature. Transitivity of confluence up to garbage (Lemma \ref{lem:transitiveconfluence}) tells us that if we establish that \(T\) is confluent up to garbage on \(\mathcal{E}\), then it is also confluent up to garbage on \(\mathcal{D}\).

\section{Backtracking-Free Language Recognition} \label{sec:languages}

In this section, we introduce a general notion of what it means to recognise a language, and what it means to be a backtracking-free specification. We then demonstrate the applicability of our earlier results by showing that there are backtracking-free specifications for the languages of labelled series-parallel graphs and extended flow diagrams, even in the absence of confluence. We thus have algorithms, specified by reduction rules, that can check membership of these languages without needing to backtrack.

\subsection{Backtracking-Free Specifications}

Given a graph transformation system and a start graph, we can think of the pair as a graph grammar, generating a graph language. If the reversed system is terminating, then membership testing is decidable, but in general, non-deterministic in the sense that a deterministic algorithm must backtrack if it produces a normal form not equal to the start graph, to determine if another derivation sequence could have reached it. It is easy to see that confluence is a sufficient condition to give determinism, however confluence is often not easily obtainable in practice. For this reason, we will consider the weaker property of confluence up to garbage on the generated language.

Using the results from the last section, it is often possible to prove local confluence up to garbage using the Generalised Critical Pair Lemma, and then, in the presence of termination and closure, use the Generalised Newman's Lemma to show confluence up to garbage. Language recognition by confluent graph reduction has been considered before by Bakewell, Plump, and Runciman, in the context of pointer structures \cite{Bakewell-Plump-Runciman03a,Bakewell-Plump-Runciman04a}, but without the concept of confluence up to garbage.

Before continuing, we must provide a formal definition of what it means to recognise a language, and that grammars satisfy our definition by considering their rules in reverse, abstracting away from grammars, with a more general definition that accounts for the fact that reduction systems may need auxiliary symbols, not in the input, in the same way grammars can use non-terminals.

\begin{definition}[Language Recognition] \label{dfn:langrec}
Let \(T = (\Sigma, \mathcal{R})\) be a GT system, \(A \subseteq \Sigma\) an input signature, and \(\mathcal{S}\) a finite set of graphs over \(\Sigma\). Then we say that \((T, \mathcal{S})\) \textit{recognises} a language \(\mathcal{L}\) over \(A\) if for all graphs \(G\) over \(A\), \([G] \in \mathcal{L}\) if and only if \(G \Rightarrow_{\mathcal{R}}^{*} S\) for some \(S \in \mathcal{S}\).
\end{definition}

\begin{theorem}[Membership Checking] \label{thm:membershiptest}
Given a \textit{grammar} \(\mathcal{G} = (\Sigma, N, \mathcal{R}, S)\), \([G] \in \mathrm{L}(\mathcal{G})\) if and only if \(G \Rightarrow_{\mathcal{R}^{-1}}^* S\) and \(G\) is terminally labelled. That is, \(((\Sigma, \mathcal{R}^{-1}), \{S\})\) recognises \(\mathrm{L}(\mathcal{G})\) over \(\Sigma \setminus N\).
\end{theorem}

\begin{proof}
The key is that rules and derivations are invertible, which means that if \(S\) can be derived from \(G\) using the reverse rules, then \(G\) can be derived from \(S\) using the original rules so is in the language. If \(S\) cannot be derived from \(G\), then \(G\) cannot be in the language since that would imply there was a derivation sequence from \(S\) to \(G\) which we could invert to give a contradiction.
\end{proof}

We are now ready to define \textit{backtracking-free specifications}, and show that such systems can test for language membership without backtracking.

\begin{definition}[Backtracking-Free Specification]
Let \(T = (\Sigma, \mathcal{R})\) be a GT system, \(A \subseteq \Sigma\) an input signature, and \(\mathcal{S}\) a finite set of graphs over \(\Sigma\). Then we say that \((T, \mathcal{S})\) is a \textit{backtracking-free specification} for a language \(\mathcal{L}\) over \(A\) if \((T, \mathcal{S})\) recognises \(\mathcal{L}\) over \(A\), \(T\) is terminating on \(\mathcal{G}(A)\), and \(T\) is confluent on \(\mathcal{L}\).
\end{definition}

\begin{theorem}
Given a backtracking-free specification \((T, \mathcal{S})\) for a language \(\mathcal{L}\) over \(A \subseteq \Sigma\) and an input graph \(G\) over \(A\), the following algorithm is correct: Compute a normal form of \(G\) by deriving successor graphs using \(T\) as long as possible. If the result graph is isomorphic to some \(S \in \mathcal{S}\), the input graph is in the language. Otherwise, the graph is not in the language.
\end{theorem}

\begin{proof}
Suppose \(G\) is not in \(\mathcal{L}\). Then, since \(T\) is terminating on \(\mathcal{G}(A)\) our algorithm must be able to find a normal form of \(G\), say \(H\), and because \(T\) recognises \(\mathcal{L}\), it must be the case that \(H\) is not isomorphic to \(S\), and so the algorithm correctly decides that \(G\) is not in \(\mathcal{L}\).

Now, suppose that \(G\) is in \(\mathcal{L}\). Then, because \(T\) is terminating, as before, we must be able to derive some normal form, \(H\). But then, since \(T\) is both confluent on \(\mathcal{L}\) and recognises \(\mathcal{L}\), it must be the case that \(H\) is isomorphic to \(S\), and so the algorithm correctly decides that \(G\) is in \(\mathcal{L}\).
\end{proof}

For the remainder of this section, we look at two examples that demonstrate how we can use our Generalised Newman's Lemma and Generalised Critical Pair Lemma to verify if we have a backtracking-free specification for a language, given a grammar that generates the language.

\subsection{Backtracking-Free Specification of Series-Parallel Graphs}

Series-parallel graphs were introduced by Duffin \cite{Duffin65a} as a model of electrical networks. A more general version of the class was introduced by Lawler \cite{Lawler78a} and Monma and Sidney \cite{Monma-Sidney79a} as a model for scheduling problems.

\begin{definition}
\textit{Series-parallel} graphs are inductively defined:
\begin{enumerate}
\item \(P\) is a series-parallel graph where \(s\) is the \textit{source} and \(t\) the \textit{sink}.
\item The class of series-parallel graphs is closed under \textit{parallel composition} and \textit{sequential composition}.
\end{enumerate}
\noindent
where \(P = \) \tikz[baseline]{\node (a) at (0.0,0.16) [draw, circle, thick, fill=black, scale=0.3] {\,}; \node (b) at (0.5,0.16) [draw, circle, thick, fill=black, scale=0.3] {\,}; \node (A) at (0.0,0.0) {\tiny{s}}; \node (B) at (0.5,0.0) {\tiny{t}}; \draw (a) edge[->,thick] (b);}, parallel composition identifies the two sources and the two sinks, and sequential composition identifies the sink of one with the source of another. Figure \ref{fig:egsp} shows an example series-parallel graph.
\end{definition}

\begin{figure}[!ht]
\centering
\begin{tikzpicture}[every node/.style={align=center}]
    \node (a) at (0.0,0.0)    [draw, circle, thick, fill=black, scale=0.3] {\,};
    \node (b) at (1.5,1.0)    [draw, circle, thick, fill=black, scale=0.3] {\,};
    \node (c) at (1.5,-1.0)   [draw, circle, thick, fill=black, scale=0.3] {\,};
    \node (d) at (3.0,1.0)    [draw, circle, thick, fill=black, scale=0.3] {\,};
    \node (e) at (3.0,-0.5)   [draw, circle, thick, fill=black, scale=0.3] {\,};
    \node (f) at (3.0,-1.5)   [draw, circle, thick, fill=black, scale=0.3] {\,};
    \node (g) at (4.5,0.0)    [draw, circle, thick, fill=black, scale=0.3] {\,};
    \node (i) at (5.25,-1.75) [draw, circle, thick, fill=black, scale=0.3] {\,};
    \node (j) at (6.0,0.0)    [draw, circle, thick, fill=black, scale=0.3] {\,};
    \node (k) at (6.0,0.0)    [draw, circle, thick, fill=black, scale=0.3] {\,};
    \node (l) at (7.5,0.0)    [draw, circle, thick, fill=black, scale=0.3] {\,};
    \node (m) at (9.0,0.0)    [draw, circle, thick, fill=black, scale=0.3] {\,};

    \draw (a) edge[->,thick,bend left=20] (b)
          (a) edge[->,thick,bend right=20] (b)
          (a) edge[->,thick] (e)
          (b) edge[->,thick] (d)
          (a) edge[->,thick] (c)
          (c) edge[->,thick] (f)
          (d) edge[->,thick] (g)
          (d) edge[->,thick,bend left=10] (m)
          (e) edge[->,thick,bend left=10] (i)
          (f) edge[->,thick] (i)
          (g) edge[->,thick] (j)
          (i) edge[->,thick,bend right=20] (m)
          (k) edge[->,thick,bend right=20] (l)
          (k) edge[->,thick,bend left=20] (l)
          (l) edge[->,thick] (m);
\end{tikzpicture}
\caption{Example series-parallel graph}
\label{fig:egsp}
\end{figure}

Duffin showed that a graph is series-parallel if and only if it can be reduced to \(P\) by a sequence of series and parallel reductions. We can rephrase this, giving a graph grammar that generates the language:

\begin{theorem}[SP Recognition \cite{Plump16a}] \label{thm:sprec}
The class of \textit{series-parallel} graphs is the language generated by grammar \(SP = ((\{\square\}, \{\square\}), (\emptyset, \emptyset), \{s, p\}, P)\).
\end{theorem}

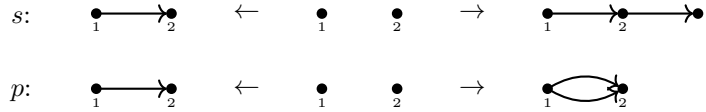
\begin{figure}[!ht]
\centering
\begin{tikzpicture}[every node/.style={align=center}]
    \node (a) at (0.0,-0.05) {$s$:};
    \node (b) at (1.0,0.0)   [draw, circle, thick, fill=black, scale=0.3] {\,};
    \node (c) at (2.0,0.0)   [draw, circle, thick, fill=black, scale=0.3] {\,};
    \node (d) at (3.0,0.0)   {$\leftarrow$};
    \node (e) at (4.0,0.0)   [draw, circle, thick, fill=black, scale=0.3] {\,};
    \node (f) at (5.0,0.0)   [draw, circle, thick, fill=black, scale=0.3] {\,};
    \node (g) at (6.0,0.0)   {$\rightarrow$};
    \node (h) at (7.0,0.0)   [draw, circle, thick, fill=black, scale=0.3] {\,};
    \node (i) at (8.0,0.0)   [draw, circle, thick, fill=black, scale=0.3] {\,};
    \node (j) at (9.0,0.0)   [draw, circle, thick, fill=black, scale=0.3] {\,};

    \node (B) at (1.0,-.18) {\tiny{1}};
    \node (C) at (2.0,-.18) {\tiny{2}};
    \node (E) at (4.0,-.18) {\tiny{1}};
    \node (F) at (5.0,-.18) {\tiny{2}};
    \node (H) at (7.0,-.18) {\tiny{1}};
    \node (J) at (8.0,-.18) {\tiny{2}};

    \draw (b) edge[->,thick] (c)
          (h) edge[->,thick] (i)
          (i) edge[->,thick] (j);

    \node (a) at (0.0,-1.05) {$p$:};
    \node (b) at (1.0,-1.0)   [draw, circle, thick, fill=black, scale=0.3] {\,};
    \node (c) at (2.0,-1.0)   [draw, circle, thick, fill=black, scale=0.3] {\,};
    \node (d) at (3.0,-1.0)   {$\leftarrow$};
    \node (e) at (4.0,-1.0)   [draw, circle, thick, fill=black, scale=0.3] {\,};
    \node (f) at (5.0,-1.0)   [draw, circle, thick, fill=black, scale=0.3] {\,};
    \node (g) at (6.0,-1.0)   {$\rightarrow$};
    \node (h) at (7.0,-1.0)   [draw, circle, thick, fill=black, scale=0.3] {\,};
    \node (i) at (8.0,-1.0)   [draw, circle, thick, fill=black, scale=0.3] {\,};

    \node (B) at (1.0,-1.18) {\tiny{1}};
    \node (C) at (2.0,-1.18) {\tiny{2}};
    \node (E) at (4.0,-1.18) {\tiny{1}};
    \node (F) at (5.0,-1.18) {\tiny{2}};
    \node (H) at (7.0,-1.18) {\tiny{1}};
    \node (I) at (8.0,-1.18) {\tiny{2}};

    \draw (b) edge[->,thick] (c)
          (h) edge[->,thick,bend left=30] (i)
          (h) edge[->,thick,bend right=30] (i);
\end{tikzpicture}
\label{fig:sp}
\caption{Series-parallel graph generation rules}
\end{figure}

By traditional critical pair analysis, one can establish that the reversed rules are confluent (Figure \ref{fig:ser-par-crit-pairs}), however, we run into a problem if we want to consider arbitrarily labelled graphs. Consider the case where the edge alphabet is of size 2, rather than size 1. The obvious modification to the rules is to use all combinations of labels in LHS graphs (Figures \ref{fig:eg-b-rules} and \ref{fig:msp}), however Hristakiev and Plump \cite{Hristakiev-Plump18a} observed that when doing the equivalent of this in GP\,2, we no longer have confluence.

\begin{figure}[!ht]
\centering
\scalebox{0.825}{
\begin{tabular}{|l|c|c|c|c|}
\hline
\multicolumn{1}{|>{\centering\arraybackslash}p{6.4cm}|}{Pair/Property} & \multicolumn{1}{>{\centering\arraybackslash}p{1.666667cm}|}{Joinable} & \multicolumn{1}{>{\centering\arraybackslash}p{1.666667cm}|}{Strongly Joinable} \\ \hline
\scalebox{0.666667}{\begin{tikzpicture}[every node/.style={align=center}]
    \node (x) at (-0.1,-0.25) {\,};
    \node (y) at (9.1,1.25)   {\,};

    \node (a) at (0.0,0.5)    [draw, circle, thick, fill=black, scale=0.3] {\,};
    \node (b) at (1.5,0.5)    [draw, circle, thick, fill=black, scale=0.3] {\,};
    \node (c) at (2.5,0.5)    {$\Leftarrow_{p^{-1}}$};
    \node (d) at (3.5,0.5)    [draw, circle, thick, fill=black, scale=0.3] {\,};
    \node (e) at (5.5,0.5)    [draw, circle, thick, fill=black, scale=0.3] {\,};
    \node (f) at (6.5,0.5)    {$\Rightarrow_{p^{-1}}$};
    \node (g) at (7.5,0.5)    [draw, circle, thick, fill=black, scale=0.3] {\,};
    \node (h) at (9.0,0.5)    [draw, circle, thick, fill=black, scale=0.3] {\,};

    \node (A) at (0.0,0.32)   {\tiny{1}};
    \node (B) at (1.5,0.32)   {\tiny{2}};
    \node (D) at (3.5,0.32)   {\tiny{1}};
    \node (E) at (5.5,0.32)   {\tiny{2}};
    \node (G) at (7.5,0.32)   {\tiny{1}};
    \node (H) at (9.0,0.32)   {\tiny{2}};

    \draw (a) edge[->,thick] (b)
          (d) edge[->,thick,bend left=30] (e)
          (d) edge[->,thick,bend right=30] (e)
          (g) edge[->,thick] (h);
\end{tikzpicture}} & \ding{51} & \ding{51} \\ \hline
\scalebox{0.666667}{\begin{tikzpicture}[every node/.style={align=center}]
    \node (x) at (-0.1,-0.25) {\,};
    \node (y) at (9.1,1.25)   {\,};

    \node (a) at (0.0,0.5)    [draw, circle, thick, fill=black, scale=0.3] {\,};
    \node (b) at (1.5,0.5)    [draw, circle, thick, fill=black, scale=0.3] {\,};
    \node (c) at (2.5,0.5)    {$\Leftarrow_{p^{-1}}$};
    \node (d) at (3.5,0.5)    [draw, circle, thick, fill=black, scale=0.3] {\,};
    \node (e) at (5.5,0.5)    [draw, circle, thick, fill=black, scale=0.3] {\,};
    \node (f) at (6.5,0.5)    {$\Rightarrow_{p^{-1}}$};
    \node (g) at (7.5,0.5)    [draw, circle, thick, fill=black, scale=0.3] {\,};
    \node (h) at (9.0,0.5)    [draw, circle, thick, fill=black, scale=0.3] {\,};

    \node (A) at (0.0,0.32)   {\tiny{1}};
    \node (B) at (1.5,0.32)   {\tiny{2}};
    \node (D) at (3.5,0.32)  {\tiny{1}};
    \node (E) at (5.5,0.32)   {\tiny{2}};
    \node (G) at (7.5,0.32)   {\tiny{1}};
    \node (H) at (9.0,0.32)   {\tiny{2}};

    \draw (a) edge[->,thick,bend left=30] (b)
          (a) edge[->,thick,bend right=30] (b)
          (d) edge[->,thick,bend left=45] (e)
          (d) edge[->,thick] (e)
          (d) edge[->,thick,bend right=45] (e)
          (g) edge[->,thick,bend left=30] (h)
          (g) edge[->,thick,bend right=30] (h);
\end{tikzpicture}} & \ding{51} & \ding{51} \\ \hline
\scalebox{0.666667}{\begin{tikzpicture}[every node/.style={align=center}]
    \node (x) at (-0.1,-0.25) {\,};
    \node (y) at (9.1,1.25)   {\,};

    \node (a) at (0.0,0.5)    [draw, circle, thick, fill=black, scale=0.3] {\,};
    \node (b) at (1.5,0.5)    [draw, circle, thick, fill=black, scale=0.3] {\,};
    \node (c) at (2.5,0.5)    {$\Leftarrow_{s^{-1}}$};
    \node (d) at (3.5,0.0)    [draw, circle, thick, fill=black, scale=0.3] {\,};
    \node (e) at (4.5,1.0)    [draw, circle, thick, fill=black, scale=0.3] {\,};
    \node (f) at (5.5,0.0)    [draw, circle, thick, fill=black, scale=0.3] {\,};
    \node (g) at (6.5,0.5)    {$\Rightarrow_{s^{-1}}$};
    \node (h) at (7.5,0.5)    [draw, circle, thick, fill=black, scale=0.3] {\,};
    \node (i) at (9.0,0.5)    [draw, circle, thick, fill=black, scale=0.3] {\,};

    \node (A) at (0.0,0.32)   {\tiny{1}};
    \node (B) at (1.5,0.32)   {\tiny{2}};
    \node (D) at (3.5,-0.18)  {\tiny{1}};
    \node (E) at (4.5,0.82)   {\tiny{2}};
    \node (F) at (5.5,-0.18)  {\tiny{3}};
    \node (H) at (7.5,0.32)   {\tiny{1}};
    \node (I) at (9.0,0.32)   {\tiny{3}};

    \draw (a) edge[->,thick,bend left=30] (b)
          (b) edge[->,thick,bend left=30] (a)
          (d) edge[->,thick] (e)
          (e) edge[->,thick] (f)
          (f) edge[->,thick] (d)
          (h) edge[->,thick,bend left=30] (i)
          (i) edge[->,thick,bend left=30] (h);
\end{tikzpicture}} & \ding{51} & \ding{51} \\ \hline
\scalebox{0.666667}{\begin{tikzpicture}[every node/.style={align=center}]
    \node (x) at (-0.1,-0.25) {\,};
    \node (y) at (9.1,1.25)   {\,};

    \node (a) at (0.0,0.5)       [draw, circle, thick, fill=black, scale=0.3] {\,};
    \node (b) at (0.75,0.5)      [draw, circle, thick, fill=black, scale=0.3] {\,};
    \node (c) at (1.5,0.5)       [draw, circle, thick, fill=black, scale=0.3] {\,};
    \node (d) at (2.5,0.5)       {$\Leftarrow_{s^{-1}}$};
    \node (e) at (3.5,0.5)       [draw, circle, thick, fill=black, scale=0.3] {\,};
    \node (f) at (4.166667,0.5)  [draw, circle, thick, fill=black, scale=0.3] {\,};
    \node (g) at (4.866667,0.5)  [draw, circle, thick, fill=black, scale=0.3] {\,};
    \node (h) at (5.5,0.5)       [draw, circle, thick, fill=black, scale=0.3] {\,};
    \node (i) at (6.5,0.5)       {$\Rightarrow_{s^{-1}}$};
    \node (j) at (7.5,0.5)       [draw, circle, thick, fill=black, scale=0.3] {\,};
    \node (k) at (8.25,0.5)      [draw, circle, thick, fill=black, scale=0.3] {\,};
    \node (l) at (9.0,0.5)       [draw, circle, thick, fill=black, scale=0.3] {\,};

    \node (A) at (0.0,0.32)      {\tiny{1}};
    \node (B) at (0.75,0.32)     {\tiny{3}};
    \node (C) at (1.5,0.32)      {\tiny{4}};
    \node (E) at (3.5,0.32)      {\tiny{1}};
    \node (F) at (4.166667,0.32) {\tiny{2}};
    \node (G) at (4.866667,0.32) {\tiny{3}};
    \node (H) at (5.5,0.32)      {\tiny{4}};
    \node (J) at (7.5,0.32)      {\tiny{1}};
    \node (K) at (8.25,0.32)     {\tiny{2}};
    \node (L) at (9.0,0.32)      {\tiny{4}};

    \draw (a) edge[->,thick] (b)
          (b) edge[->,thick] (c)
          (e) edge[->,thick] (f)
          (f) edge[->,thick] (g)
          (g) edge[->,thick] (h)
          (j) edge[->,thick] (k)
          (k) edge[->,thick] (l);
\end{tikzpicture}} & \ding{51} & \ding{51} \\ \hline
\end{tabular}
}
\caption{Series-parallel critical pair analysis}
\label{fig:ser-par-crit-pairs}
\end{figure}

\begin{definition}
The class of \textit{labelled series-parallel graphs} (LSPs) is all series-parallel graphs, but with arbitrary edge labels chosen from \(\Sigma_{E} = \{a, b\}\).
\end{definition}

The GT system with the 7 rules from Figures \ref{fig:eg-b-rules} and \ref{fig:msp} has 26 non-isomorphic critical pairs. 16 of the critical pairs are conflicts between the sequential reduction rules (Figure \ref{fig:critpairs1}) and the remaining 10 are conflicts between the parallel reduction rules (Figure \ref{fig:critpairspar}). The non-joinable pairs confirm we no longer have confluence, however the fact that the language of \textit{labelled series-parallel graphs} is closed under the rules, the rules are terminating, and all the non-garbage critical pairs are strongly joinable, allows us to conclude the rules are confluent up to garbage on the language of \textit{labelled series-parallel graphs}.

\begin{figure}[!ht]
\centering
\begin{tikzpicture}[every node/.style={align=center}]
    \node (a) at (0.0,1.20)   {$p_1$:};
    \node (b) at (1.0,1.25)   [draw, circle, thick, fill=black, scale=0.3] {\,};
    \node (d) at (2.0,1.25)   [draw, circle, thick, fill=black, scale=0.3] {\,};
    \node (e) at (3.0,1.25)   {$\leftarrow$};
    \node (f) at (4.0,1.25)   [draw, circle, thick, fill=black, scale=0.3] {\,};
    \node (g) at (5.0,1.25)   [draw, circle, thick, fill=black, scale=0.3] {\,};
    \node (h) at (6.0,1.25)   {$\rightarrow$};
    \node (i) at (7.0,1.25)   [draw, circle, thick, fill=black, scale=0.3] {\,};
    \node (j) at (8.0,1.25)   [draw, circle, thick, fill=black, scale=0.3] {\,};

    \node (B) at (1.0,1.07)   {\tiny{1}};
    \node (D) at (2.0,1.07)   {\tiny{2}};
    \node (F) at (4.0,1.07)   {\tiny{1}};
    \node (G) at (5.0,1.07)   {\tiny{2}};
    \node (I) at (7.0,1.07)   {\tiny{1}};
    \node (J) at (8.0,1.07)   {\tiny{2}};

    \draw (b) edge[->,thick, bend left=30] node[above] {a} (d)
          (b) edge[->,thick, bend right=30] node[below] {a} (d)
          (i) edge[->,thick] node[above] {a} (j);

    \node (a) at (0.0,-0.30)  {$p_2$:};
    \node (b) at (1.0,-0.25)  [draw, circle, thick, fill=black, scale=0.3] {\,};
    \node (d) at (2.0,-0.25)  [draw, circle, thick, fill=black, scale=0.3] {\,};
    \node (e) at (3.0,-0.25)  {$\leftarrow$};
    \node (f) at (4.0,-0.25)  [draw, circle, thick, fill=black, scale=0.3] {\,};
    \node (g) at (5.0,-0.25)  [draw, circle, thick, fill=black, scale=0.3] {\,};
    \node (h) at (6.0,-0.25)  {$\rightarrow$};
    \node (i) at (7.0,-0.25)  [draw, circle, thick, fill=black, scale=0.3] {\,};
    \node (j) at (8.0,-0.25)  [draw, circle, thick, fill=black, scale=0.3] {\,};

    \node (B) at (1.0,-0.43)  {\tiny{1}};
    \node (D) at (2.0,-0.43)  {\tiny{2}};
    \node (F) at (4.0,-0.43)  {\tiny{1}};
    \node (G) at (5.0,-0.43)  {\tiny{2}};
    \node (I) at (7.0,-0.43)  {\tiny{1}};
    \node (J) at (8.0,-0.43)  {\tiny{2}};

    \draw (b) edge[->,thick, bend left=30] node[above] {a} (d)
          (b) edge[->,thick, bend right=30] node[below] {b} (d)
          (i) edge[->,thick] node[above] {a} (j);

    \node (a) at (0.0,-1.80)  {$p_3$:};
    \node (b) at (1.0,-1.75)  [draw, circle, thick, fill=black, scale=0.3] {\,};
    \node (d) at (2.0,-1.75)  [draw, circle, thick, fill=black, scale=0.3] {\,};
    \node (e) at (3.0,-1.75)  {$\leftarrow$};
    \node (f) at (4.0,-1.75)  [draw, circle, thick, fill=black, scale=0.3] {\,};
    \node (g) at (5.0,-1.75)  [draw, circle, thick, fill=black, scale=0.3] {\,};
    \node (h) at (6.0,-1.75)  {$\rightarrow$};
    \node (i) at (7.0,-1.75)  [draw, circle, thick, fill=black, scale=0.3] {\,};
    \node (j) at (8.0,-1.75)  [draw, circle, thick, fill=black, scale=0.3] {\,};

    \node (B) at (1.0,-1.93)  {\tiny{1}};
    \node (D) at (2.0,-1.93)  {\tiny{2}};
    \node (F) at (4.0,-1.93)  {\tiny{1}};
    \node (G) at (5.0,-1.93)  {\tiny{2}};
    \node (I) at (7.0,-1.93)  {\tiny{1}};
    \node (J) at (8.0,-1.93)  {\tiny{2}};

    \draw (b) edge[->,thick, bend left=30] node[above] {b} (d)
          (b) edge[->,thick, bend right=30] node[below] {b} (d)
          (i) edge[->,thick] node[above] {a} (j);
\end{tikzpicture}
\caption{Parallel LSP reduction rules}
\label{fig:msp}
\end{figure}
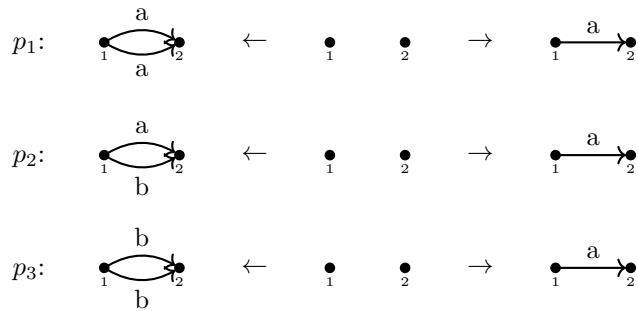

\begin{figure}[!ht]
\centering
\scalebox{0.825}{
\begin{tabular}{|l|c|c|c|c|}
\hline
\multicolumn{1}{|>{\centering\arraybackslash}p{6.4cm}|}{Pair/Property} & \multicolumn{1}{>{\centering\arraybackslash}p{1.666667cm}|}{Joinable} & \multicolumn{1}{>{\centering\arraybackslash}p{1.666667cm}|}{Strongly Joinable} & \multicolumn{1}{>{\centering\arraybackslash}p{1.666667cm}|}{Non-Garbage} \\ \hline
\scalebox{0.666667}{\begin{tikzpicture}[every node/.style={align=center}]
    \node (x) at (-0.1,-0.25) {\,};
    \node (y) at (9.1,1.25)   {\,};

    \node (a) at (0.0,0.333333)    [draw, circle, thick, fill=black, scale=0.3] {\,};
    \node (b) at (1.5,0.333333)    [draw, circle, thick, fill=black, scale=0.3] {\,};
    \node (c) at (2.5,0.333333)    {$\Leftarrow_{p_1}$};
    \node (d) at (3.5,0.333333)    [draw, circle, thick, fill=black, scale=0.3] {\,};
    \node (e) at (5.5,0.333333)    [draw, circle, thick, fill=black, scale=0.3] {\,};
    \node (f) at (6.5,0.333333)    {$\Rightarrow_{p_1}$};
    \node (g) at (7.5,0.333333)    [draw, circle, thick, fill=black, scale=0.3] {\,};
    \node (h) at (9.0,0.333333)    [draw, circle, thick, fill=black, scale=0.3] {\,};

    \node (A) at (0.0,0.153333)   {\tiny{1}};
    \node (B) at (1.5,0.153333)   {\tiny{2}};
    \node (D) at (3.5,0.153333)   {\tiny{1}};
    \node (E) at (5.5,0.153333)   {\tiny{2}};
    \node (G) at (7.5,0.153333)   {\tiny{1}};
    \node (H) at (9.0,0.153333)   {\tiny{2}};

    \draw (a) edge[->,thick] node[above] {a} (b)
          (d) edge[->,thick,bend left=30] node[above] {a} (e)
          (d) edge[->,thick,bend right=30] node[above] {a} (e)
          (g) edge[->,thick] node[above] {a} (h);
\end{tikzpicture}} & \ding{51} & \ding{51} & \ding{51} \\ \hline
\scalebox{0.666667}{\begin{tikzpicture}[every node/.style={align=center}]
    \node (x) at (-0.1,-0.25) {\,};
    \node (y) at (9.1,1.25)   {\,};

    \node (a) at (0.0,0.333333)    [draw, circle, thick, fill=black, scale=0.3] {\,};
    \node (b) at (1.5,0.333333)    [draw, circle, thick, fill=black, scale=0.3] {\,};
    \node (c) at (2.5,0.333333)    {$\Leftarrow_{p_1}$};
    \node (d) at (3.5,0.333333)    [draw, circle, thick, fill=black, scale=0.3] {\,};
    \node (e) at (5.5,0.333333)    [draw, circle, thick, fill=black, scale=0.3] {\,};
    \node (f) at (6.5,0.333333)    {$\Rightarrow_{p_1}$};
    \node (g) at (7.5,0.333333)    [draw, circle, thick, fill=black, scale=0.3] {\,};
    \node (h) at (9.0,0.333333)    [draw, circle, thick, fill=black, scale=0.3] {\,};

    \node (A) at (0.0,0.153333)   {\tiny{1}};
    \node (B) at (1.5,0.153333)   {\tiny{2}};
    \node (D) at (3.5,0.153333)   {\tiny{1}};
    \node (E) at (5.5,0.153333)   {\tiny{2}};
    \node (G) at (7.5,0.153333)   {\tiny{1}};
    \node (H) at (9.0,0.153333)   {\tiny{2}};

    \draw (a) edge[->,thick,bend left=30] node[above] {a} (b)
          (a) edge[->,thick,bend right=30] node[above] {a} (b)
          (d) edge[->,thick,bend left=45] node[above] {a} (e)
          (d) edge[->,thick] node[above] {a} (e)
          (d) edge[->,thick,bend right=45] node[above] {a} (e)
          (g) edge[->,thick,bend left=30] node[above] {a} (h)
          (g) edge[->,thick,bend right=30] node[above] {a} (h);
\end{tikzpicture}} & \ding{51} & \ding{51} & \ding{51} \\ \hline
\scalebox{0.666667}{\begin{tikzpicture}[every node/.style={align=center}]
    \node (x) at (-0.1,-0.25) {\,};
    \node (y) at (9.1,1.25)   {\,};

    \node (a) at (0.0,0.333333)    [draw, circle, thick, fill=black, scale=0.3] {\,};
    \node (b) at (1.5,0.333333)    [draw, circle, thick, fill=black, scale=0.3] {\,};
    \node (c) at (2.5,0.333333)    {$\Leftarrow_{p_1}$};
    \node (d) at (3.5,0.333333)    [draw, circle, thick, fill=black, scale=0.3] {\,};
    \node (e) at (5.5,0.333333)    [draw, circle, thick, fill=black, scale=0.3] {\,};
    \node (f) at (6.5,0.333333)    {$\Rightarrow_{p_2}$};
    \node (g) at (7.5,0.333333)    [draw, circle, thick, fill=black, scale=0.3] {\,};
    \node (h) at (9.0,0.333333)    [draw, circle, thick, fill=black, scale=0.3] {\,};

    \node (A) at (0.0,0.153333)   {\tiny{1}};
    \node (B) at (1.5,0.153333)   {\tiny{2}};
    \node (D) at (3.5,0.153333)   {\tiny{1}};
    \node (E) at (5.5,0.153333)   {\tiny{2}};
    \node (G) at (7.5,0.153333)   {\tiny{1}};
    \node (H) at (9.0,0.153333)   {\tiny{2}};

    \draw (a) edge[->,thick,bend left=30] node[above] {a} (b)
          (a) edge[->,thick,bend right=30] node[above] {b} (b)
          (d) edge[->,thick,bend left=45] node[above] {a} (e)
          (d) edge[->,thick] node[above] {a} (e)
          (d) edge[->,thick,bend right=45] node[above] {b} (e)
          (g) edge[->,thick,bend left=30] node[above] {a} (h)
          (g) edge[->,thick,bend right=30] node[above] {a} (h);
\end{tikzpicture}} & \ding{51} & \ding{51} & \ding{51} \\ \hline
\scalebox{0.666667}{\begin{tikzpicture}[every node/.style={align=center}]
    \node (x) at (-0.1,-0.25) {\,};
    \node (y) at (9.1,1.25)   {\,};

    \node (a) at (0.0,0.333333)    [draw, circle, thick, fill=black, scale=0.3] {\,};
    \node (b) at (1.5,0.333333)    [draw, circle, thick, fill=black, scale=0.3] {\,};
    \node (c) at (2.5,0.333333)    {$\Leftarrow_{p_1}$};
    \node (d) at (3.5,0.333333)    [draw, circle, thick, fill=black, scale=0.3] {\,};
    \node (e) at (5.5,0.333333)    [draw, circle, thick, fill=black, scale=0.3] {\,};
    \node (f) at (6.5,0.333333)    {$\Rightarrow_{p_2}$};
    \node (g) at (7.5,0.333333)    [draw, circle, thick, fill=black, scale=0.3] {\,};
    \node (h) at (9.0,0.333333)    [draw, circle, thick, fill=black, scale=0.3] {\,};

    \node (A) at (0.0,0.153333)   {\tiny{1}};
    \node (B) at (1.5,0.153333)   {\tiny{2}};
    \node (D) at (3.5,0.153333)   {\tiny{1}};
    \node (E) at (5.5,0.153333)   {\tiny{2}};
    \node (G) at (7.5,0.153333)   {\tiny{1}};
    \node (H) at (9.0,0.153333)   {\tiny{2}};

    \draw (a) edge[->,thick,bend left=30] node[above] {a} (b)
          (a) edge[->,thick,bend right=30] node[above] {b} (b)
          (d) edge[->,thick,bend left=45] node[above] {a} (e)
          (d) edge[->,thick] node[above] {a} (e)
          (d) edge[->,thick,bend right=45] node[above] {b} (e)
          (g) edge[->,thick,bend left=30] node[above] {a} (h)
          (g) edge[->,thick,bend right=30] node[above] {a} (h);
\end{tikzpicture}} & \ding{51} & \ding{51} & \ding{51} \\ \hline
\scalebox{0.666667}{\begin{tikzpicture}[every node/.style={align=center}]
    \node (x) at (-0.1,-0.25) {\,};
    \node (y) at (9.1,1.25)   {\,};

    \node (a) at (0.0,0.333333)    [draw, circle, thick, fill=black, scale=0.3] {\,};
    \node (b) at (1.5,0.333333)    [draw, circle, thick, fill=black, scale=0.3] {\,};
    \node (c) at (2.5,0.333333)    {$\Leftarrow_{p_2}$};
    \node (d) at (3.5,0.333333)    [draw, circle, thick, fill=black, scale=0.3] {\,};
    \node (e) at (5.5,0.333333)    [draw, circle, thick, fill=black, scale=0.3] {\,};
    \node (f) at (6.5,0.333333)    {$\Rightarrow_{p_2}$};
    \node (g) at (7.5,0.333333)    [draw, circle, thick, fill=black, scale=0.3] {\,};
    \node (h) at (9.0,0.333333)    [draw, circle, thick, fill=black, scale=0.3] {\,};

    \node (A) at (0.0,0.153333)   {\tiny{1}};
    \node (B) at (1.5,0.153333)   {\tiny{2}};
    \node (D) at (3.5,0.153333)   {\tiny{1}};
    \node (E) at (5.5,0.153333)   {\tiny{2}};
    \node (G) at (7.5,0.153333)   {\tiny{1}};
    \node (H) at (9.0,0.153333)   {\tiny{2}};

    \draw (a) edge[->,thick,bend left=30] node[above] {a} (b)
          (a) edge[->,thick,bend right=30] node[above] {b} (b)
          (d) edge[->,thick,bend left=45] node[above] {a} (e)
          (d) edge[->,thick] node[above] {b} (e)
          (d) edge[->,thick,bend right=45] node[above] {b} (e)
          (g) edge[->,thick,bend left=30] node[above] {a} (h)
          (g) edge[->,thick,bend right=30] node[above] {b} (h);
\end{tikzpicture}} & \ding{51} & \ding{51} & \ding{51} \\ \hline
\scalebox{0.666667}{\begin{tikzpicture}[every node/.style={align=center}]
    \node (x) at (-0.1,-0.25) {\,};
    \node (y) at (9.1,1.25)   {\,};

    \node (a) at (0.0,0.333333)    [draw, circle, thick, fill=black, scale=0.3] {\,};
    \node (b) at (1.5,0.333333)    [draw, circle, thick, fill=black, scale=0.3] {\,};
    \node (c) at (2.5,0.333333)    {$\Leftarrow_{p_2}$};
    \node (d) at (3.5,0.333333)    [draw, circle, thick, fill=black, scale=0.3] {\,};
    \node (e) at (5.5,0.333333)    [draw, circle, thick, fill=black, scale=0.3] {\,};
    \node (f) at (6.5,0.333333)    {$\Rightarrow_{p_2}$};
    \node (g) at (7.5,0.333333)    [draw, circle, thick, fill=black, scale=0.3] {\,};
    \node (h) at (9.0,0.333333)    [draw, circle, thick, fill=black, scale=0.3] {\,};

    \node (A) at (0.0,0.153333)   {\tiny{1}};
    \node (B) at (1.5,0.153333)   {\tiny{2}};
    \node (D) at (3.5,0.153333)   {\tiny{1}};
    \node (E) at (5.5,0.153333)   {\tiny{2}};
    \node (G) at (7.5,0.153333)   {\tiny{1}};
    \node (H) at (9.0,0.153333)   {\tiny{2}};

    \draw (a) edge[->,thick,bend left=30] node[above] {a} (b)
          (a) edge[->,thick,bend right=30] node[above] {b} (b)
          (d) edge[->,thick,bend left=45] node[above] {a} (e)
          (d) edge[->,thick] node[above] {b} (e)
          (d) edge[->,thick,bend right=45] node[above] {b} (e)
          (g) edge[->,thick,bend left=30] node[above] {a} (h)
          (g) edge[->,thick,bend right=30] node[above] {b} (h);
\end{tikzpicture}} & \ding{51} & \ding{51} & \ding{51} \\ \hline
\scalebox{0.666667}{\begin{tikzpicture}[every node/.style={align=center}]
    \node (x) at (-0.1,-0.25) {\,};
    \node (y) at (9.1,1.25)   {\,};

    \node (a) at (0.0,0.333333)    [draw, circle, thick, fill=black, scale=0.3] {\,};
    \node (b) at (1.5,0.333333)    [draw, circle, thick, fill=black, scale=0.3] {\,};
    \node (c) at (2.5,0.333333)    {$\Leftarrow_{p_2}$};
    \node (d) at (3.5,0.333333)    [draw, circle, thick, fill=black, scale=0.3] {\,};
    \node (e) at (5.5,0.333333)    [draw, circle, thick, fill=black, scale=0.3] {\,};
    \node (f) at (6.5,0.333333)    {$\Rightarrow_{p_3}$};
    \node (g) at (7.5,0.333333)    [draw, circle, thick, fill=black, scale=0.3] {\,};
    \node (h) at (9.0,0.333333)    [draw, circle, thick, fill=black, scale=0.3] {\,};

    \node (A) at (0.0,0.153333)   {\tiny{1}};
    \node (B) at (1.5,0.153333)   {\tiny{2}};
    \node (D) at (3.5,0.153333)   {\tiny{1}};
    \node (E) at (5.5,0.153333)   {\tiny{2}};
    \node (G) at (7.5,0.153333)   {\tiny{1}};
    \node (H) at (9.0,0.153333)   {\tiny{2}};

    \draw (a) edge[->,thick,bend left=30] node[above] {a} (b)
          (a) edge[->,thick,bend right=30] node[above] {b} (b)
          (d) edge[->,thick,bend left=45] node[above] {a} (e)
          (d) edge[->,thick] node[above] {b} (e)
          (d) edge[->,thick,bend right=45] node[above] {b} (e)
          (g) edge[->,thick,bend left=30] node[above] {a} (h)
          (g) edge[->,thick,bend right=30] node[above] {a} (h);
\end{tikzpicture}} & \ding{51} & \ding{51} & \ding{51} \\ \hline
\scalebox{0.666667}{\begin{tikzpicture}[every node/.style={align=center}]
    \node (x) at (-0.1,-0.25) {\,};
    \node (y) at (9.1,1.25)   {\,};

    \node (a) at (0.0,0.333333)    [draw, circle, thick, fill=black, scale=0.3] {\,};
    \node (b) at (1.5,0.333333)    [draw, circle, thick, fill=black, scale=0.3] {\,};
    \node (c) at (2.5,0.333333)    {$\Leftarrow_{p_2}$};
    \node (d) at (3.5,0.333333)    [draw, circle, thick, fill=black, scale=0.3] {\,};
    \node (e) at (5.5,0.333333)    [draw, circle, thick, fill=black, scale=0.3] {\,};
    \node (f) at (6.5,0.333333)    {$\Rightarrow_{p_3}$};
    \node (g) at (7.5,0.333333)    [draw, circle, thick, fill=black, scale=0.3] {\,};
    \node (h) at (9.0,0.333333)    [draw, circle, thick, fill=black, scale=0.3] {\,};

    \node (A) at (0.0,0.153333)   {\tiny{1}};
    \node (B) at (1.5,0.153333)   {\tiny{2}};
    \node (D) at (3.5,0.153333)   {\tiny{1}};
    \node (E) at (5.5,0.153333)   {\tiny{2}};
    \node (G) at (7.5,0.153333)   {\tiny{1}};
    \node (H) at (9.0,0.153333)   {\tiny{2}};

    \draw (a) edge[->,thick,bend left=30] node[above] {a} (b)
          (a) edge[->,thick,bend right=30] node[above] {b} (b)
          (d) edge[->,thick,bend left=45] node[above] {a} (e)
          (d) edge[->,thick] node[above] {b} (e)
          (d) edge[->,thick,bend right=45] node[above] {b} (e)
          (g) edge[->,thick,bend left=30] node[above] {a} (h)
          (g) edge[->,thick,bend right=30] node[above] {a} (h);
\end{tikzpicture}} & \ding{51} & \ding{51} & \ding{51} \\ \hline
\scalebox{0.666667}{\begin{tikzpicture}[every node/.style={align=center}]
    \node (x) at (-0.1,-0.25) {\,};
    \node (y) at (9.1,1.25)   {\,};

    \node (a) at (0.0,0.333333)    [draw, circle, thick, fill=black, scale=0.3] {\,};
    \node (b) at (1.5,0.333333)    [draw, circle, thick, fill=black, scale=0.3] {\,};
    \node (c) at (2.5,0.333333)    {$\Leftarrow_{p_3}$};
    \node (d) at (3.5,0.333333)    [draw, circle, thick, fill=black, scale=0.3] {\,};
    \node (e) at (5.5,0.333333)    [draw, circle, thick, fill=black, scale=0.3] {\,};
    \node (f) at (6.5,0.333333)    {$\Rightarrow_{p_3}$};
    \node (g) at (7.5,0.333333)    [draw, circle, thick, fill=black, scale=0.3] {\,};
    \node (h) at (9.0,0.333333)    [draw, circle, thick, fill=black, scale=0.3] {\,};

    \node (A) at (0.0,0.153333)   {\tiny{1}};
    \node (B) at (1.5,0.153333)   {\tiny{2}};
    \node (D) at (3.5,0.153333)   {\tiny{1}};
    \node (E) at (5.5,0.153333)   {\tiny{2}};
    \node (G) at (7.5,0.153333)   {\tiny{1}};
    \node (H) at (9.0,0.153333)   {\tiny{2}};

    \draw (a) edge[->,thick] node[above] {a} (b)
          (d) edge[->,thick,bend left=30] node[above] {b} (e)
          (d) edge[->,thick,bend right=30] node[above] {b} (e)
          (g) edge[->,thick] node[above] {a} (h);
\end{tikzpicture}} & \ding{51} & \ding{51} & \ding{51} \\ \hline
\scalebox{0.666667}{\begin{tikzpicture}[every node/.style={align=center}]
    \node (x) at (-0.1,-0.25) {\,};
    \node (y) at (9.1,1.25)   {\,};

    \node (a) at (0.0,0.333333)    [draw, circle, thick, fill=black, scale=0.3] {\,};
    \node (b) at (1.5,0.333333)    [draw, circle, thick, fill=black, scale=0.3] {\,};
    \node (c) at (2.5,0.333333)    {$\Leftarrow_{p_3}$};
    \node (d) at (3.5,0.333333)    [draw, circle, thick, fill=black, scale=0.3] {\,};
    \node (e) at (5.5,0.333333)    [draw, circle, thick, fill=black, scale=0.3] {\,};
    \node (f) at (6.5,0.333333)    {$\Rightarrow_{p_3}$};
    \node (g) at (7.5,0.333333)    [draw, circle, thick, fill=black, scale=0.3] {\,};
    \node (h) at (9.0,0.333333)    [draw, circle, thick, fill=black, scale=0.3] {\,};

    \node (A) at (0.0,0.153333)   {\tiny{1}};
    \node (B) at (1.5,0.153333)   {\tiny{2}};
    \node (D) at (3.5,0.153333)   {\tiny{1}};
    \node (E) at (5.5,0.153333)   {\tiny{2}};
    \node (G) at (7.5,0.153333)   {\tiny{1}};
    \node (H) at (9.0,0.153333)   {\tiny{2}};

    \draw (a) edge[->,thick,bend left=30] node[above] {a} (b)
          (a) edge[->,thick,bend right=30] node[above] {b} (b)
          (d) edge[->,thick,bend left=45] node[above] {b} (e)
          (d) edge[->,thick] node[above] {b} (e)
          (d) edge[->,thick,bend right=45] node[above] {b} (e)
          (g) edge[->,thick,bend left=30] node[above] {b} (h)
          (g) edge[->,thick,bend right=30] node[above] {a} (h);
\end{tikzpicture}} & \ding{51} & \ding{51} & \ding{51} \\ \hline
\end{tabular}
}
\caption{Critical pair analysis of parallel rules}
\label{fig:critpairspar}
\end{figure}

\begin{theorem}[Backtracking-Free LSP Specification]
Let \(\Sigma = (\{\square\}, \{a, b\})\), \(T = (\Sigma,\) \(\{s_1, s_2, s_3, s_4, p_1, p_2, p_3\}\)), \(P_a = \) \tikz[baseline]{\node (a) at (0.0,0.1) [draw, circle, thick, fill=black, scale=0.3] {\,}; \node (b) at (0.5,0.1) [draw, circle, thick, fill=black, scale=0.3] {\,}; \node (C) at (0.2,0.2) {\tiny{a}}; \draw (a) edge[->,thick] (b);} and \(P_b = \) \tikz[baseline]{\node (a) at (0.0,0.1) [draw, circle, thick, fill=black, scale=0.3] {\,}; \node (b) at (0.5,0.1) [draw, circle, thick, fill=black, scale=0.3] {\,}; \node (C) at (0.2,0.2) {\tiny{b}}; \draw (a) edge[->,thick] (b);}. Then \((T, \{P_a, P_b\})\) is a \textit{backtracking-free specification} for the \textit{labelled series-parallel graphs} over \(\Sigma\).
\end{theorem}

\begin{proof}
We denote by \(\mathcal{L}\) the language of all labelled series-parallel graphs. Our rules are structurally the same as the unlabelled rules, so because our LHS graphs are arbitrarily labelled, language recognition of \(\mathcal{L}\) over \(\Sigma\) follows from Theorem \ref{thm:sprec}. Formally, our above discussion used Corollary \ref{cor:ngcritpairlem} to establish that \(T\) is confluent up to garbage on \(\mathcal{L}\), as required.
\end{proof}

Finally, we remark that this construction generalises for arbitrary edge label alphabets, and not just those of size \(2\). The number of conflicts is simply much larger, however the critical pair analysis will always conclude in the same way. Thus, we have shown that the obvious generalisation of the series-parallel reduction rules to a non-trivial edge labelling set admits a back-tracking free specification, even though the system is not confluent.

\subsection{Backtracking-Free Specification of Extended Flow Diagrams}
\label{subsec:flow_diagrams}

In 1976, Farrow, Kennedy and Zucconi presented \textit{semi-structured flow graphs}, defining a grammar with confluent reduction rules \cite{Farrow-Kennedy-Zucconi76a}. Plump has considered a restricted version of this language: \textit{extended flow diagrams} (EFDs) \cite{Plump05a}. The reduction rules for \textit{extended flow diagrams} are a backtracking-free specification for the EFDs, despite not being confluent.

Throughout this subsection, we will use a shorthand notation for rules, where we assume all interface graphs contain no edges, and any node that appears in the interface graph will be labelled by a subscript number on both sides of the rule, writing only the left-hand side and right-hand side graphs. We also highlight persistent nodes within critical pairs in blue, for ease of reading. The colouring has no special meaning, other than that.

\begin{figure}[!ht]
\centering
\includegraphics[totalheight=6.0cm]{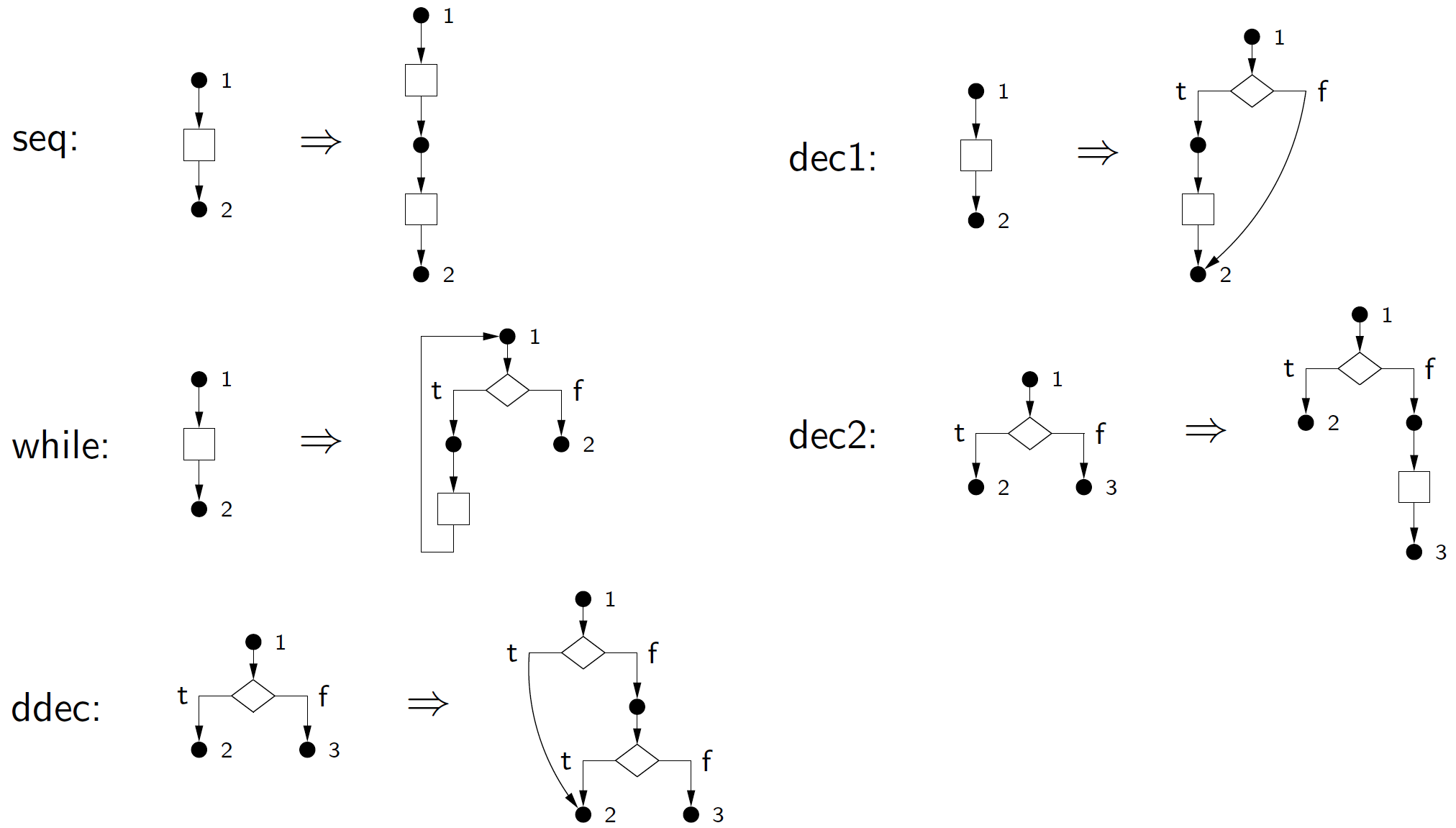}
\caption{EFD grammar rules}
\label{fig:efd-rules}
\end{figure}

We now define extended flow diagrams using a grammar:

\begin{definition}
The language of \textit{extended flow diagrams} is generated by the grammar \(\mathrm{EFD} = (\Sigma, N, \mathcal{R}, S)\) where \(\Sigma_V = \{\bullet, \square, \Diamond\}\), \(\Sigma_E = \{t, f, \square\}\), \(N_V = N_E = \emptyset\) (Figure \ref{fig:efd-rules}), \(\mathcal{R} = \{seq, while, ddec, dec1, dec2\}\), and \(S = \) \tikz[baseline]{ \tikzstyle{ann} = [draw=none,fill=none,right] \node (a) at (0.0,0.1) [draw, circle, thick, fill=black, scale=0.3] {}; \node[rectangle] (b) at (0.5,0.1) [minimum height=0.2cm,minimum width=0.2cm,draw] {}; \node (c) at (1.0,0.1) [draw, circle, thick, fill=black, scale=0.3] {}; \draw (a) edge[->,thick] (b) (b) edge[->,thick] (c);}.
\end{definition}

Before we show that reversing these rules admits a backtracking-free specification, we first need the following fact:

\begin{lemma} \label{lem:efdcycles}
Every \textit{directed cycle} in an EFD contains a \(t\)-labelled edge
\end{lemma}

\begin{proof}
By induction.
\end{proof}

\begin{theorem}[Backtracking-Free EFD Specification]
Let \(T = (\Sigma, \mathcal{R}^{-1})\). Then \((T, \{S\})\) is a \textit{backtracking-free specification} for \(\mathrm{L}(\mathrm{EFD})\) over \(\Sigma\).
\end{theorem}

\begin{figure}[!ht]
\centering
\includegraphics[totalheight=3.2cm]{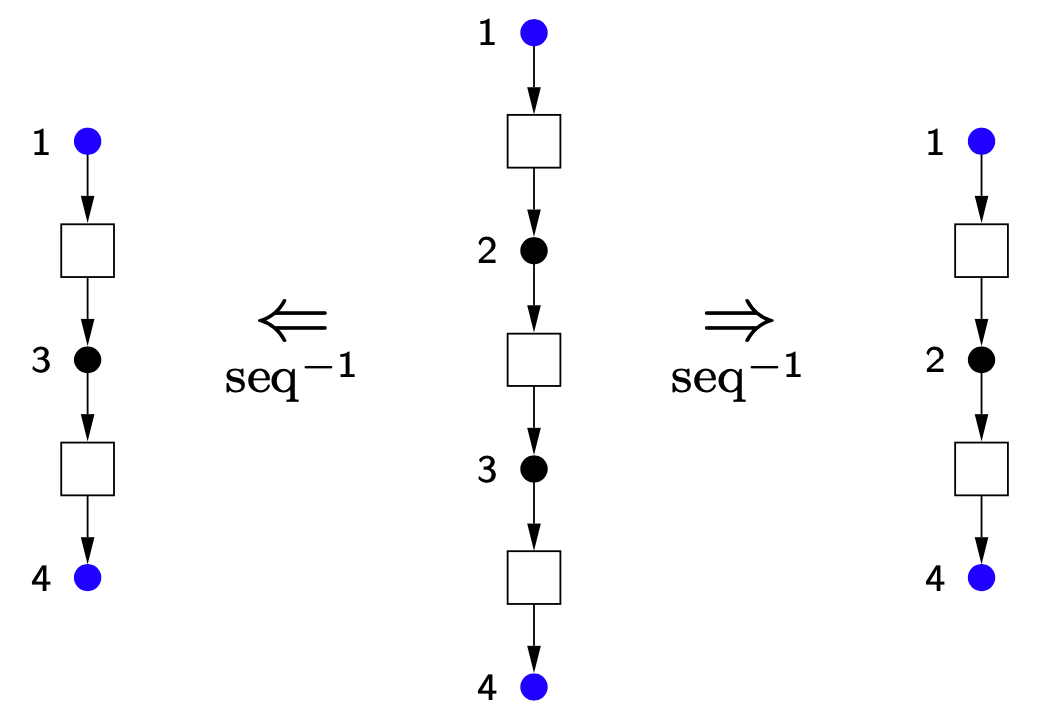}
\caption{EFD critical pair 1}
\label{fig:efd-pair-1}
\end{figure}

\begin{figure}[!ht]
\centering
\includegraphics[totalheight=3.0cm]{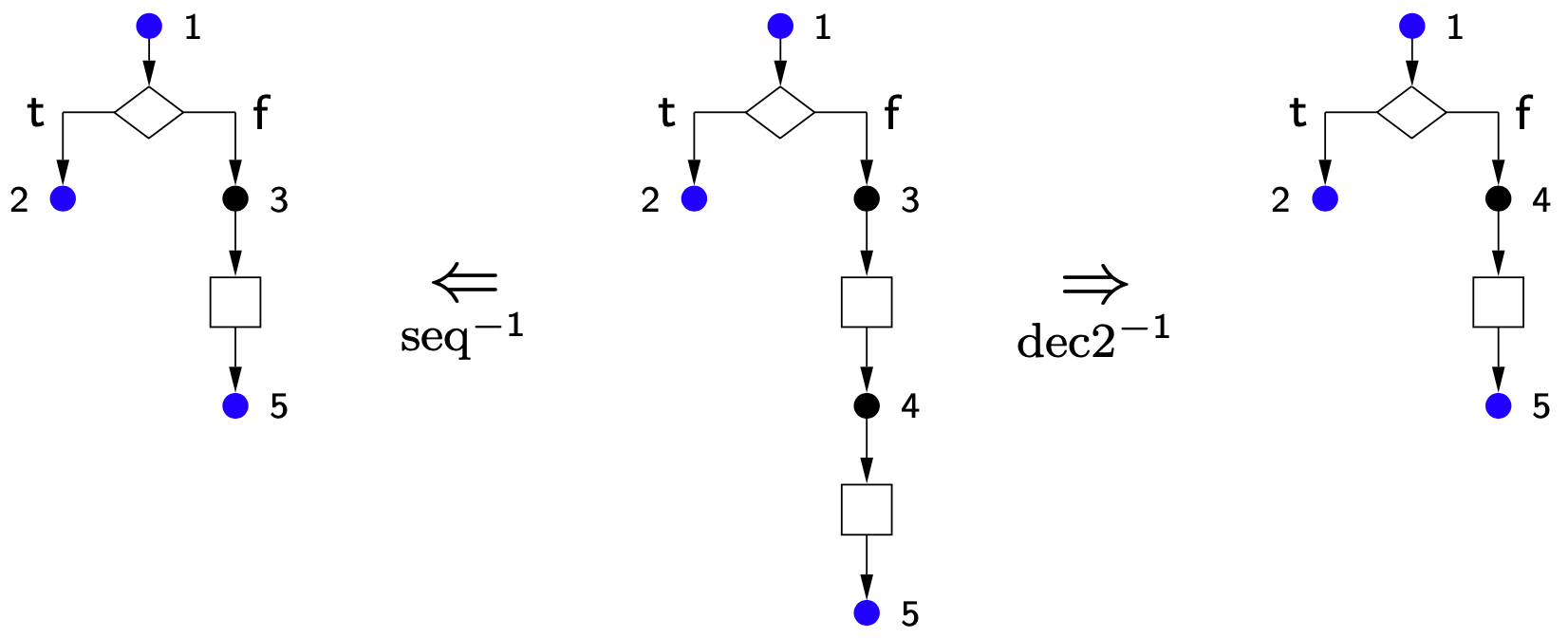}
\caption{EFD critical pair 3}
\label{fig:efd-pair-3}
\end{figure}

\begin{figure}[!ht]
\centering
\includegraphics[totalheight=2.2cm]{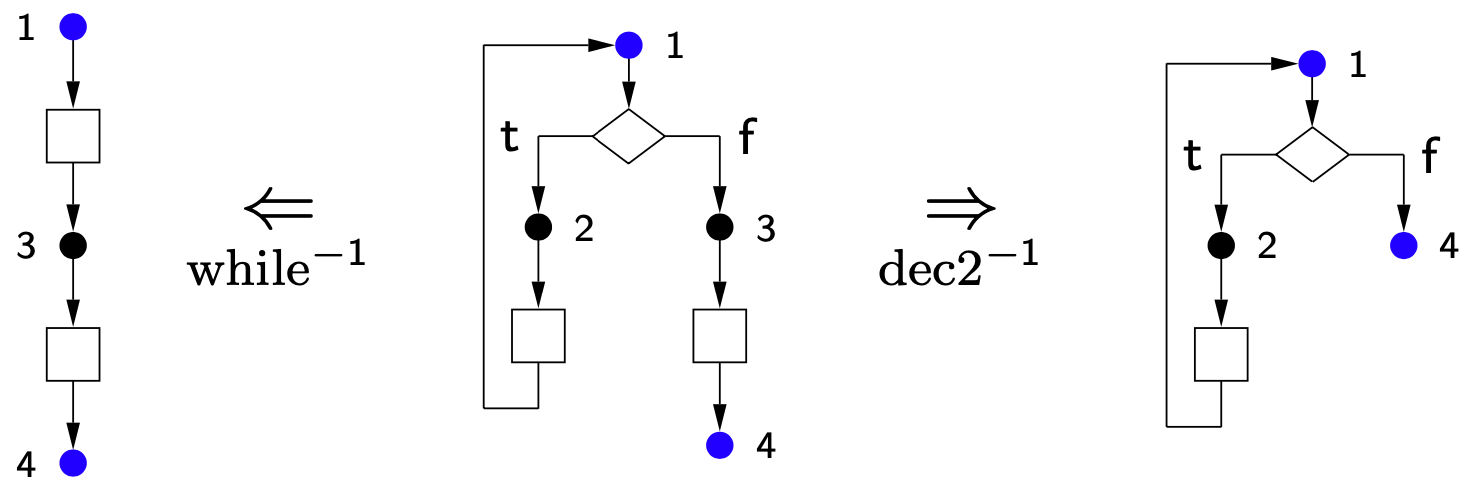}
\caption{EFD critical pair 6}
\label{fig:efd-pair-6}
\end{figure}

\begin{figure}[!ht]
\centering
\includegraphics[totalheight=2.7cm]{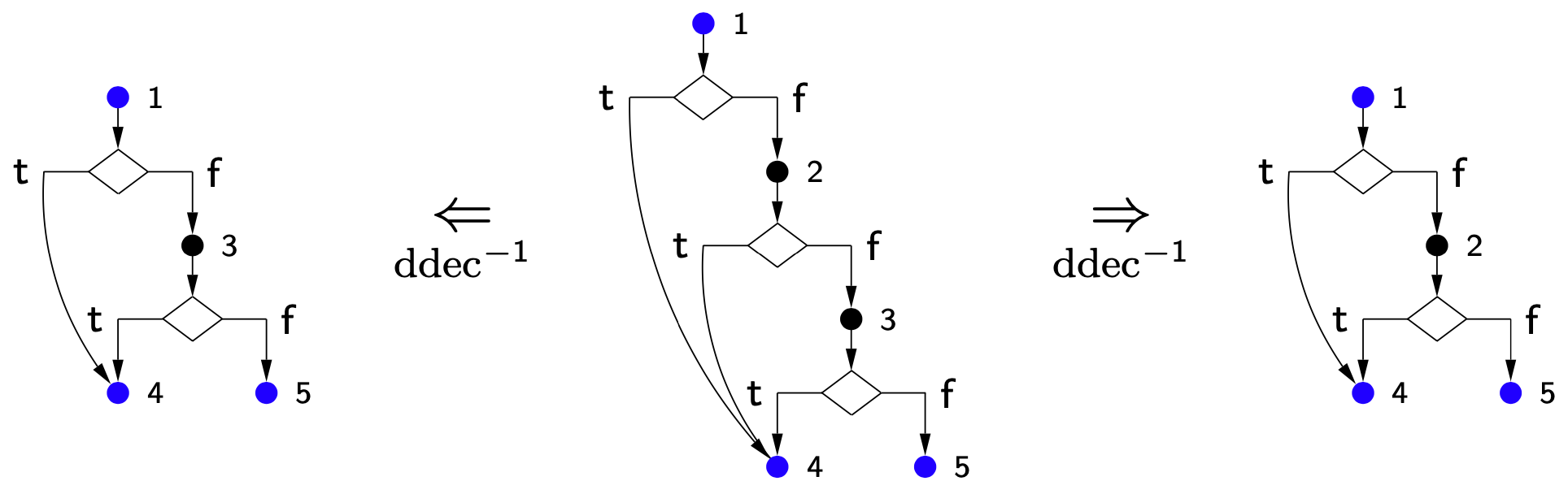}
\caption{EFD critical pair 7}
\label{fig:efd-pair-7}
\end{figure}

\begin{figure}[!ht]
\centering
\includegraphics[totalheight=2.85cm]{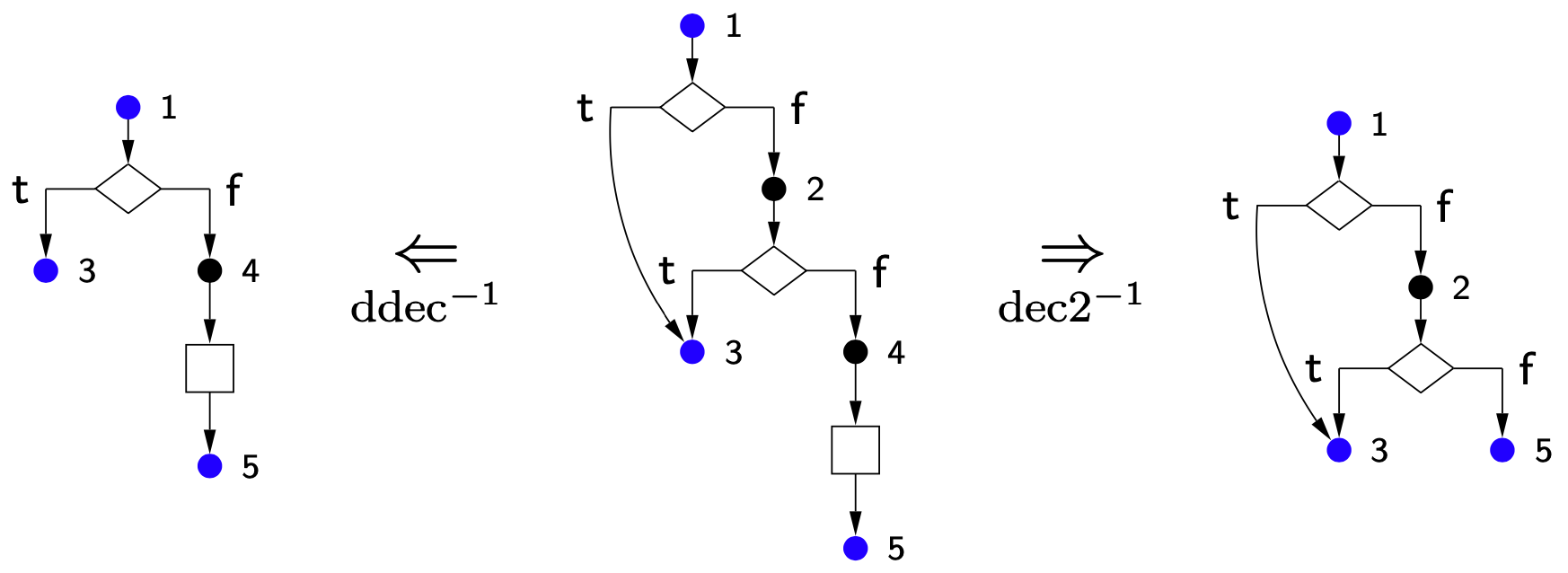}
\caption{EFD critical pair 9}
\label{fig:efd-pair-9}
\end{figure}

\begin{figure}[!ht]
\centering
\includegraphics[totalheight=2.4cm]{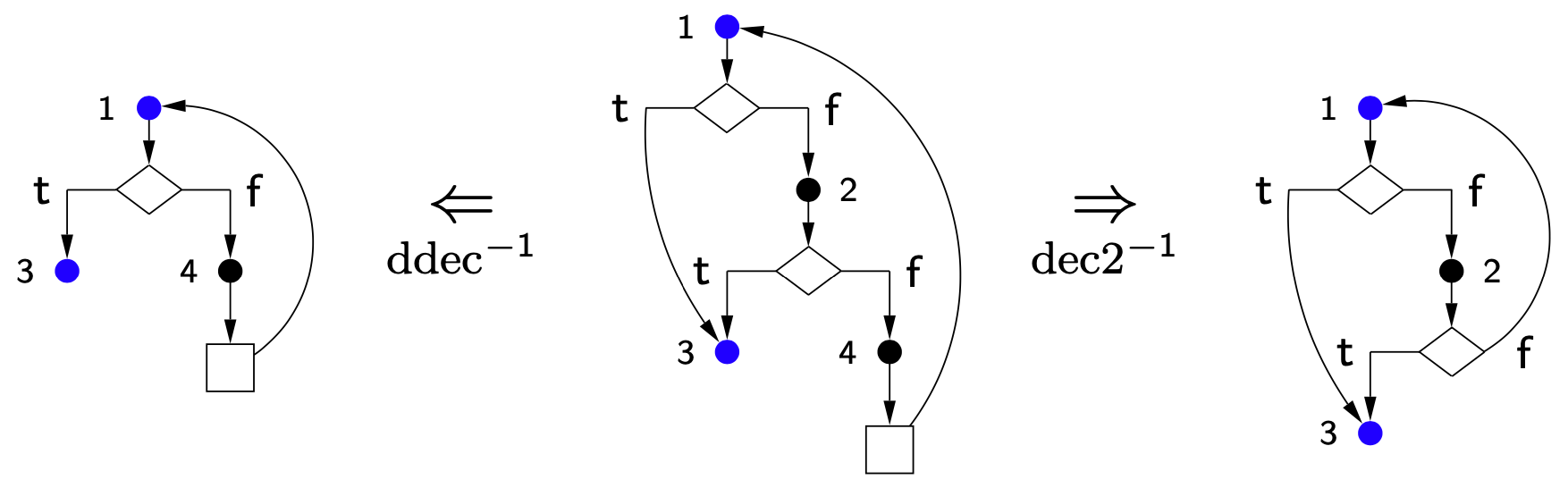}
\caption{EFD critical pair 10}
\label{fig:efd-pair-10}
\end{figure}

\begin{proof}
By Theorem \ref{thm:membershiptest}, \(T\) recognises \(\mathrm{L}(\mathrm{EFD})\) over \(\Sigma\), and one can see that it is terminating since each rule is size reducing. We now proceed by performing critical pair analysis on \(T\).

There are ten non-isomorphic critical pairs:
\begin{enumerate}
    \item The pair exactly as in Figure \ref{fig:efd-pair-1};
    \item The pair in Figure \ref{fig:efd-pair-1} with nodes \(1\) and \(4\) identified;
    \item The pair exactly as in Figure \ref{fig:efd-pair-3};
    \item The pair in Figure \ref{fig:efd-pair-3} with nodes \(1\) and \(5\) identified;
    \item The pair in Figure \ref{fig:efd-pair-3} with nodes \(2\) and \(5\) identified;
    \item The pair exactly as in Figure \ref{fig:efd-pair-6};
    \item The pair exactly as in Figure \ref{fig:efd-pair-7};
    \item The pair in Figure \ref{fig:efd-pair-7} with nodes \(1\) and \(5\) identified;
    \item The pair exactly as in Figure \ref{fig:efd-pair-9};
    \item The pair exactly as in Figure \ref{fig:efd-pair-10};
\end{enumerate}

Pairs 1 through 9 are strongly joinable, and pair 10 is not joinable. Now observe that Lemma \ref{lem:efdcycles} tells us that EFDs cannot contain such cycles. With this knowledge, we define \(\mathcal{D}\) to be all graphs such that directed cycles contain at least one \(t\)-labelled edge (over \(\Sigma\)).

Clearly, \(\mathcal{D}\) is subgraph closed, and then by our Generalised Critical Pair Lemma (Theorem \ref{thm:ngcritpairlem}), we have that \(T\) is locally confluent on \(\mathcal{D}\). Next, it is easy to see that \(\mathcal{D}\) is closed under \(T\), so we can use Generalised Newman's Lemma (Theorem \ref{thm:newmangarbage}) to conclude confluence on \(\mathcal{D}\) and thus, by Lemma \ref{lem:transitiveconfluence}, \(T\) is confluent on \(\mathrm{L}(\mathrm{EFD})\).

Thus, \(T\) is a backtracking-free specification for \(\mathrm{L}(\mathrm{EFD})\) over \(\Sigma\), as required.
\end{proof}

\section{Subcommutativity} \label{sec:subcommuativity}

In this section, we study critical pair analysis with a view to establish subcommutativity up to garbage, rather than confluence up to garbage, as previously in Section \ref{sec:confluence}. We have already introduced subcommutativity in Section \ref{sec:preliminaries} and subcommutativity up to garbage in Section \ref{sec:confluence}.

We start by giving the main result of this section:

\begin{theorem} \label{thm:subcomcritpairlem}
Let \(T = (\Sigma, \mathcal{R})\) and \(\mathcal{D} \subseteq \mathcal{G}(\Sigma)\). If all \(T\)'s \(\mathcal{D}\)-\textit{non-garbage critical pairs} are \textit{strongly subcommutative}, then \(T\) is \textit{subcommutative up to garbage} on \(\mathcal{D}\).
\end{theorem}

\begin{proof}
Easy modification of the proof of the original theorem (Theorem \ref{thm:ngcritpairlem}).
\end{proof}

\begin{corollary}
Let \(T = (\Sigma, \mathcal{R})\) and \(\mathcal{D} \subseteq \mathcal{G}(\Sigma)\). If all \(T\) has no \(\mathcal{D}\)-\textit{non-garbage critical pairs}, then \(T\) is \textit{subcommutative up to garbage} on \(\mathcal{D}\).
\end{corollary}

Just as before, it suffices to only analyse the non-isomorphic critical pairs. A notable difference, however, is that closure and termination are no longer needed to check for joinability, since we are only looking for strong subcommutativity of critical pairs. As noted in Subsection \ref{subsec:confluence_garbage} however, closedness is required in order for subcommutativity up to garbage to imply confluence up to garbage.

Revisiting our examples in Section \ref{sec:languages}, all 4 of the critical pairs of the series-parallel reduction system area actually strongly subcommutative, all 18 non-garbage critical pairs of the labelled series-parallel reduction rules are strongly subcommutative, and so are the 9 non-garbage critical pairs of the extended flow diagram reduction system.

We finish this subsection with a simple example demonstrating termination is not a requirement to establish subcommutative up to garbage.

\begin{example} \label{eg:sub}
Let \(\mathcal{D}\) be the language of discrete graphs and \(T\) be the GT system with the three rules in Figure \ref{fig:eg-sub-rules}. There are two non-isomorphic critical pairs, all of which are garbage (Figure \ref{fig:eg-sub-crit-pairs}), which allow us to immediately conclude subcommutativity of \(T\) up to garbage on \(\mathcal{D}\) (Theorem \ref{thm:subcomcritpairlem}). Notice \(\mathcal{D}\) is closed under \(T\), which means \(T\) is also confluent to garbage on \(\mathcal{D}\) (Lemma \ref{lem:confimpl}).

Notice these rules aren't terminating, even up to garbage on discrete graphs, thus naive machine checking for strong joinability of these pairs would not terminate, however we only need to check for subcommutativity. 
\end{example}

\begin{figure}[!ht]
\centering
\begin{tikzpicture}[every node/.style={align=center}]
    \node (a) at (1.0,-0.05)  {$r_1$:};
    \node (b) at (2.5,0.0)    [draw, circle, thick, fill=black, scale=0.3] {\,};
    \node (c) at (3.5,0.0)    {$\leftarrow$};
    \node (d) at (4.5,0.0)    [draw, circle, thick, fill=black, scale=0.3] {\,};
    \node (e) at (5.5,0.0)    {$\rightarrow$};
    \node (f) at (6.5,0.0)    [draw, circle, thick, fill=black, scale=0.3] {\,};
    \node (g) at (7.5,0.0)    [draw, circle, thick, fill=black, scale=0.3] {\,};

    \node (B) at (2.5,-0.18)  {\tiny{1}};
    \node (D) at (4.5,-0.18)  {\tiny{1}};
    \node (F) at (6.5,-0.18)  {\tiny{1}};

    \node (a) at (1.0,-1.05)  {$r_2$:};
    \node (b) at (2.5,-1.0)   [draw, circle, thick, fill=black, scale=0.3] {\,};
    \node (c) at (3.5,-1.0)   {$\leftarrow$};
    \node (d) at (4.5,-1.0)   {$\emptyset$};
    \node (e) at (5.5,-1.0)   {$\rightarrow$};
    \node (f) at (6.5,-1.0)   [draw, circle, thick, fill=black, scale=0.3] {\,};
    \node (f) at (7.5,-1.0)   [draw, circle, thick, fill=black, scale=0.3] {\,};

    \node (a) at (1.0,-2.05)  {$r_3$:};
    \node (b) at (2.5,-2.0)   [draw, circle, thick, fill=black, scale=0.3] {\,};
    \node (c) at (3.5,-2.0)   {$\leftarrow$};
    \node (d) at (4.5,-2.0)   {$\emptyset$};
    \node (e) at (5.5,-2.0)   {$\rightarrow$};
    \node (f) at (6.5,-2.0)   {$\emptyset$};

    \draw (b) edge[->,thick,loop left] (b);
\end{tikzpicture}
\caption{Rules for Example \ref{eg:sub}}
\label{fig:eg-sub-rules}
\end{figure}
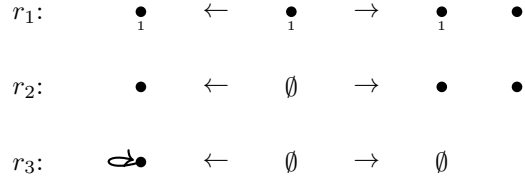

\begin{figure}[!ht]
\centering
\scalebox{0.825}{
\begin{tabular}{|l|c|c|c|c|}
\hline
\multicolumn{1}{|>{\centering\arraybackslash}p{6.25cm}|}{Pair/Property} & \multicolumn{1}{>{\centering\arraybackslash}p{2.5cm}|}{Strongly Subcommutative} & \multicolumn{1}{>{\centering\arraybackslash}p{2.5cm}|}{Non-Garbage} \\ \hline
\scalebox{0.75}{\begin{tikzpicture}[every node/.style={align=center}]
    \node (x) at (-0.5,-1.0) {\,};
    \node (y) at (6.8,0.1)   {\,};

    \node (b) at (0.0,-0.5)  [draw, circle, thick, fill=black, scale=0.3] {\,};
    \node (c) at (1.0,-0.5)  [draw, circle, thick, fill=black, scale=0.3] {\,};
    \node (d) at (2.0,-0.5)  {$\Leftarrow_{r_1}$};
    \node (e) at (3.5,-0.5)   [draw, circle, thick, fill=black, scale=0.3] {\,};
    \node (h) at (5.0,-0.5)  {$\Rightarrow_{r_2}$};
    \node (i) at (6.0,-0.5)   [draw, circle, thick, fill=black, scale=0.3] {\,};
    \node (j) at (7.0,-0.5)  [draw, circle, thick, fill=black, scale=0.3] {\,};

    \node (A) at (0.0,-0.68)   {\tiny{1}};
    \node (E) at (3.5,-0.68)   {\tiny{1}};
\end{tikzpicture}} & \ding{51} & \ding{51} \\ \hline
\scalebox{0.75}{\begin{tikzpicture}[every node/.style={align=center}]
    \node (x) at (-0.5,-1.0) {\,};
    \node (y) at (6.8,0.1)   {\,};

    \node (b) at (0.0,-0.5)  [draw, circle, thick, fill=black, scale=0.3] {\,};
    \node (c) at (1.0,-0.5)  [draw, circle, thick, fill=black, scale=0.3] {\,};
    \node (d) at (2.0,-0.5)  {$\Leftarrow_{r_2}$};
    \node (e) at (3.5,-0.5)   [draw, circle, thick, fill=black, scale=0.3] {\,};
    \node (h) at (5.0,-0.5)  {$\Rightarrow_{r_3}$};
    \node (j) at (6.5,-0.5)  {$\emptyset$};

    \node (A) at (0.0,-0.68)   {\tiny{1}};
    \node (E) at (3.5,-0.68)   {\tiny{1}};

    \draw (b) edge[->,thick,loop left] (b)
          (e) edge[->,thick,loop left] (e);
\end{tikzpicture}} & \ding{55} & \ding{55} \\ \hline
\end{tabular}
}
\caption{Critical pair analysis for Example \ref{eg:sub}}
\label{fig:eg-sub-crit-pairs}
\end{figure}

\section{Conclusion and Future Work} \label{sec:conclusion}

In this paper we have introduced local confluence, confluence, subcommutativity, and termination up to garbage for DPO graph transformation systems, and shown that Newmann's Lemma and Plump's Critical Pair Lemma can be generalised, providing us with checkable conditions for confluence and subcommutativity up to garbage, using only critical pairs. Of course, confluence up to garbage of terminating graph transformation systems is undecidable in general, however, now we can detect more positive cases of confluence up to garbage using non-garbage critical pair analysis, where we previously would have been unable to draw a conclusion due to non-strong joinability of some critical pairs.

We have directly applied our results to recognition of languages, looking specifically at the class of extended flow diagrams and the class of labelled series-parallel graphs. We have backtracking-free algorithms that apply reduction rules as long as possible, with correctness established via non-garbage critical pair analysis. We also anticipate there to be other applications, since there are many other reasons one would want to show confluence up to garbage, such as considering GT systems as computing functions where we restrict the domain \cite{Habel-Muller-Plump01a}. Indeed, one might only be interested in the non-garbage critical pairs themselves, and classification of conflicts \cite{Lambers-Born-Orejas-Struber-Taentzer18a,Lambers-Kosiol-Struber-Taentzer19a}.

\subsection{Generalisations}

Our results also work if we relax the injectivity requirement of the \(K \to R\) morphism in rules. One should note that the two equivalent definitions of parallel independence we have in Subsection \ref{subsec:confluence_checking} were specialised for injective rules only. More generally, two direct derivations \(H_1 \Leftarrow_{r_1,g_1} G \Rightarrow_{r_2,g_2} H_2\) are \textit{parallelly independent} if there are morphisms \(L_1 \to D_2\) and \(L_2 \to D_1\) such that \(L_1 \to D_2 \to G = L_1 \to G\), \(L_2 \to D_1 \to G = L_2 \to G\), \(L_1 \to D_1 \to H_2\) is injective, and \(L_2 \to D_1 \to H_1\) is injective \cite{Habel-Muller-Plump01a}.

Our results also work in the setting of hypergraph transformation, as well as just graph transformation, with almost identical proofs. We think it is extremely likely our results hold for any \(\mathcal{M}\)-adhesive system with the usual restrictions \cite{Ehrig-Golas-Habel-Lambers-Orejas12a}, by modification of the original proof of the Generalised Critical Lemma from Campbell's BSc Thesis \cite{Campbell19a}, which operates by showing completeness of the non-garbage critical pairs. In an unpublished report, we have also shown that these results hold for graph transformation with relabelling \cite{Campbell-Plump19a}, which is important, since the setting is not \(\mathcal{M}\)-adhesive \cite{Habel-Plump12a} and is the graph transformation framework used by GP\,2 \cite{Plump12a,Bak15a}.

\subsection{Future Work}

Confluence analysis of GT systems (and related systems) still remains a generally under-explored area. One obvious piece of future work is to investigate the connection to the work by Lambers, Ehrig and Orejas on \textit{essential critical pairs} \cite{Lambers-Ehrig-Orejas08a} and the continued work by others including Born and Taentzer \cite{Lambers-Born-Orejas-Struber-Taentzer18a}. That said, all of our examples exhibit only essential critical pairs, so non-essential critical pair analysis only has the effect of slowing down the analysis.

It is also not obvious if there is a relation between confluence up to garbage and graphs satisfying negative constraints \cite{Lambers09a}. Moreover, developing a stronger version of the Generalised Critical Pair Lemma that allows for the detection of persistent nodes that need not be identified in the joined graph would allow conclusions of confluence up to garbage where it was previously not determined.

Future work also includes developing further checkable sufficient conditions under which one can decide if a graph is in the subgraph closure of a language, beyond those in Subsection \ref{subsec:critpairgen}. Finally, applying our theory in a rooted context and to GP\,2 is future work \cite{Bak15a}. It is likely that the theory will be applicable there, since program preconditions correspond exactly to non-garbage input, and so it is only natural to be interested in confluence up to garbage, rather than confluence. We would also expect there to be analogues of our results for other kinds of rewriting systems such as string and term rewriting.

\bibliography{ms}

\end{document}